\documentclass[12pt]{article}
\pdfoutput=1
\usepackage{jheppub}
\usepackage{amsmath}
\usepackage{bm}
\usepackage{slashed}
\usepackage{graphicx}
\usepackage[table]{xcolor} 
\usepackage{subcaption}
\usepackage{ytableau}
\usepackage{youngtab}
\usepackage{ytableau}

\newlength{\dummysp}
\newcommand{\beq}{\begin{eqnarray}}
\newcommand{\eeq}{\end{eqnarray}}

\newcommand{\gappeq}{\mathrel{\rlap {\raise.5ex\hbox{$>$}}
{\lower.5ex\hbox{$\sim$}}}}
\newcommand{\lappeq}{\mathrel{\rlap{\raise.5ex\hbox{$<$}}
{\lower.5ex\hbox{$\sim$}}}}

\newcommand{\ben}{\begin{enumerate}}
\newcommand{\een}{\end{enumerate}}

\newcommand{\bit}{\begin{itemize}}
\newcommand{\eit}{\end{itemize}}

\def\[{\left [}
\def\]{\right ]}
\def\({\left (}
\def\){\right )}

\def\Z{{\mathbb Z}}

\title{Anomalies on ALE spaces and phases of gauge theory}
   
 \author{Mohamed M. Anber}
  
\affiliation{Centre for Particle Theory, Department of Mathematical Sciences, Durham University, South Road, Durham DH1 3LE, UK}

\emailAdd{mohamed.anber@durham.ac.uk}  

\abstract{

{\flushleft{W}}e show that certain ’t Hooft anomalies not detected by standard closed four-dimensional probes can become visible when a quantum field theory is placed on asymptotically locally Euclidean (ALE) spaces. As a concrete example, we use the Eguchi--Hanson (EH) space, whose defining features are its nontrivial second cohomology generated by the self-intersecting two-sphere and its asymptotic boundary $\mathbb{RP}^3$, which carries torsion and thus furnishes additional cohomological data absent on conventional backgrounds. For a theory with symmetry $G_1\times G_2$, we turn on background flux for $G_1$ and probe potential anomalies by performing a global $G_2$ transformation; the resulting anomaly is captured by a five-dimensional mapping torus. The anomaly receives contributions from the four-dimensional characteristic classes on EH space as well as from the $\eta$-invariant associated with the $\mathbb{RP}^3$ boundary. The anomaly detected in this way imposes additional constraints on asymptotically free gauge theories with fixed trivial asymptotic boundary conditions. In particular, infrared composite spectra that match anomalies on standard closed manifolds may nevertheless fail to reproduce the EH anomaly, and thus cannot by themselves furnish a complete symmetry-preserving infrared realization.
}

\begin{document}

\maketitle

\flushbottom

\section{Introduction}
’t~Hooft anomalies are obstructions to promoting global symmetries of a quantum field theory (QFT) 
to gauge symmetries. Because they remain invariant under renormalization group flow, they provide 
powerful, exact constraints on the nonperturbative dynamics of a theory. In practice, anomalies are 
essential for understanding how symmetries are realized in the infrared regime of asymptotically free 
theories, for establishing and testing dualities, and for characterizing topological phases of 
matter. With the formulation of generalized symmetries \cite{Gaiotto:2014kfa}, ’t~Hooft anomalies associated with 
higher-form and noninvertible symmetries have become standard tools: they allow one to derive genuinely 
new predictions and to recast many classic results in a more robust, symmetry-centric framework; see \cite{Brennan:2023mmt,Shao:2023gho} for reviews.

Anomalies appear as non-trivial phases in the partition function when the 
background gauge fields for a global symmetry $G$ undergo gauge transformations. Such anomalies 
 are captured by topological terms in one 
higher dimension. Mathematically, these topological terms are elements of appropriate cohomology 
groups or, more generally, generalized cohomology (bordism) groups.  
For a $4$-dimensional QFT with global symmetry $G$, the anomaly is classified by the bordism group
$
\Omega_{5}^{\mathcal S}(BG)
$, where $BG$ is the classifying space of $G$ and $\mathcal S$ specifies the relevant tangential 
structure (such as $\mathrm{Spin}$, $\mathrm{Spin}^c$,  or $\mathrm{Pin}^{\pm}$) \cite{Dai:1994kq,Witten:2019bou}.

In concrete theories, one probes these anomalies by placing
the $4$-dimensional QFT on suitably chosen manifolds $X_{4}$ that admit nontrivial
background gauge fields, and by studying how the partition function transforms under
gauge transformations or global symmetry operations.
Standard examples of $X_{4}$ include spin manifolds such as $S^{4}$ (the one–point compactification of 
$\mathbb{R}^{4}$), $S^{2}\!\times\! S^{2}$, and $\mathbb{T}^{4}$.  
The latter two manifolds possess nonvanishing second cohomology and therefore permit fractional 
background fluxes for $1$-form symmetries.  
One may also employ non-spin manifolds such as $\mathbb{CP}^{2}$.  
Although $\mathbb{CP}^{2}$ does not admit a spin structure, an appropriate fractional monopole flux 
can be introduced to render fermions globally well-defined, and its nontrivial second cohomology 
likewise enables one to probe anomalies involving $1$-form symmetries.  
By placing the theory on $S^4$, $\mathbb{T}^{4}$, $S^{2}\!\times\! S^{2}$, and $\mathbb{CP}^{2}$ and turning 
on the most general fluxes compatible with the matter content, one can detect a wide class of 
generalized ’t~Hooft anomalies associated with both $0$-form and $1$-form global symmetries \cite{Anber:2019nze,Anber:2020gig}.

In this work, we show that certain anomalies, which may not be detected on the closed manifolds mentioned above, become visible once the theory is placed on asymptotically locally Euclidean (ALE) spaces \cite{Kronheimer:1989zs,PeterBKronheimer:1990zmj}. An ALE space is a complete, non-compact Riemannian 4-manifold $({\cal M},g)$ such that, outside a compact set, ${\cal M}$ is diffeomorphic to $(\mathbb{R}^4 \setminus B)/\Gamma$ for some finite $\Gamma \subset SO(4)$ acting freely on $S^3$, and in the corresponding coordinates at infinity the metric approaches the flat metric, $g_{ij} = \delta_{ij} + \mathcal{O}(r^{-\tau})$, $\partial_k g_{ij} = \mathcal{O}(r^{-1-\tau})$ as $r \to \infty$ for some $\tau>0$. We solely focus on the simplest example, the Eguchi--Hanson (EH) space \cite{Eguchi:1978xp,Eguchi:1978gw}.

The EH manifold (instanton) is (anti)-self-dual solution to the vacuum Euclidean Einstein equations. It is a simply connected, spin manifold, with second cohomology 
$H^{2}(\mathrm{EH},\mathbb{Z}) \cong \mathbb{Z}$ generated by the exceptional $2$-sphere (bolt), whose 
self-intersection number is $-2$.  It also supports a unique (anti)-self-dual harmonic $2$-form, enabling us to turn on a background $U(1)$ gauge field localized at the bolt and decays rapidly at infinity. 
For any integer charge $q_{\Psi}\in\mathbb{Z}$, there exists a globally defined line bundle 
$L^{q_{\Psi}}$ associated to  $U(1)$, and Dirac fermions may be defined as 
sections of $S\otimes L^{q_{\Psi}}$, where $S$ denotes the spin bundle.  

Although the bulk cohomology of EH space is torsion-free, its asymptotic boundary 
$\partial\mathrm{EH}=S^3/\mathbb Z_2 \simeq \mathbb{RP}^{3}$ carries torsion 
$H_{1}(\mathbb{RP}^{3},\mathbb{Z}) \cong \mathbb{Z}_{2}$.  
As a consequence, the bulk flux seen by a charged fermion determines a $\mathbb{Z}_{2}$ holonomy at 
infinity.  
An EH instanton with background flux resembles a BPST instanton in that it localizes the field that decays to the vacuum at infinity. However, unlike a BPST instanton, it also enjoys a refined cohomology: the interplay between the nontrivial bulk $H^{2}$ and the torsion boundary data distinguishes EH from 
the standard closed $4$–manifolds commonly used for anomaly detection.

We consider a QFT with microscopic gauge group $\mathcal{G}$ and global symmetry group $G_{\rm global}$.
For the purposes of our anomaly, we single out a subgroup
$
  G_1 \;\subset\; \mathcal{G} \times G_{\rm global}
$, 
which may contain both (subgroups of) the microscopic gauge group and purely global factors, and an abelian
global symmetry $G_2 \subset G_{\rm global}$. 
For instance, in a vector-like gauge theory, one may choose 
$G_{1}\equiv SU(N)\times U(1)/\mathbb{Z}_{p}$, where $SU(N)$ is the gauge group and $U(1)$ is 
the baryon number symmetry and $\mathbb Z_p$ is a subgroup of the center $\mathbb Z_N$, which depends on the fermion representation ${\cal R}$ under $SU(N)$, while $G_2$ is a discrete chiral symmetry acting on the fermions. Because the EH space has nontrivial second cohomology, we can introduce background 
fluxes along the Cartan generators of $G_{1}$, with matter charges valued modulo $\mathbb{Z}_{p}$.  These background fields are localized at the bolt and decay rapidly to the vacuum (the bundles become flat) at the boundary of EH space, while the holonomy at the boundary as seen by a Dirac fermion depends on the representation ${\cal R}$.
To test for a mixed ’t~Hooft anomaly, we place the theory on the EH background, turn on these 
$G_{1}$ fluxes, and then perform a $G_{2}$ transformation.  
The resulting phase in the partition function diagnoses a mixed 
$G_{1}$–$G_{2}$–gravity anomaly.

To compute the anomaly, we construct the five-dimensional mapping torus 
$Y_5 = \mathrm{EH} \times [0,1]/r$, where $r$ is the identification
$(x,1) \sim (u\cdot x,0)$, which implements a $G_2$ symmetry twist by 
$u = e^{i2\pi\beta_\tau} \in G_2$.  Here $x = (p, \mathcal{A}_{G_1}, \mathcal{B}_{G_2})$, 
with $p \in \mathrm{EH}$ a spacetime point, and $\mathcal{A}_{G_1}$ and 
$\mathcal{B}_{G_2}$ denoting the background bundle data for the symmetries 
$G_1$ and $G_2$, respectively.  
The anomaly of the $4$-D QFT is given by the 
$\eta$-invariant of the Dirac operator on $Y_5$, 
$\mathcal{A} = e^{i2\pi\,\eta(\slashed D_{Y_5})}$.  
A direct evaluation of the $\eta$-invariant yields 
$\mathcal{A} = e^{-i2\pi\beta_\tau\,\mathcal{I}^{(\mathrm{EH})}}$, 
where $\mathcal{I}^{(\mathrm{EH})}$ is the number of normalizable fermion 
zero modes on EH space in the background of the $G_1$ fluxes.

To compute $\mathcal{I}^{(\mathrm{EH})}$, we employ three complementary methods that all yield the same results.  
First, we solve the Dirac equation on the EH background in the presence of the $G_{1}$ 
fluxes and directly count the normalizable zero modes.  Such zero modes are all localized near the bolt and decay rapidly at infinity. 
Second, we use the Atiyah–Patodi–Singer (APS) index theorem, which expresses the number of zero modes 
in terms of bulk characteristic classes together with the $\eta$-invariant associated with the 
$\mathbb{RP}^{3}$ boundary of the EH space.  This boundary $\eta$-invariant encodes torsion-sensitive global information and is what allows EH backgrounds to detect anomaly data not seen by the standard closed-manifold probes mentioned above. 

Both of the above-mentioned approaches require constructing the weights of the fermion representation 
$\mathcal{R}$ explicitly.
To avoid this step and to circumvent the need for an explicit computation of the $\eta$-invariant, in a few cases, we can also employ a third method based on a capping construction: we glue the EH space 
to an orientation-reversed copy $\mathrm{EH}'$ carrying suitable background fluxes, chosen such that 
the resulting closed manifold is smooth and the number of the zero modes on the original EH space is preserved.

The index $\mathcal{I}^{(\mathrm{EH})}$ is a
topological invariant\footnote{This is the case since the boundary Dirac operator defined on $\mathbb{RP}^3$ does not admit zero modes; see the discussion in Section \ref{The Atiyah--Patodi--Singer (APS) index theorem}. Then, the index $\mathcal{I}^{(\mathrm{EH})}$ does not change as we vary the bulk metric and the gauge fields.} since it depends only on the topological data of the
EH space and the background bundles, and is insensitive to any
smooth deformation of the geometry (including varying the bolt size) or of the
gauge fields that preserve the prescribed asymptotic structure. Then, the anomaly
$
\mathcal{A}=e^{-\,i2\pi\beta\,\mathcal{I}^{(\mathrm{EH})}}
$
cannot be removed by local counterterms. 

We define the QFT on EH space by fixing the $3$-D boundary theory at asymptotic infinity to be trivial, and all anomaly constraints are derived within this asymptotic definition of the theory.
Because anomalies are invariant under renormalization-group flow, any proposed
infrared (IR) scenario must reproduce exactly the same anomaly when placed in
the same ultraviolet (UV) background configuration.
Since the key point is that the UV definition of the theory requires the gauge fields
to become flat in the asymptotic region of the EH space, we assume that no $3$-D topological sector or other independent degrees of freedom are dynamically generated on the asymptotic $\mathbb{RP}^3$ boundary in the IR. Equivalently, both the UV and IR are defined with the same trivial $3$-D boundary theory. This is a reasonable assumption since the IR theory is not expected to generate new physical degrees of
freedom localized near the boundary at infinity: such states would have no
natural origin in a local $4$-dimensional quantum field theory and would be
incompatible with the prescribed UV asymptotics.  Therefore, {\em under this assumption}, the EH anomaly
must be saturated entirely by the genuine IR degrees of freedom supported in
the interior of the spacetime.  Within this asymptotic definition of the theory, this requirement constrains admissible IR phases and rules out scenarios in which the bulk IR degrees of freedom fail to reproduce the EH anomaly.

We demonstrate, via several examples, that the EH anomaly imposes new constraints on the IR spectrum of both vector-like and chiral gauge theories. We first apply this anomaly to vector-like theories with discrete chiral symmetry. Under the common assumption that such theories generate a condensate, the EH anomaly leads to cases in which the entire discrete chiral symmetry is forced to break down to fermion number.
We then analyze a broad class of theories (vector-like and chiral) that admit massless composite fermions in the infrared, in the sense that the composites match 't Hooft anomalies on standard closed manifolds while preserving the global symmetries. Using the EH anomaly as a diagnostic, we find that these putative composites often fail to reproduce the EH anomaly and therefore cannot by themselves constitute the correct low-energy spectrum. Moreover, when combined with the usual anomalies computed on standard closed $4$-manifolds, the EH anomaly yields additional, more stringent constraints on the set of admissible IR degrees of freedom.

A particularly illuminating example is provided by the vector-like $SU(5)$ gauge theory with a single Dirac fermion in the $2$-index antisymmetric representation, a theory that possesses a $(U(1)\times\mathbb{Z}_3)/\mathbb Z_5$ global symmetry. In this case, one can compare the anomaly detected on the EH background with the anomaly constraints encoded by the five-dimensional spin bordism group for $U(5)\times \mathbb{Z}_3$ backgrounds. The candidate infrared composite fermions match the anomalies captured by these closed-manifold cobordism invariants, but they nevertheless fail to reproduce the anomaly phase on the EH space. This suggests that the EH background can impose a stronger anomaly constraint than the closed-manifold probes. Assuming trivial asymptotic boundary conditions, the example therefore illustrates how EH anomalies can further restrict candidate symmetry-preserving infrared phases.

We emphasize that the EH anomalies identified in this work arise from explicit examples rather than from a complete classification of anomalies on ALE spaces. Our analysis shows that EH backgrounds can reveal anomaly constraints beyond those detected by the conventional closed-manifold probes, but it does not amount to a systematic classification of all such anomalies or of all possible ALE backgrounds on which they may be realized. Developing a broader framework for classifying anomalies on noncompact spaces of this type, and clarifying their relation to conventional cobordism-based classifications, remains an interesting problem for future work.

This paper is organized as follows. Section~\ref{EHSPACEALLTHAT} contains all of the geometric and field–theoretic material required for the anomaly computation. We begin with a brief review of the Eguchi--Hanson (EH) space, after which we describe how to turn on background fluxes for an $SU(N)\times U(1)/\mathbb{Z}_p$ bundle. We then analyze fermions transforming in a general representation ${\cal R}$ of $SU(N)$ in this background, together with the global consistency conditions they must satisfy. Subsequently, we solve the Dirac equation to obtain the fermion zero modes in the presence of the background flux. We show that the counting of zero modes can also be obtained from the Atiyah–Patodi–Singer (APS) index theorem. The APS index involves an $\eta$-invariant, whose computation requires the full set of weights of ${\cal R}$. To avoid this complication, we conclude the section by introducing a method for capping the EH space, which in a few cases allows us to compute the index without explicitly enumerating all the weights.
In Section~\ref{The EH space as a probe of anomalies}, we evaluate the anomaly on the EH space using both the mapping-torus construction and standard field-theoretic techniques, demonstrating that the two approaches agree. We also discuss how this anomaly differs from those obtained on conventional closed manifolds, we give a qualitative description of how the anomaly can be saturated as we vary the bolt size, and conclude the section by giving a Hilbert-space interpretation of the anomaly.
Finally, Sections~\ref{ApplicationsI} and~\ref{ApplicationII} illustrate applications of the EH anomaly in various vector-like and chiral gauge theories. In particular, we show how the EH anomaly constrains patterns of chiral symmetry breaking and the structure of infrared composites. These sections also provide numerous concrete examples that clarify the abstract ideas developed earlier in the paper. In particular, in Section~\ref{More on the EH anomaly in the 2-index antisymmetric theory}, we examine the cobordism anomalies of the $SU(5)$ vector-like gauge theory with a Dirac fermion in the $2$-index antisymmetric representation, and show that while these anomalies can be matched by candidate infrared composites, the same composites nevertheless fail to reproduce the EH anomaly. We conclude in Section \ref{Future directions} by offering directions for future studies. 

Many technical details are relegated to the appendices. Appendix~\ref{app:LieAlgebra} reviews our notation and basic facts about Lie algebras, while Appendix~\ref{secverma-bases} briefly introduces the Verma--basis method used to construct the weights of arbitrary representations of $SU(N)$. In Appendix~\ref{sec:fermions-EH} we solve the Dirac equation on EH space in the presence of a background $U(1)$ gauge field and explicitly construct the corresponding zero modes. Appendix~\ref{anomaliesonT4CP2} reviews the anomalies on $\mathbb{T}^4$ and $\mathbb{CP}^2$ which, together with the EH anomaly, are used to constrain the dynamics of the gauge theories studied in the main text. Finally, Appendix~\ref{app:bordismU5Z3} computes the bordism group $\Omega^{\rm spin}(B(U(5)\times Z_3))$ relevant to the $SU(5)$ example discussed in Section~\ref{More on the EH anomaly in the 2-index antisymmetric theory}.

\section{Eguchi--Hanson space, background fluxes, and fermions}
\label{EHSPACEALLTHAT}

\subsection{Eguchi--Hanson space: a crash course}
Eguchi--Hanson (EH) instantons are (anti-)self-dual solutions of the Euclidean Einstein equations with vanishing cosmological constant~\cite{Eguchi:1978xp,Eguchi:1978gw}.  They are hyperk\"ahler
$4$-manifolds that resolves the $\mathbb{C}^2/\mathbb{Z}_2$ orbifold
singularity.
The EH line element can be expressed elegantly using the left-invariant $1$-forms 
$\{\sigma_x,\sigma_y,\sigma_z\}$ on $S^3\cong SU(2) $: 
\begin{equation}\label{EH_metric}
ds^2_{\text{EH}} \;=\; f^2(r)\,dr^2 \;+\; r^2\bigl(\sigma_x^2+\sigma_y^2+f^{-2}(r)\,\sigma_z^2 \bigr),
\end{equation}
where the deformation factor is
\begin{equation}\label{f_factor}
f^{-2}(r) \;=\; 1 - \left(\frac{a}{r}\right)^4.
\end{equation}
The left-invariant $1$-forms are given in terms of Hopf coordinates $(\theta,\varphi,\psi)$ by
\begin{align}\label{sigmas}\nonumber
\sigma_x &= \tfrac{1}{2}\bigl(\sin\psi\, d\theta - \sin\theta\cos\psi\, d\varphi \bigr), \\\nonumber
\sigma_y &= \tfrac{1}{2}\bigl(-\cos\psi\, d\theta - \sin\theta\sin\psi\, d\varphi \bigr), \\
\sigma_z &= \tfrac{1}{2}\bigl(d\psi + \cos\theta\, d\varphi \bigr).
\end{align}
Here, $r$ is the radial coordinate, and $a>0$ is a parameter of dimension length which resolves the geometry.  
For $a=0$, one recovers flat $\mathbb{R}^4$, while for $a>0$ a smooth $2$-sphere (the ``bolt'', or the exceptional $2$-sphere) emerges at $r=a$, making the space geodesically complete.  
The coordinate ranges are
\begin{equation}
a \leq r < \infty, \qquad 
0 \leq \theta \leq \pi, \qquad 
0 \leq \varphi < 2\pi, \qquad 
0 \leq \psi < 2\pi.
\end{equation}
In particular, the $\psi$ coordinate has period $2\pi$, i.e. half the usual $4\pi$ period on $S^3$.  
This reduction ensures regularity at $r=a$ and implies that the EH space is asymptotically locally Euclidean (ALE):  
\begin{equation}
\partial(\text{EH}) = S^3/\mathbb{Z}_2 \;\simeq\; \mathbb{RP}^3.
\end{equation}
Thus, at infinity the EH manifold approaches $\mathbb{R}^4/\mathbb{Z}_2$, making it the simplest example of a nontrivial ALE space.

The Euler characteristic $\chi$ and Hirzebruch signature $\tau$ of the EH space can be expressed using the Atiyah--Patodi--Singer (APS) formulas\footnote{\label{conventionfootnote} In this work, we mostly use the notation and conventions of \cite{Eguchi:1980jx}. However, we note that our convention for the Euler characteristic differs by an overall minus sign from that used in \cite{Eguchi:1980jx}. This difference arises because the orientation of our tangential frame is opposite to the orientation of the coordinate frame $(r,\theta,\varphi,\psi)$. Our choice matches the convention adopted in \cite{Anber:2025gvb} and has the practical advantage of avoiding the need to carry an overall minus sign when applying the index theorem. With our orientation, the signature is $\tau = 1$ and the first Pontryagin number is $p = 3$, in contrast with $\tau = -1$ and $p = -3$ obtained in \cite{Eguchi:1978xp,Eguchi:1978gw} where the coordinate and orthonormal frames are taken to have the same orientation. In our convention, the topological charge of the $U(1)$ gauge field (\ref{Q_def}) is positive, while the gravitational topological charge (\ref{QGCHARGE}) is negative. Again, these signs would have been reversed had we adopted a convention where both the orthonormal and coordinate frames had the same orientation. For a detailed description of the geometry of the EH space, we refer the reader to Appendix~A of \cite{Anber:2025gvb}.}:
\begin{eqnarray}\label{TINV}\nonumber
\chi
&=&- \frac{1}{32\pi^2}\int_{\mathrm{EH}} \epsilon_{abcd}\, R_{ab}\wedge R_{cd}
\;+\; \frac{1}{32\pi^2}\int_{\mathbb{RP}^3} \epsilon_{abcd}\!
\left( 2\,\Theta_{ab}\wedge R_{cd}
- \frac{4}{3}\,\Theta_{ab}\wedge \Theta_{ce}\wedge \Theta_{ed} \right),\\[6pt]
\tau
&=& -\frac{1}{24\pi^2}\int_{\mathrm{EH}} \mathrm{tr}\, R\wedge R
\;+\; \frac{1}{24\pi^2}\int_{\mathbb{RP}^3} \mathrm{tr}\, \Theta\wedge R
\;-\; \eta_s(\mathbb{RP}^3),
\end{eqnarray}
where the Latin indices $a,b,c,d$ run over $0,1,2,3$ and denote the components in the locally flat (tangent) frame, $R$ is the curvature $2$-form, $\Theta$ is the second fundamental form, and $\eta_s(\mathbb{RP}^3)$ denotes the APS $\eta$-invariant of the signature operator on the boundary, which vanishes for EH space. 
Evaluating these expressions for the EH geometry gives
$
\chi = 2, \tau = 1
$ \cite{Eguchi:1978xp,Eguchi:1978gw}. 
The Euler characteristic $\chi=2$ is consistent with the presence of a single bolt $2$-sphere. The Hirzebruch signature $\tau$ measures the difference between the number of anti-self-dual and self-dual harmonic $2$-forms, which in the EH case means there is a unique harmonic $2$-form. It is given by  
\begin{equation}\label{form of K}
{\cal K}=d\left(\frac{\sigma_z a^2}{2\pi r^2}\right)=-\frac{a^2}{2\pi r^3}dr\wedge (d\psi+\cos\theta d\varphi)+\frac{a^2}{4\pi r^2}\sin\theta d\varphi\wedge d\theta\,,
\end{equation} where the choice of the numerical coefficient $1/(2\pi)$ fixes the normalization of ${\cal K}$ such that
\begin{equation}\label{K_norm}
   \int_{S^2} {\cal K} \;=\; 1,
\end{equation}
and $S^2$ is the bolt of the EH space spanned by the coordinates $\theta,\varphi$.  
One can embed an abelian gauge field into this $2$-form by taking the $U(1)$ 
field strength to be
\begin{equation}\label{F_def}
   F_{(1)} \;=\; -\,2\pi\,{\cal C}\,{\cal K},
\end{equation}
with ${\cal C}$ a constant. From (\ref{K_norm}), we readily find that the field strength $F_{(1)}$ is localized at the bolt and decays rapidly, and it becomes a flat bundle at the boundary. One can easily verify that  $F_{(1)}$  satisfies 
\begin{eqnarray}
  \int_{S^2} F_{(1)}=2\pi {\cal C}\,,
\end{eqnarray}
imposing the quantization condition on ${\cal C}\in \mathbb Z$. Due to the self-duality of $F_{(1)}$, the gauge field does not backreact on the geometry: thus, the EH metric accompanied by the gauge field satisfies the Einstein-Maxwell equations.

Consider now a small deformation $S^{2\prime}$ of the bolt which intersects $S^2$ transversely.  
The self-intersection number of the bolt is encoded in the wedge product of ${\cal K}$,  
\begin{equation}\label{K_wedge_K}
   \int_{\text{EH}} {\cal K}\wedge{\cal K} \;=\; \tfrac{1}{2},
\end{equation}
consistent with the known result that the bolt has self-intersection number\footnote{We note that the factor of $1/2$ in (\ref{K_wedge_K}) arises from the chosen parametrization of the EH metric and the normalization \eqref{K_norm}.} $(-2)$ 
\cite{Yuille:1987vw,Kronheimer:1989zs,Bianchi:1996zj}.
The associated topological charge of the abelian field is then (see Footnote \ref{conventionfootnote})
\begin{equation}\label{Q_def}
   Q_{(1)} \;=\; \frac{1}{8\pi^2}\int_{\text{EH}} F_{(1)}\wedge F_{(1)}
   \;=\; \tfrac{1}{4}\,{\cal C}^2,
\end{equation}
which is fractional. The extra factor of $\tfrac{1}{4}$ serves as a \emph{refinement}: it lowers the Dirac index and accordingly modifies the anomaly phase. As we will show, this refinement imposes stringent constraints on the IR chiral condensates needed for anomaly matching\footnote{For comparison, on $\mathbb{CP}^2$, a nonspin manifold, one finds instead $\int_{\mathbb {CP}^1\subset\mathbb{CP}^2}{\cal K}=1$,
$\int_{\mathbb{CP}^2}{\cal K}\wedge{\cal K}=1$, and thus, $Q_{(1)} =\tfrac{1}{2}\,{\cal C}^2$;  see Appendix \ref{anomaliesonT4CP2}  and reference
\cite{Anber:2020gig}.}.

In summary, the EH space supports smooth, self-dual $U(1)$ gauge fields constructed from 
its unique normalizable self-dual harmonic $2$-form. These gauge backgrounds leave the geometry 
unchanged, yet they carry nontrivial flux and topological charge, and will be crucial in 
determining the boundary conditions and fermionic spectrum on the EH background. Readers who are interested in a more detailed description of the geometry are referred to the original literature \cite{Eguchi:1978xp,Eguchi:1978gw} or Appendix~A of \cite{Anber:2025gvb}.

\subsection{General background fluxes}

We now turn to the construction of general background gauge fields, specifically, $SU(N)\times U(1)/\mathbb Z_p$ bundles, with $p = \mathrm{gcd}(N,n)$ and $n$ to be defined below, localized at the bolt of the EH space. 
A key ingredient is the unique normalizable self-dual harmonic $2$-form, ${\cal K}$, which allows us to turn on a background gauge field without backreacting on the geometry. 
In this construction, we restrict to gauge fields aligned with the Cartan subalgebra of $SU(N)$ as well as with the abelian $U(1)$ factor. In the examples we consider in this paper, the use of both gauge fields is enough to render the bundles consistent in the presence of fermions charged under both $SU(N)$ and $U(1)$.  The corresponding gauge connections  take the form
\begin{equation}\label{mainfieldsfun}
A_{SU(N)} \;=\; m_{(N)}\, \bm H_{\Box}\!\cdot\!\bm \nu_{a} \; \frac{\sigma_z\, a^2}{r^2},
\qquad 
A_{U(1)} \;=\; k\, \frac{\sigma_z\, a^2}{r^2},
\end{equation}
where we used the defining representation to express the gauge connection of $SU(N)$.
Here, $m_{(N)}$ is an arbitrary integer, which must be chosen carefully so that the correct quantization conditions of the fields are respected on the $S^2$ bolt. In most cases, we can set $m_{(N)}=1$.  The vector matrices $\bm H_{\Box}=(H^1,H^2,\ldots,H^{N-1})_{\Box}$ denotes the Cartan generators of $SU(N)$ in the defining representation; each $H_{\Box}^i$, $i=1,2,..,N-1$, is a diagonal $N\times N$ matrix consisting of the weights $\nu_{\,a}^i\,, a=1,2,..,N$ in the defining representation. Alternatively, we write the weights of the defining representation as the vectors $\bm \nu_{a}$ $(a=1,2,\dots,N)$. Without loss of generality, we set $a=1$ in (\ref{mainfieldsfun}) and identify $\bm \nu_1$ with the fundamental weight, $\bm \nu_1=\bm w_1$, and hence we find that the gauge field $A_{SU(N)}$ is expressed as the diagonal $N\times N$ matrix  (the Lie algebra and conventions are reviewed in Appendix \ref{app:LieAlgebra}):
\begin{eqnarray}\label{ASUNCBULK}
A_{SU(N)} \;=\; m_{(N)}\, \mbox{diag}_{N}\left(1-\frac{1}{N},-\frac{1}{N},...,-\frac{1}{N}\right) \frac{\sigma_z\, a^2}{r^2}\,,
\end{eqnarray}
and we used the identity $\bm \nu_a\cdot\bm \nu_b=\delta_{ab}-\frac{1}{N}$ (we use a normalization such that the simple roots satisfy $\bm \alpha^2=2$). Furthermore, we shall argue below (when studying the global consistency conditions of Dirac fermions) that $k$ that appears in $A_{U(1)}$ must be taken to be  $k=m_{N}\frac{n}{N} + {\cal C}$, and ${\cal C}$ is an arbitrary integer. Thus,
\begin{eqnarray}\label{abelian background needed}
A_{U(1)} =\left(m_{(N)}\frac{n}{N} + {\cal C}\right) \frac{\sigma_z\, a^2}{r^2}\,.
\end{eqnarray}
Here, $n$ is a positive integer defined mod $N$, and we shall show below that it is the $N$-ality of the fermion representation.

 Borrowing the result (\ref{F_def}), we find that the field strengths $F_{(N)}=dA_{SU(N)}$ and $F_{(1)}=dA_{U(1)}$ satisfy the following quantization conditions when integrated over the EH bolt:
\begin{eqnarray}\label{fractional fluxes}
\int_{S^2}F_{(N)}\in \frac{2\pi\mathbb Z^{N-1}}{N}\,,\quad \int_{S^2}F_{(1)}\in \frac{2\pi n\mathbb Z}{N}\,,
\end{eqnarray}
with topological charges given by
\begin{eqnarray}\label{SUNU1BTC}\nonumber
Q_{(N)}&=&\frac{1}{8\pi^2}\int_{\text{EH}}\mbox{tr}_\Box F_{(N)}\wedge F_{(N)}=\frac{m_{(N)}^2}{4}\left(1-\frac{1}{N}\right)\,,\\
Q_{(1)}&=&\frac{1}{4}\left(m_{(N)}\frac{n}{N}+{\cal C}\right)^2\,.
\end{eqnarray}

Since we shall consider fermions in an arbitrary representation ${\cal R}$, which can be represented by a $\dim{\cal R}\times 1$ coloumn vector (see Eq. \ref{the dirac coloumn}), it is convenient to also cast the gauge connection $A_{SU(N)}$ using the weights of ${\cal R}$. One replaces (\ref{mainfieldsfun}) with
\begin{equation}\label{mainfields}
A_{SU(N)} \;=\; m_{(N)}\, \bm H_{\cal R}\!\cdot\!\bm \nu_{a} \; \frac{\sigma_z\, a^2}{r^2},
\qquad 
A_{U(1)} \;=\;k\, \frac{\sigma_z\, a^2}{r^2},
\end{equation}
so that a fermion $\Psi$ in representation ${\cal R}$ couples via the covariant derivative to the total gauge field
\begin{equation}\label{atotal}
A_t \;=\; A_{SU(N)} \,+\, A_{U(1)}\, I_{\dim{\cal R}},
\end{equation}
with $A_{SU(N)}$ given by (\ref{mainfields}).
Now, $\bm H_{\cal R}=(H^1,H^2,\ldots,H^{N-1})_{\cal R}$ denotes the Cartan generators of $SU(N)$ in representation ${\cal R}$; each $H_{{\cal R}}^i$, $i=1,2,..,N-1$, is a diagonal $\dim{\cal R}\times \dim{\cal R}$ matrix consisting of the weights $\mu_{{\cal R}\,a}^i\,, a=1,2,..,\dim{\cal R}$, in representation ${\cal R}$\footnote{We note that in what follows, we will make use of both representations of the $SU(N)$ gauge connection, the defining representation
given in~\eqref{mainfieldsfun} and the general representation given in~\eqref{mainfields}, and switch between them as needed, 
including in the discussion of fermions in a general representation ${\cal R}$. 
At each step, we will indicate explicitly which representation is being used.}.

To compute the weights $\bm \mu_{{\cal R}\,a}$ of a representation ${\cal R}$, we employ the method of Verma basis \cite{:/content/aip/journal/jmp/27/3/10.1063/1.527222}, reviewed in Appendix~\ref{secverma-bases}.  
This construction provides a systematic procedure to generate all weights, starting from the highest weight $\bm\mu_h$ of the representation ${\cal R}$.  
The highest weight is expressed in terms of the Dynkin labels $(m_1,m_2,\ldots,m_{N-1})\in\mathbb{Z}_{\geq 0}^{\,N-1}$ as
\begin{equation}\label{hiestwe}
\bm\mu_h = \sum_{a=1}^{N-1} m_a \bm w_a,
\end{equation}
where $\bm w_a$ denote the fundamental weights of $su(N)$ algebra.  
From $\bm\mu_h$, the full weight system is obtained by successive subtractions of simple roots, following the Verma-basis construction, which can be easily algorithmically implemented.  

We now turn to the determination of the constant $k$ that multiplies $A_{U(1)}$ in (\ref{mainfields}). 
Our starting point is the relation between the fundamental weights $\bm w_a$ and the weights 
$\bm \nu_a$ of the defining representation: 
\begin{equation}\label{weightsfund}
   \bm w_a = \sum_{b=1}^a \bm \nu_b, 
   \qquad a=1,2,\ldots,N-1.
\end{equation}
The highest weight $\bm \mu_h$ of a representation ${\cal R}$ is the first entry in 
$\bm H_{\cal R}$, see~\eqref{mainfields}. 
Using $\bm\nu_1=\bm w_1$, substitution of~\eqref{weightsfund} into~\eqref{hiestwe}, 
together with the identity $\bm \nu_a \cdot \bm \nu_b = \delta_{ab}-1/N$, yields
\begin{equation}\label{almost}
   \bm \mu_h \cdot \bm w_1 
   = \sum_{a=1}^{N-1} m_a \;-\; \frac{\sum_{a=1}^{N} a m_a}{N}.
\end{equation}
The second term can be written in terms of the $N$-ality of the representation, $n$,  defined as
$\sum_{a=1}^{N-1} a m_a \;\text{mod}\; N$.  
Thus, we have
\begin{equation}
   \frac{\sum_{a=1}^{N} a m_a}{N} \;=\; \frac{n}{N} + {\cal C}, 
   \qquad {\cal C}\in\mathbb{Z}.
\end{equation}

As we shall discuss, a fermion can be globally defined on the EH space in the presence of an abelian 
background field if and only if each entry of the matrix $A_t$ in~\eqref{atotal} is an integer.  
To obtain the strongest constraint on $k$, we set $m_{(N)}=1$.  
From the analysis above, the first entry of $A_{SU(N)}$ consists of an integer plus the fractional part 
$\tfrac{n}{N}$.  
Therefore, the integrality of $A_t$ requires setting (restoring the general value of $m_{(N)}$)
\begin{equation}\label{valuesofk}
   k \;=\; m_{(N)}\frac{n}{N} + {\cal C}, 
   \qquad {\cal C}\in\mathbb{Z}
\end{equation}
in the baryon-number background gauge field $A_{U(1)}$. 
What about the other entries in $A_t$?  
Since all weights in a representation ${\cal R}$ are generated by subtracting simple roots 
$\bm\alpha_a$ ($a=1,\ldots,N-1$), and since $\bm\alpha_a \cdot \bm w_1 = \delta_{a1}$, these subtractions are integers that do not affect the fractional contribution in~\eqref{almost}.  
We conclude that the abelian background field (\ref{abelian background needed})
is sufficient to ensure the integrality of the entries of $A_t$, and as a result, that the fermions are globally well-defined in the background gauge field $A_t$. Both $m_{(N)}$ and ${\cal C}$ are arbitrary integers, and they should be carefully chosen to obtain the most refined anomaly, as we shall discuss in several examples.

In summary, we have turned on fluxes in 
$SU(N)\times U(1)/\mathbb Z_p$, with $p = \mathrm{gcd}(N,n)$. 
We will argue that, in the presence of these background fluxes, the fermions remain globally well-defined. 
The corresponding topological charges are given by (\ref{SUNU1BTC}), 
where the factor of $1/4$ originates from the self-intersection of the bolt. 
Such refined topological charges can give rise to refined anomalies, 
thereby imposing additional constraints on the infrared dynamics of asymptotically free theories.

\subsection{Fermions, global consistency, and zero modes}

Now, we turn to the discussion of fermions and how they can be globally well-defined in EH spaces in the background of gauge fluxes. A Dirac fermion in a representation $\mathcal{R}$ can be written as the $\dim{\cal R}\times 1$  column vector
\begin{equation}\label{the dirac coloumn}
\Psi \;=\; 
\begin{pmatrix}
\Psi_{1} \\[4pt]
\Psi_{2} \\[2pt]
\vdots \\[2pt]
\Psi_{\dim \mathcal{R}}
\end{pmatrix}\,,
\end{equation}
which couples to the total gauge field $A_t$ in (\ref{atotal}) through the covariant derivative 
$\slashed{D}$. The Dirac equation then takes the form
\begin{equation}\label{maindiraceq}
\slashed{D}\,\Psi \;=\; 0, 
\qquad 
\slashed{D} \;=\; \gamma^a E_{a}^{\;\mu}
\Bigl( \left(\partial_\mu + \tfrac{1}{2}\,\Omega_\mu\right)I_{\dim {\cal R}} + i A_{t\,\mu} \Bigr)\,,
\end{equation}
where $E_{a}^{\;\mu}$ are the inverse vielbeins and  $\Omega_\mu$ is the spin-connection contribution,
\begin{equation}
\Omega_\mu \;=\; \omega_{ab\mu}\,\Sigma^{ab}, 
\qquad 
\Sigma^{ab} \;=\; \tfrac{1}{4}\,[\gamma^a,\gamma^b]\,,
\end{equation}
$\omega_{ab\mu}$ is the spin connection $1$-form,  $\gamma^a$  are the gamma matrices, and the indices $a=0,1,2,3$ and $\mu=r,\theta,\varphi,\psi$ label the orthonormal frame and curved space coordinates, respectively. The Dirac fermion sees the full gauge field $A_t$ in representation ${\cal R}$, now given by the diagonal $\dim{\cal R}\times \dim{\cal R}$ matrix with integral entries:
\begin{eqnarray}\label{finalAt}
A_t=\mbox{diag}_{\;\dim{\cal R}}\left({\cal C}_1,..., {\cal C}_{\dim{\cal R}}\right) \frac{\sigma_z\, a^2}{r^2} \,,\quad {\cal C}_a\in \mathbb Z\,.
\end{eqnarray}
The exact values of ${\cal C}_a$ depend on $m_{(N)}$, ${\cal C}$, and the representation ${\cal R}$.
The integrality of the entries of $A_t$ relies on the choice of the abelian background specified in~(\ref{abelian background needed}). 
In what follows, we shall argue that, unless every entry of $A_t$ is an integer, the fermion fields cannot be globally well defined on the EH space.

A Dirac spinor $\Psi(x)$ must be continuous along any contractible path; this path has to define an element of the covering group of $\mathrm{SO}(4)$, namely $\mathrm{Spin}(4)$, unambiguously \cite{Back:1978zf}. 
Consider the parallel transport of $\Psi$, charged under $A_t$, along the contractible path $\ell_{r \approx a}$ connecting $(r \approx a,\theta=0,\varphi=0,\psi)$ and $(r \approx a,\theta=0,\varphi=0,\psi+2\pi)$ (or any path near $r\approx a$ with constant $\theta,\varphi$) near the bolt\footnote{\label{nearthebolt}To analyze the behavior of the EH metric (\ref{EH_metric}) near $r \approx a$, it is convenient to introduce the variable $u^2 = r^2 \left(1 - \left(\frac{a}{r}\right)^4\right)$. In terms of $u$, the EH metric becomes
$
ds^2_{EH} = du^2/\left(1+a^4/r^4\right)^2 + \frac{u^2}{4}\left(\pm d\psi + \cos\theta\, d\varphi\right)^2 + \frac{r^2}{4}\left(d\theta^2 + \sin^2\theta\, d\varphi^2\right)\,.
$
In the limit $r \to a$, we have $u \to 0$, and the metric reduces to
$
\lim_{r \to a} ds^2_{EH} \cong \frac{du^2}{4} + \frac{u^2}{4}\left(\pm d\psi + \cos\theta\, d\varphi\right)^2 + \frac{a^2}{4}\left(d\theta^2 + \sin^2\theta\, d\varphi^2\right)\,,
$
demonstrating that the geometry is regular at $r=a$. For fixed $(\theta,\varphi)$, the induced metric is $\tfrac{1}{4}(du^2 + u^2 d\psi^2)$, which describes flat $\mathbb{R}^2$ provided $\Delta\psi = 2\pi$, ensuring the absence of conical singularities. Thus, a loop near $r \approx a$ at fixed $(\theta,\varphi)$ lies in the $(\psi,u)$ plane and is contractible (in the geometrical sense).}. Along this path, the fermion acquires both gauge and gravitational holonomies:
\begin{align}\label{PARALLTABUL}
\Psi(r,\theta,\varphi,\psi+2\pi) &= \exp\left[i\int_{\ell_{r \approx a}} A_t\right]\,
\exp\left[-\tfrac{1}{2}\int_{\ell_{r\approx a}}\omega_{ab}\Sigma^{ab}\right]\Psi(r,\theta,\varphi,\psi)\,,
\end{align}
where $\omega_{ab}$ are spin connection $1$-forms (see their definition in Appendix \ref{sec:fermions-EH}), $\Sigma^{ab}=[\gamma^a,\gamma^b]/4$, and $\gamma^a$ are the Dirac matrices. The Latin letters $a,b$ run over $0,1,2,3$.
Simple calculations show 
\begin{eqnarray}\nonumber
\exp\left[-\tfrac{1}{2}\int_{\ell_{r\approx a}}\omega_{ab}\Sigma^{ab}\right] 
&=& I_4\,,\\
\exp\left[i\int_{\ell_{r\approx a}} A_t\right] &=&\exp\left[i\pi \mbox{diag}_{\;\dim{\cal R}}\left({\cal C}_1,..., {\cal C}_{\dim{\cal R}}\right)\right]\,.
\end{eqnarray}
Thus, the gravitational holonomy is trivial along $\ell_{r \approx a}$, reflecting the fact that EH spaces are spin. Only the gauge holonomy is generally nontrivial, so that
\begin{align}\label{PARALLTA2BUL}
\text{at } r \approx a:\quad \Psi_a(r,\theta,\varphi,\psi+2\pi) =(-1)^{{\cal C}_a}\,\Psi_a(r,\theta,\varphi,\psi)\,,\quad a=1,2,...,\dim{\cal R}\,.
\end{align}
Thus, every component in $\Psi$ satisfies either periodic or anti-periodic boundary conditions. In the special case of a single $U(1)$ background gauge field with flux $\int_{S^2}F_{(1)}=2\pi{\cal C}$, the boundary condition (\ref{PARALLTA2BUL}) reduces to 
\begin{eqnarray}
 \Psi(r \approx a,\theta,\varphi,\psi+2\pi) =(-1)^{{\cal C}}\,\Psi(r \approx a,\theta,\varphi,\psi)\,.
\end{eqnarray}

It is important to emphasize that, owing to the Abelian nature of $A_t$, each component 
$\Psi_a$ (with $a = 1,2,\ldots,\dim \mathcal{R}$) evolves independently and therefore 
possesses its own total angular momentum. In a Hamiltonian formulation, where we define the theory on constant-$r$ slices,  this independence follows from the fact that the Hamiltonian is diagonal in color space, and each diagonal element commutes with the 
total angular momentum operator:
$
  [H, J^2] = 0 \, .
$
When solving the Dirac equation~\eqref{maindiraceq} in the Abelian background~\eqref{finalAt}, 
each color component $\Psi_a$ carries a total angular momentum $j_a$ that is either an integer 
or a half-integer. The global consistency of $\Psi_a$ is ensured only if its total angular momentum 
is compatible with the boundary condition~\eqref{PARALLTA2BUL}. In particular, we find:
\begin{itemize}
  \item For ${\cal C}_a$ even, only integer values of the total angular momentum are allowed.
  \item For ${\cal C}_a$ odd, only half-integer values of the total angular momentum are allowed.
\end{itemize}
This is explained in detail in Appendix \ref{sec:fermions-EH}.
Most importantly, if ${\cal C}_a$ is fractional, then no value of the angular momentum 
satisfies the boundary condition near the bolt,
$
  \Psi_a( r \approx  a,\theta,\varphi,\psi) 
  = e^{i\pi {\cal C}_a}\, \Psi_a( r \approx a,\theta,\varphi,\psi+2\pi) \, .
$
This explains why ${\cal C}_a$ must be integral; otherwise, the fermionic wavefunctions 
become ill-defined on the EH space. This also explains the reason behind turning on the $A_{U(1)}$ background in (\ref{abelian background needed}): it is needed to render the fermions well-defined in EH space. 

The explicit solution of the Dirac equation, together with the boundary conditions, 
determines the number of normalizable zero modes for each color component $\Psi_a$; 
see Appendix~\ref{sec:fermions-EH}. It is useful to emphasize certain features of these modes. 
Decomposing $\Psi_a$ into Weyl components, we write
\begin{equation}
\Psi_a \;=\;
\begin{bmatrix}
\lambda_{a\,\alpha}\\[2pt]
\bar\chi_a^{\dot\alpha}
\end{bmatrix}, 
\qquad 
\alpha,\dot\alpha = 1,2 \,,
\end{equation}
where $\alpha$ and $\dot\alpha$ label the two spin components. 
Normalizability requires
\begin{equation}\label{eq:normbulk}
\int |\Psi_a|^2\, r^3 \, dr \wedge \sigma_x \wedge \sigma_y \wedge \sigma_z < \infty \, .
\end{equation}
Because EH space is (anti)-self-dual, only the $\lambda_{a\,\alpha}$ components admit 
normalizable zero modes. A detailed analysis shows that consistent solutions take the form
\begin{equation}
\lambda_{a\,1}= g_{a\,1}(r)\,|j_a,m'=j_a,m\rangle, 
\qquad 
\lambda_{a\,2}= g_{a\,2}(r)\,|j_a,m'=-j_a,m\rangle, 
\qquad |m|\leq j_a \,,
\end{equation}
where $|j_a,m',m\rangle$ are the Wigner $D$-functions. Importantly, only one of the two radial 
profiles, $g_{a\,1}$ or $g_{a\,2}$, is well-behaved at the bolt and decays rapidly at infinity, 
depending on the sign of ${\cal C}_a$.\footnote{If ${\cal C}_a=0$, no normalizable zero 
mode exists for $\Psi_a$.}  
For instance, when ${\cal C}_a>0$, normalizable modes arise from $g_{a\,2}$ for angular 
momenta satisfying ${\cal C}_a>2j_a$, with
\begin{equation}
g_{a\,2}(r) =\frac{K}{r}(r^2-a^2)^{\frac{{\cal C}_a-2j_a-2}{4}}
(r^2+a^2)^{-\frac{{\cal C}_a+2j_a+2}{4}},
\qquad j_a \geq 0, \quad {\cal C}_a > 2j_a,
\end{equation}
where $K$ is a normalization constant.
Near the bolt, $g_{a\,2}^2 \sim (r-a)^{({\cal C}_a-2j_a-2)/2}$, so the integral 
\eqref{eq:normbulk} converges provided ${\cal C}_a>2j_a$. 
At large $r$, $r^3 g_{a\,2}^2 \sim r^{-4j_a-3}$, which is normalizable for all $j_a\geq 0$.

The crucial point is that all zero modes are localized near the bolt and decay 
rapidly towards infinity. Thus, no mode escapes to the asymptotic region. 
This fact is essential for anomaly matching: if zero modes extended to infinity, one 
might mistakenly attribute them to the IR spectrum, thereby obscuring the proper UV--IR 
matching of anomalies.

The counting of the normalizable zero modes is as follows:
\begin{itemize}
  \item If ${\cal C}_a = 2p_a$ is even, the holonomy at the bolt is trivial, 
  which restricts $j_a$ to integer values. In this case, the number of zero modes is
  \begin{equation}\label{evenC}
    {\cal I}_a = p_a^2 \, .
  \end{equation}

  \item If ${\cal C}_a = 2p_a + 1$ is odd, the holonomy contributes a $(-1)$ phase, 
  which restricts $j_a$ to half-integer values. In this case, the number of zero modes is
  \begin{equation}\label{oddC}
    {\cal I}_a = p_a(p_a+1) \, .
  \end{equation}
\end{itemize}
The total number of fermionic zero modes is obtained by summing over all the zero modes of the
color components $\Psi_a$ with $a = 1,2,\ldots,\dim \mathcal{R}$. Thus,
\begin{eqnarray}\label{explicit}
{\cal I}_{\cal R}=\sum_{a=1}^{\dim \cal R}{\cal I}_a\,.
\end{eqnarray}

Before closing this section, let us repeat the problem of parallel transport in the 
asymptotic region $r \to \infty$, where the EH space approaches the 
boundary $\mathbb{RP}^3$. This discussion will be directly relevant in the following 
section, when we turn to the Atiyah--Patodi--Singer (APS) index theorem. 
Consider the path $\ell_{\infty}$ defined by the shift $\psi \mapsto \psi + 2\pi$ at 
fixed $(\theta,\varphi)$ and $r \to \infty$. Since $\mathbb{RP}^3$ is not simply connected, 
this loop is noncontractible. In fact, the first homology group of $\mathbb{RP}^3$ is
$
H_1(\mathbb{RP}^3,\mathbb{Z}) \;\cong\; \mathbb{Z}_2 \,,
$
so there exists exactly one nontrivial $1$-cycle, of $\mathbb{Z}_2$ torsion type.
For fermions transported around $\ell_{\infty}$, one finds
\begin{equation}
\exp\!\left[-\tfrac{1}{2}\int_{\ell_{\infty}} \omega_{ab}\,\Sigma^{ab}\right]
= -\,\gamma^5, 
\qquad \gamma^5 \equiv \gamma^0\gamma^1\gamma^2\gamma^3 \, .
\end{equation}
The situation for the gauge holonomy $e^{i\int_{\ell_{\infty}} A_t}$ is subtler. Because 
the connection $A_t$ falls off rapidly, one might expect a trivial result $(+1)$; 
however, such reasoning ignores the possibility that topological data from the bulk 
is encoded at the boundary. To evaluate the holonomy correctly, one applies Stokes’ theorem:
$
\int_{\ell_{\infty}} A_t = \int_{\Sigma} F_t,
$
where $\Sigma$ is a two-dimensional surface bounded by $\ell_{\infty}$ on 
$\mathbb{RP}^3$ and smoothly capping off at the bolt in the interior. Topologically, 
$\Sigma$ resembles a cigar: it extends from the boundary and closes at the bolt, 
where the $\psi$-circle shrinks to zero size (see Footnote \ref{nearthebolt} to help with the visualization). Using the explicit form of $F_t$ (see (\ref{form of K})) and 
integrating over $0 \leq \psi < 2\pi$, $a \leq r < \infty$, one finds
\begin{equation}\label{mainhol}
\exp\!\left[i\int_{\ell_\infty} A_t\right] 
= \exp\!\left[i\int_{\Sigma} F_t\right] 
= \exp\!\left[i\pi\,\mathrm{diag}_{\,\dim{\cal R}}
\!\left({\cal C}_1,\ldots,{\cal C}_{\dim{\cal R}}\right)\right] \,.
\end{equation}
It follows that the boundary condition satisfied by the fermions is
\begin{equation}\label{inftyBC}
\Psi_a(r \to \infty, \theta, \varphi, \psi+2\pi) 
= (-1)^{{\cal C}_a+1}\,\gamma^5\,
\Psi_a(r \to \infty, \theta, \varphi, \psi)\,,\quad a=1,2,...,\dim{\cal R} .
\end{equation}
The overall sign appearing in Eq.~\eqref{inftyBC} is decisive for the value of the 
$\eta$-invariant, and hence for the APS index, as will be explained in the next section.

\subsection{The Atiyah--Patodi--Singer (APS) index theorem}
\label{The Atiyah--Patodi--Singer (APS) index theorem}

The counting of the zero modes can also be obtained by applying the Atiyah--Patodi--Singer (APS) index theorem \cite{Atiyah:1975jf,Atiyah:1976qjr,Atiyah:1976jg,Eguchi:1980jx} of the Dirac operator, which gives the number of chiral-zero modes of a twisted spin complex (twisted by a gauge bundle $V$) in the background space ${\cal M}$ with boundaries\footnote{We note that the sign in front of the term $\frac{T_{\cal R}}{8\pi^2}\int_{{\cal M}}\mbox{tr}F\wedge F$ in \cite{Eguchi:1980jx} is opposite to our sign since they use anti-hermitian gauge fields.}:
\begin{eqnarray}\label{APSINDEXTHEO}\nonumber
{\cal I}_{\cal R}&=&\frac{T_{\cal R}}{8\pi^2}\left[\int_{\cal M}\mbox{tr}_\Box F\wedge F
\right]+ \frac{\dim{\cal R}}{24\cdot 8\pi^2}\left[\int_{\cal M}\mbox{tr}R\wedge R-\int_{\partial\cal M}\mbox {tr}\, \Theta\wedge R\right]-\eta_V(\partial{\cal M})\,.\\
\end{eqnarray}
The generalization to the case of two bundles is straightforward. 
In particular, in our setup, we turn on background fields for both 
$SU(N)$ and $U(1)$.
Here, $R$ is the curvature $2$-form, $\Theta$ is the second fundamental form computed at the boundary of $\cal M$, and $\eta$ is the $\eta$-inavriant of the Dirac operator defined as
\begin{eqnarray}\label{etainvdef}
\eta_V(\partial {\cal M})\equiv\frac{1}{2}\left(\mbox{lim}_{T\rightarrow 0}\sum_{\lambda \neq 0}\frac{\mbox{sign}(\lambda)}{|\lambda|^{T}}+h_V(\partial{\cal M})\right)\,.
\end{eqnarray}
The subscript $V$ means that, in general, the $\eta$-invariant depends on the vector bundle $V$, and the sum is over all the non-zero eigenvalues of the Dirac operator in the presence of the gauge-bundle twist computed at the boundary of $\cal M$. The term $h_V$ is the harmonic correction: it is given by the dimension of the kernel of the Dirac operator on the boundary, $h_V=\mbox{dim ker}\slashed D(\partial {\cal M})$, which accounts for the missing zero eigenvalues not accounted for in the first term in $\eta_V$. The term $\int_{\partial\cal M}\mbox{tr}\, \Theta\wedge R$ is the gravitational Chern-Simons boundary term, which vanishes for the EH space\footnote{This term can contribute when the metric is not a product metric near the boundary. No analogous correction is required for the vector bundle piece \cite{Eguchi:1980jx}.}. Moreover, the harmonic correction $h_V(\partial{\cal M})=0$ in our case  \cite{Eguchi:1978gw}; more on this point will follow momentarily. 

The topological charges of the $SU(N)$ and $U(1)_{B}$ bundles are given by (\ref{SUNU1BTC}), while the gravitational backgrounds give the gravitational topological charge (see Footnote \ref{conventionfootnote})
\begin{eqnarray}\label{QGCHARGE}
Q_G= \frac{1}{24\cdot 8\pi^2}\int_{\scriptsize\mbox{EH}}\mbox{tr}R\wedge R=-\frac{1}{8}\,.
\end{eqnarray}
 The $\eta$-invariant gives a non-trivial contribution at the boundary $\mathbb {RP}^3$ of EH space. In the case of a single $U(1)$ bundle \begin{equation}A={\cal C}\frac{a^2\sigma_z}{r^2}\,,\end{equation}
 the value of $\eta$ depends on whether the $U(1)$ charge ${\cal C}$ is even or odd:\footnote{The explcit computation of the $\eta$-invariant was carried out in Appendix C in \cite{Anber:2025gvb}.}
\begin{eqnarray}\label{etavalrp3}\nonumber
\eta(\mathbb{RP}^3)&=&-\frac{1}{8}\,, \quad {\cal C}\in 2\mathbb Z\,,\\
\eta(\mathbb{RP}^3)&=&+\frac{1}{8}\,, \quad {\cal C}\in 2\mathbb Z+1\,.
\end{eqnarray}
The two different values of $\eta$ are attributed to the fact that there are two choices of twisting by a flat line bundle $\{+1,-1\}$, which are selected by whether ${\cal C}$ is even or odd.

Now, a critical remark is in order. The APS construction for computing the index on a manifold with boundary is based on applying non-local boundary conditions \cite{Atiyah:1975jf,Atiyah:1976qjr,Atiyah:1976jg}. It is convenient to truncate the EH space at a large radius $R$ and impose APS boundary conditions on the induced boundary
$
\mathbb{RP}^3
$. 
For the background metric and flat holonomy that we consider, an explicit spectral analysis of the twisted $3$-dimensional Dirac operator on $\mathbb{RP}^3$ shows that it has no zero modes\footnote{The eigenvalues of the twisted Dirac operator on the noncanonical $\mathbb{RP}^3$ with metric $ds^2=\sigma_x^2+\sigma_y^2+f^{-2}\sigma_z^2$ and $U(1)$ gauge field $A=\beta\sigma_z$ are given by \cite{Pope:1981jx}: $\lambda=\frac{1}{2}f^{-1}\pm f\sqrt{(2s+1-\beta)^2+4f^{-2}(j-s)(j+s+1)}$, where $-j\leq s\leq j-1$, and $\lambda=f(2j+1\mp\beta)+f^{-1}/2$. The degeneracies of these eigenvalues are $2j+1$, and the total angular momentum $j$ takes integer or half-integer values $j=0,1/2,1,3/2,... $. The constant radial slices, $r=R$, in the EH metric is the noncanonical metric on $\mathbb{RP}^3$ (modulo an overall conformal factor), where $f^{-2}=1-a^4/R^4$ and $\beta={\cal C}a^2/R^2$. For large enough values of ${\cal R}$, one can see that the eigenvalues do not cross zero. In particular, at asymptotic infinity, so that $R\rightarrow \infty$ and $\beta\rightarrow 0$, the eignevalues become $\lambda_+=3/2+m$ with degenerecy $(m+1)(m+2)$ and $\lambda_-=-1/2-m$ with degenercey $m(m+1)$ for $m=0,1,..$.
}. By standard continuity properties of Dirac spectra on compact manifolds, this implies that for all sufficiently large $R$, the twisted Dirac operator on $\mathbb{RP}^3$ remains invertible: its eigenvalues vary smoothly with $R$, but none crosses zero in the asymptotic region. This means that the space of the positive eigenmodes and the space of the negative eigenmodes of the boundary Dirac operators are stable under deformation, and no jumps in their dimensions take place.
In sequence, the APS index 
is independent of the precise choice of truncation radius $R$, and is stable under smooth deformations of the bulk metric and gauge field that preserve the boundary data (the metric and the holonomy class on $\mathbb{RP}^3$). Had the Dirac operator on the boundary contained zero modes, this would invalidate the index from being a topological invariant \cite{Witten:2015aba}.

Now, we continue with the application of the APS index theorem. When applying the APS index to a nonabelian bundle, one also must pay special attention to the holonomy on $\mathbb{RP}^3$. 
Since the holonomy is defined modulo $2$, the holonomy of a well-defined general gauge bundle 
(such that fermions can be globally defined) takes the form, e.g., 
\begin{equation}\label{holrp3}
  \mathrm{Hol}(A_t) \;\equiv\; \exp\!\left(i \int_{\ell_\infty} A_t\right) 
  \;=\; \mathrm{diag}_{\dim\cal R}\left((-1)^{{\cal C}_1},...,(-1)^{{\cal C}_{\dim{\cal R}}} \right)\,.
\end{equation}
We define the \emph{signature} of the holonomy matrix to be the number of $(+1)$ eigenvalues 
minus the number of $(-1)$ eigenvalues, i.e.
\begin{equation}
  \mathrm{sgn}(\mathrm{Hol}) \;=\; \#(+1) - \#(-1)\,.
\end{equation}
Since gauge transformations act on the holonomy by conjugation, $\mbox{Hol}\rightarrow U\;\mbox{Hol}\;U^{-1}$, the multiset of eigenvalues
of the holonomy matrix is preserved. In particular, if the holonomy has eigenvalues 
$\{ \lambda_1, \dots, \lambda_n \}$, these are unchanged (up to reordering) under gauge transformations. Thus,  $ \mathrm{sgn}(\mathrm{Hol}) $ is gauge invariant. 
With this definition, and using (\ref{SUNU1BTC}, \ref{etavalrp3}), the APS index gives the total number of normalizable zero modes as
\begin{eqnarray}\label{indexmastermix}\nonumber
  \mathcal{I}_{\cal R} 
  \;&=&\; \frac{T_{\mathcal{R}}}{8\pi^2} \int_{\scriptsize\mbox{EH}} \mathrm{tr}_\Box\, F_{(N)} \wedge F_{(N)}  
      \;+\frac{\dim{\mathcal{R}}}{8\pi^2}\int_{\scriptsize\mbox{EH}}F_{(1)} \wedge F_{(1)} + \frac{-\,\dim {\mathcal{R}} + \mathrm{sgn}(\mathrm{Hol})}{8}\\
      &=&\frac{m_{(N)}^2T_{\mathcal{R}}}{4}\left(1-\frac{1}{N}\right)+\frac{\dim {\mathcal{R}} }{4}\left(m_{(N)}\frac{n}{N}+{\cal C}\right)^2+ \frac{-\,\dim {\mathcal{R}} + \mathrm{sgn}(\mathrm{Hol})}{8}\,.
\end{eqnarray}
At this level, the computation of $\mbox{sgn(Hol)}$ requires the explicit computation of all $\{{\cal C}_a\}$, which are obtained using the Verma basis method.
In all the examples we examined, we confirmed that the direct counting 
of fermion zero modes in (\ref{explicit}) agrees precisely with the result of the APS index. In the next section, we shall devise a method that, in certain cases, depending on ${\cal R}$, enables us to compute  $\mbox{sgn(Hol)}$ directly in terms of group theory information. 

\subsection{Gluing EH spaces and the $\eta$-invariant}
\label{Gluing construction}

In this paper, we are interested in anomalies that are not seen on manifolds such as $S^{4}$, $\mathbb{T}^{4}$, $S^2\times S^2$, or $\mathbb{CP}^{2}$. Our main claim is that the EH space furnishes refined cohomological information which leads to new anomaly structures and constraints on quantum field theories. A special feature of EH space is the presence of a boundary at asymptotic infinity, which induces a boundary contribution to the Dirac index in the form of the $\eta$-invariant.

\begin{figure}[h] 
   \centering
   \includegraphics[width=5in]{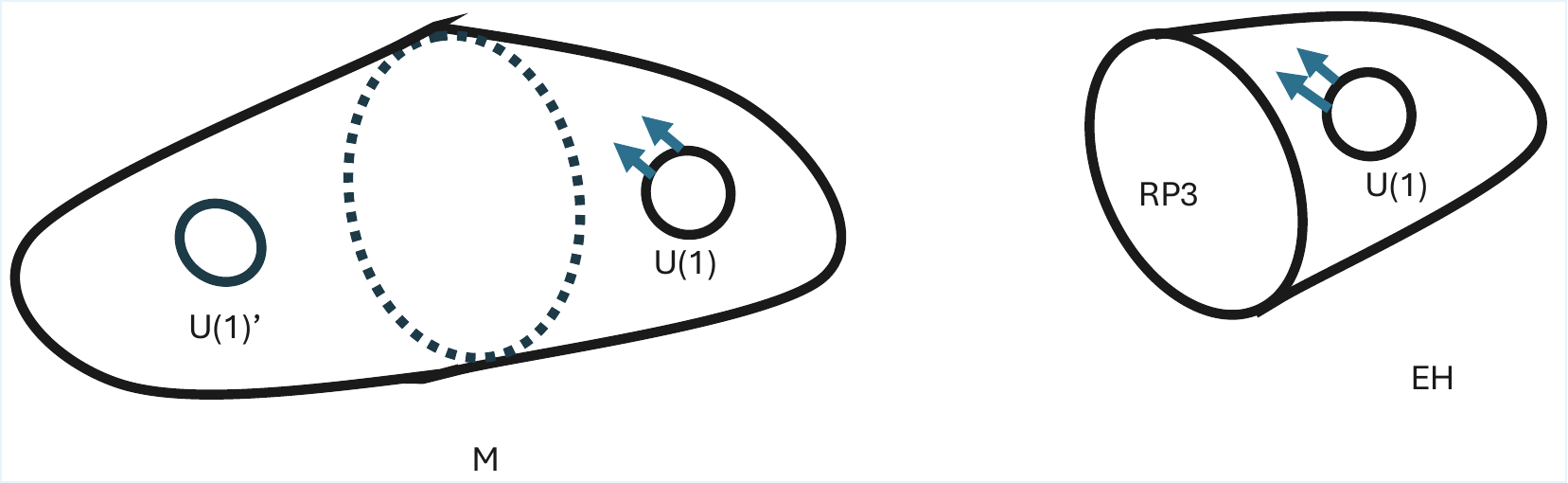} 
  \caption{Sewing EH (on the right) and $\mbox{EH}'$ (an orientation reversal of EH but with a different $U(1)$ flux) via their $\mathbb{RP}^3$ boundary, to give ${\cal M}=\mbox{EH}\cup\mbox{EH}'$ (on the left). The sewing must happen such that the gauge bundles over both spaces match smoothly via $\mathbb{RP}^3$, so that we avoid a $\mathbb {Z} _2$ jump in the holonomy seen by the fermions. The sewing must also not change the number of fermion zero modes (represented by the arrows) on the original EH space.} 
   \label{Msewing}
\end{figure}

To better understand this feature and to compute the index without the need to compute $\eta$ explicitly (in a few cases), we construct a closed $4$-manifold ${\cal M}$ by gluing EH to a copy with reversed orientation, which we denote by $\mathrm{EH}'$. Thus,
$
\mathcal{M} =\mathrm{EH} \cup \mathrm{EH}'
$.
The space $\mathrm{EH}'$ has the opposite orientation to $\mathrm{EH}$, but is not simply a geometric mirror image: in general, it carries different background fluxes for the gauge fields. The construction of $\mathrm{EH}'$ is chosen so that (i) there is no topological obstruction to gluing along the common $\mathbb{RP}^{3}$ boundary, and (ii) the resulting closed manifold $\mathcal{M}$ has the same Dirac index as $\mathrm{EH}$. Equivalently, this gluing procedure caps off the boundary of the original EH space without altering the number of chiral zero modes supported near the bolt of the $\mathrm{EH}$ space; see Figure \ref{Msewing}.

We begin by considering the simplest setting of a single $U(1)$ background gauge field. 
Let us turn on field strengths $F_{(1)}$ on $\mathrm{EH}$ and $F_{(1)}'$ on $\mathrm{EH}'$ such that
$\frac{1}{2\pi}\int_{S^2} F_{(1)} = \mathcal{C}$ and $\frac{1}{2\pi}\int_{S'^2} F_{(1)}' = \mathcal{C}'$, 
where $S^2$ and $S'^2$ denote the bolt $2$-spheres of $\mathrm{EH}$ and $\mathrm{EH}'$, respectively. 
These gauge field configurations are localized near the bolts and decay rapidly toward the boundary. 
However, as emphasized above, the induced flat connection on the common boundary $\mathbb{RP}^3$ depends only on the flux modulo $2$. 
Therefore, when gluing $\mathrm{EH}$ and $\mathrm{EH}'$ along their shared boundary, we must require that the associated boundary holonomies seen by the fermions (assuming that the fermion has a unit charge under $U(1)$) agree:
\begin{eqnarray}\label{gluingcond}
(-1)^{\mathcal{C}} = (-1)^{\mathcal{C}'} 
\qquad \Longleftrightarrow \qquad 
\mathcal{C} \equiv \mathcal{C}' \pmod{2}.
\end{eqnarray}
This condition is necessary to avoid a topological obstruction to the gluing; without it, the background $U(1)$ bundle would not extend smoothly across the interface.

Next, we apply the index theorem to the closed manifold 
${\cal M}= \mathrm{EH} \cup \mathrm{EH}'$. 
Consider a Dirac fermion $\Psi$ on ${\cal M}$ with unit charge under the background gauge field 
$F_{(1)}^{({\cal M})}$, which should be understood as the field strengths 
$F_{(1)}$ and $F_{(1)}'$ supported near the bolts of $\mathrm{EH}$ and $\mathrm{EH}'$, respectively, and rapidly decaying away from them. 
The number of chiral zero modes of $\Psi$ is given by the index theorem:
\begin{eqnarray}
{\cal I}^{(\cal M)}
&=&\frac{1}{8\pi^2}\int_{\mathrm{EH} \cup \mathrm{EH}'}F_{(1)}^{({\cal M})}\wedge F_{(1)}^{({\cal M})}
+\frac{1}{24\cdot 8\pi^2}\int_{\mathrm{EH} \cup \mathrm{EH}'}\mathrm{tr}\, R^{(\cal M)}\wedge R^{(\cal M)}\,,
\end{eqnarray}
where $\partial{\cal M}=\emptyset$, so there is no contribution from an $\eta$-invariant. The curvature $2$-form $R^{({\cal M})}$ restricts to $R$ on $\mathrm{EH}$ and $R'$ on $\mathrm{EH}'$ away from the gluing region. 
Splitting the integral into the two components, one involving EH, the other $\mbox{EH}'$ with the opposite orientation to its orientation in ${\cal M}= \mathrm{EH} \cup \mathrm{EH}'$ (this gives both EH and $\mbox{EH}'$ the same relative orientation), we obtain
\begin{eqnarray}
{\cal I}^{(\cal M)}
&=&\int_{\mathrm{EH}}\frac{F_{(1)}\wedge F_{(1)}}{8\pi^2}
+\int_{\mathrm{EH}} \frac{\mathrm{tr}\,R\wedge R}{24\cdot 8\pi^2}
-\int_{\mathrm{EH}'}\frac{F'_{(1)}\wedge F'_{(1)}}{8\pi^2}
-\int_{\mathrm{EH}'}\frac{\mathrm{tr}\,R'\wedge R'}{24\cdot 8\pi^2}\,,
\end{eqnarray}
which may be rearranged by adding and subtracting $\eta(\mathbb{RP}^3)$:
\begin{eqnarray}\nonumber
{\cal I}^{(\cal M)}
&=&\int_{\mathrm{EH}}\frac{F_{(1)}\wedge F_{(1)}}{8\pi^2}
+\int_{\mathrm{EH}} \frac{\mathrm{tr}\,R\wedge R}{24\cdot 8\pi^2}
-\eta(\mathbb{RP}^3)
\\ \nonumber
&&-\left(\int_{\mathrm{EH}'}\frac{F'_{(1)}\wedge F'_{(1)}}{8\pi^2}
+\int_{\mathrm{EH}'}\frac{\mathrm{tr}\, R'\wedge R'}{24\cdot 8\pi^2}
-\eta(\mathbb{RP}^3)\right)
\\
&=&{\cal I}^{(\scriptsize\mathrm{EH})}-{\cal I}^{(\scriptsize\mathrm{EH}')}\,,
\end{eqnarray}
where we emphasize that $\eta(\mathbb{RP}^3)$ depends only on the metric and holonomy on $\mathbb{RP}^3$. 
In our construction, the Dirac index on $\mathrm{EH}'$ must vanish, ${\cal I}_{\scriptsize\mathrm{EH}'}=0$, since ${\cal M}$ must reproduce the index of the original $\mathrm{EH}$ space. Therefore,
\begin{eqnarray}\label{niceeta}\nonumber
\eta(\mathbb{RP}^3)
&=&\int_{\mathrm{EH}'}\frac{F'_{(1)}\wedge F'_{(1)}}{8\pi^2}
+\int_{\mathrm{EH}'}\frac{\mathrm{tr}\, R'\wedge R'}{24\cdot 8\pi^2}
\\
&=&\frac{{\cal C}'^2}{4}-\frac{1}{8}\,.
\end{eqnarray}
From our earlier analysis, there are no localized zero modes on $\mathrm{EH}'$ when ${\cal C}'=0,1$; see~(\ref{evenC}), (\ref{oddC}). 
Together with the gluing condition~(\ref{gluingcond}), this implies we must choose ${\cal C}'=0$ for even ${\cal C}$ and ${\cal C}'=1$ for odd ${\cal C}$. Consequently,
\begin{eqnarray}\label{etaevenodd}\nonumber
\eta(\mathbb{RP}^3)&=&-\frac{1}{8}\,, \qquad {\cal C}\in 2\mathbb Z\,,\\
\eta(\mathbb{RP}^3)&=&+\frac{1}{8}\,, \qquad {\cal C}\in 2\mathbb Z+1\,,
\end{eqnarray}
in agreement with~(\ref{etavalrp3}) obtained directly from the definition~(\ref{etainvdef}), as demonstrated in Appendix~C of~\cite{Anber:2025gvb}. We therefore conclude that
\begin{eqnarray}\label{oneomretime}
{\cal I}^{({\cal M}=\mathrm{EH} \cup \mathrm{EH}')}={\cal I}^{(\mathrm{EH})}
= \begin{cases}
\displaystyle \frac{{\cal C}^2}{4}\,, & {\cal C}\in 2\mathbb Z\,,\\[6pt]
\displaystyle \frac{{\cal C}^2-1}{4}\,, & {\cal C}\in 2\mathbb Z+1\,,
\end{cases}
\end{eqnarray}
which shows that ${\cal M}=\mathrm{EH} \cup \mathrm{EH}'$ supports the same number of chiral fermion zero modes (all of the same chirality) as the original $\mathrm{EH}$ space. A byproduct of this construction is a bulk expression for the $\eta$-invariant, allowing us to compute it without referring to boundary spectral data.

We now argue, though not rigorously, that ${\cal M}$ should be understood as $S^2\times S^2$; thus, we have the equivalance ${\cal M}\cong S^2\times S^2$ \cite{Milnor1958SimplyConnected4Manifolds,https://doi.org/10.1112/jlms/s1-39.1.141,Hawking:1979pi}. First, the Euler characteristics of $\mathrm{EH}$ and $\mathrm{EH}'$ are equal, $\chi(\mathrm{EH})=\chi(\mathrm{EH}')=2$, since $\chi$ is insensitive to orientation. Therefore,
$
\chi({\cal M}) 
= \chi(\mathrm{EH}) + \chi(\mathrm{EH}') -\chi(\mathrm{EH}\cap\mathrm{EH}')
= 4 
= \chi(S^2\times S^2).
$
Next, the signature satisfies
$
\tau({\cal M})
= \tau(\mathrm{EH}) - \tau(\mathrm{EH}')
= 0,
$
again matching the signature of $S^2\times S^2$. Thus, ${\cal M}$ shares the same Euler characteristic and signature as $S^2\times S^2$.

We further compute the Dirac index on ${\cal M}$ and show that it coincides with the index on $S^2\times S^2$ equipped with appropriate $U(1)$ bundles. Recall the flux quantization conditions on the bolt spheres of $\mathrm{EH}$ and $\mathrm{EH}'$,
\begin{equation}\label{sphersbegin}
\int_{S^2(\mathrm{EH})} F_{(1)} = 2\pi {\cal C}, 
\qquad
\int_{S^2(\mathrm{EH}')} F'_{(1)} = 2\pi {\cal C}'.
\end{equation}
Introduce two $2$-cycles $S^2_A$ and $S^2_B$ in ${\cal M}$ as the linear combinations
\begin{equation}\label{linearcomS}
S^2(\mathrm{EH}) = S^2_A + S^2_B,
\qquad
S^2(\mathrm{EH}') = S^2_A - S^2_B,
\end{equation}
so that the fluxes through $S^2_A$ and $S^2_B$ become
\begin{equation}\label{flux quan}
\int_{S^2_A} F^{(A)}_{(1)} = 2\pi \frac{{\cal C}+{\cal C}'}{2},
\qquad
\int_{S^2_B} F^{(B)}_{(1)} = 2\pi \frac{{\cal C}-{\cal C}'}{2}.
\end{equation}
It follows that the topological charge on $S^2_A\times S^2_B$ is
\begin{equation}\label{sphersend}
\frac{1}{8\pi^2}\int_{S^2_A\times S^2_B} 
F^{}_{(1)}\wedge F^{}_{(1)}
= \frac{{\cal C}^2-{\cal C}'^2}{4}.
\end{equation}
Finally, using the gluing condition 
${\cal C}'=0$ for even ${\cal C}$ and 
${\cal C}'=1$ for odd ${\cal C}$, 
we recover the Dirac index in \eqref{oneomretime}. Also, with this choice, the correct quantization conditions in (\ref{flux quan}) on $S_A^2$ and $S_B^2$ are reproduced. 
Hence, the Dirac index on ${\cal M}$ agrees with that on $S^2\times S^2$ with the corresponding quantized fluxes.

An important point must be emphasized. On the EH space, one can consistently define Dirac operators for multiple fermion species with arbitrary $U(1)$ charges $q_\Psi \in \mathbb Z$; both even and odd charges are allowed. The fermions are sections of the bundle $S\otimes L^{q_\Psi}$, where $S$ is the spin bundle and $L$ is the $U(1)$ line bundle.  However, the gluing construction that produces the closed manifold ${\cal M}$ while preserving the fermion zero-mode content of the original EH space requires all fermions to experience the same $\mathbb{Z}_2$ holonomy at the boundary, namely $(-1)^{q_\Psi {\cal C}}$. 
Consequently, if ${\cal C}$ is even, both even- and odd-charged fermions can be simultaneously accommodated on ${\cal M}$. In contrast, when ${\cal C}$ is odd, all fermions must have the same charge parity (either all even or all odd); otherwise, the boundary holonomies would differ across $\mathrm{EH}$ and $\mathrm{EH}'$, preventing a consistent gluing. 
This subtlety is essential for discussing 't~Hooft anomaly matching. Since ${\cal M}\cong S^2\times S^2$, anomaly matching on ${\cal M}$ does not produce new constraints (since the anomalies on $S^2\times S^2$ are well understood, and they are the same anomalies obtained on $\mathbb T^4$; these are $0$-form and $1$-form anomalies). In contrast, anomaly matching on EH does yield new obstructions: when ${\cal C}$ is odd, EH allows the coexistence of fermions with even and odd $U(1)$ charges, a situation that ${\cal M}$ cannot accommodate under smooth gluing. Thus, EH provides strictly stronger anomaly-detection power than its closed extension ${\cal M}$.

Having clarified this point, we now generalize the construction of ${\cal M}$ to include $SU(N)\times U(1)/\mathbb Z_p$, with $p = \mathrm{gcd}(N,n)$, background fields.  This, in certain cases, will allow us to obtain a compact expression for the Dirac index on the EH space without explicitly computing $\mathrm{sgn}(\mathrm{Hol})$ in~(\ref{indexmastermix}) using the Verma basis. 

We employ the $SU(N)$ and $U(1)$ background gauge fields introduced in~(\ref{mainfields}), with $k$ given by~(\ref{valuesofk}), and take $\bm{\nu}_a=\bm{\nu}_1$. As an illustrative example, we consider a fermion in the $2$-index antisymmetric representation 
${\cal R}=\tiny\yng(1,1)$ of $SU(N)$, which has 
$
\dim {\cal R} = \frac{N(N-1)}{2}, T_{\cal R}=N-2,
$
together with unit $U(1)$ charge. We set $m_{(N)}=1$ and ${\cal C}=1$.

Depending on the parity of ${\cal C}$, the holonomy at the $\mathbb{RP}^3$ boundary seen by the fermion is
\begin{eqnarray}\label{2indexhol}\nonumber
\exp\left[i\int_{\ell_\infty}A_t\right]
= \begin{cases}
\displaystyle \mathrm{diag}_{\,N(N-1)/2}\!\left(\underbrace{-1,-1,-1,\ldots,-1}_{N-1},+1,..,+1\right)\,, & {\cal C}\in 2\mathbb Z\,,\\[6pt]
\displaystyle \mathrm{diag}_{\,N(N-1)/2}\!\left(\underbrace{1,1,1,\ldots,1}_{N-1},-1,..,-1\right)\,, & {\cal C}\in 2\mathbb Z+1\,,
\end{cases}\\
\end{eqnarray}
and we have explicitly verified this pattern using the Verma basis.

On the $\mathrm{EH}'$ space, we turn on the background
\begin{eqnarray}\label{2indexfluxes}
A'_{SU(N)}=\bm H_{\tiny\yng(1,1)}\cdot \bm\nu_1\,\frac{\sigma_z a^2}{r^2}\,,\qquad 
A'_{U(1)}=\left(\frac{2}{N}+{\cal C}'\right)\frac{\sigma_z a^2}{r^2}\,,
\end{eqnarray}
with the choice 
\begin{eqnarray}
{\cal C}' = 
\begin{cases}
0\,, & {\cal C}\in 2\mathbb{Z}\,,\\
-1\,, & {\cal C}\in 2\mathbb{Z}+1\,.
\end{cases}
\end{eqnarray}
The resulting $U(N)$ background on $\mathrm{EH}'$ is 
$A'_t = A'_{SU(N)} + I_{N(N-1)/2} A'_{U(1)}$, explicitly
\begin{eqnarray}\label{2indat}
A'_{t}
= \begin{cases}
\displaystyle\mathrm{diag}_{\,N(N-1)/2}\!\left(\underbrace{1,1,1,\ldots,1}_{N-1},0,..,0\right)\frac{\sigma_z a^2}{r^2}\,, & {\cal C}\in 2\mathbb Z\,,\\[6pt]
\displaystyle \mathrm{diag}_{\,N(N-1)/2}\!\left(\underbrace{0,0,0,\ldots,0}_{N-1},-1,..,-1\right)\frac{\sigma_z a^2}{r^2}\,, & {\cal C}\in 2\mathbb Z+1\,.
\end{cases}
\end{eqnarray}
This specific background produces no localized zero modes on $\mathrm{EH}'$, i.e. 
${\cal I}_{\scriptsize\mathrm{EH}'}=0$, and simultaneously reproduces the boundary holonomy~\eqref{2indexhol}. Repeating the argument leading to~(\ref{niceeta}) and using (\ref{SUNU1BTC}), we obtain (recall we are using ${\cal C}=1$, and thus, we must take ${\cal C}'=-1$)
\begin{eqnarray}\nonumber
\eta(\mathbb{RP}^3)&=&\frac{T_{\mathcal{R}}}{8\pi^2} \int_{\scriptsize\mathrm{EH}'} \mathrm{tr}_\Box\, F'_{(N)} \wedge F'_{(N)}  
      \;+\;\frac{\dim{\mathcal{R}}}{8\pi^2}\int_{\scriptsize\mathrm{EH}'}F'_{(1)} \wedge F'_{(1)} 
      \;+\;\frac{\dim {\cal R}}{24\cdot 8\pi^2}\int_{\mathrm{EH}'}\mathrm{tr}\, R'\wedge R'\\
      &=&\frac{(N-1)(N-4)}{16}\,.
\end{eqnarray}
Substituting this into~(\ref{APSINDEXTHEO}), we immediately find
\begin{eqnarray}\label{index2ind}
{\cal I}^{({\cal M})}={\cal I}^{(\scriptsize\mathrm{EH})}=N-1\,,
\end{eqnarray}
which we have also verified independently using the Verma basis construction of all weights.

In fact, this result admits a natural generalization to general ${\cal R}$ (assuming there exists ${\cal C'}$ that enables smooth sewing without changing the index) by exploiting the equivalence
$
 \mathcal{M} \cong S^2 \times S^2
$, 
and following the same logic employed between Eqs.~\eqref{sphersbegin} and \eqref{sphersend}.
Using the linear combinations of the two $2$–spheres $S_A^2$ and $S_B^2$ introduced in Eq.~\eqref{linearcomS}, the $SU(N)$ and $U(1)$ gauge fields obey the flux quantization conditions (we set $m_{(N)}=1$; also compare with (\ref{fractional fluxes}))
\begin{align}
\frac{1}{2\pi}\!\int_{S^2_A}\!\!\! F_{(N)}^{(A)} &= \, \mathbf{H}_{\Box} \!\cdot\! \boldsymbol{\nu}_1 , &
\frac{1}{2\pi}\!\int_{S^2_A}\!\!\! F_{(1)}^{(A)} &= 
    \frac{\mathcal{C}+\mathcal{C}'}{2}+\frac{ n}{N}, \nonumber \\[6pt]
\frac{1}{2\pi}\!\int_{S^2_B}\!\!\! F_{(N)}^{(B)} &= 0 , &
\frac{1}{2\pi}\!\int_{S^2_B}\!\!\! F_{(1)}^{(B)} &= 
    \frac{\mathcal{C}-\mathcal{C}'}{2}. \label{fluxB}
\end{align}
Here we recall that $\mathcal{C}'=-1$ when $\mathcal{C}$ is odd, and $\mathcal{C}'=0$ when $\mathcal{C}$ is even, so that we don't alter the number of the fermion zero modes on EH space.  
The fields on $S^2_A$ correspond to the average of the fluxes on $\mathrm{EH}$ and $\mathrm{EH}'$, while those on $S^2_B$ correspond to the half-difference.  
Both $S_A^2$ and $S_B^2$ therefore carry fluxes compatible with the cocycle conditions, ensuring that the full $U(N)$ field strength
\begin{equation}
    F_t = F_{(N)} + F_{(1)}\, I_{\dim \mathcal{R}},
\end{equation}
as seen by a fermion in ${\cal R}$, satisfies the proper quantization:
\begin{equation}
    \int_{S_A^2} F_t^{(A)} \in 2\pi \mathbb{Z}, \qquad
    \int_{S_B^2} F_t^{(B)} \in 2\pi \mathbb{Z}.
\end{equation}

From Eq.~\eqref{fluxB}, the corresponding topological charges on $S^2\times S^2$ follow immediately:
\begin{eqnarray}
    Q_{(N)} = 0,\quad
    Q_{(1)} = 
    \left(\frac{\mathcal{C}+\mathcal{C}'}{2} + \frac{ n}{N}\right)
    \left(\frac{\mathcal{C}-\mathcal{C}'}{2}\right).
\end{eqnarray}
The Dirac index on the EH space, equal to that on $\mathcal{M}\cong S^2\times S^2$ by construction, is therefore
\begin{equation}\label{masterMformula}
\mathcal{I}_{\mathcal{R}}^{(\mathrm{EH})}
= \mathcal{I}_{\mathcal{R}}^{(\mathcal{M})}
= \dim \mathcal{R}
\left(\frac{\mathcal{C}+\mathcal{C}'}{2} + \frac{ n}{N}\right)
\left(\frac{\mathcal{C}-\mathcal{C}'}{2}\right),
\end{equation}
a result that we verified independently using the Verma basis. 
A particularly simple case is obtained by choosing $\mathcal{C}=1$, for which $\mathcal{C}'=-1$, yielding
\begin{equation}
\mathcal{I}_{\mathcal{R}}^{(\mathrm{EH})}
= \mathcal{I}_{\mathcal{R}}^{(\mathcal{M})}
= \frac{n}{N} \, \dim \mathcal{R}.
\end{equation}

We emphasize, again, that the construction of $\mathcal{M}$ is merely a computational device that, in some cases, enables the evaluation of the index without requiring an explicit calculation of the $\eta$-invariant. This construction should not be interpreted as providing a universal geometric \emph{capping} of the EH space. In particular, there is no single manifold $\mathrm{EH}'$ that can consistently cap $\mathrm{EH}$ for \emph{all} fermionic charge assignments while {\em preserving} the Dirac index of each fermion species, since different fermion representations induce distinct boundary holonomies, which in general cannot be extended to the same bulk filling. Moreover, we must emphasize that, depending on ${\cal R}$, there are cases in which we cannot find a cap that preserves the number of zero modes on the original space. In such cases, we cannot apply (\ref{masterMformula}). Admittedly, it is unclear to us, a priori, without computing all the weights, which representations admit capping and which do not.

 \section{The EH space as a probe of anomalies}
 \label{The EH space as a probe of anomalies}

Here, we begin by reviewing the diagnosis of anomalies from two complementary perspectives: (i) the mapping torus construction, and (ii) a field-theoretic approach that characterizes anomalies through the non-invariance of the partition function under symmetry transformations in the presence of non-trivial background fields. Next, we present the anomaly in the EH space and explain how this anomaly differs qualitatively from the previously known anomalies on compact manifolds.

\subsection{Diagnosis of anomalies}
\label{Diagnosis of anomalies}

We consider a QFT ${\cal Q}$ with fermions charged under a symmetry group
\begin{eqnarray}
G = \frac{G_1 \times G_2 \times \cdots}{H}\,,
\end{eqnarray}
where both $G_1$ and $G_2$ are continuous or discrete abelian groups, while $H$ is a discrete quotient. We introduce a background flux (a $2$-form curvature) associated with $G_1$---taken to be either a single $U(1)$ factor or generated by the Cartan elements of a $U(N)$ group---supported at a $2$-cycle of $X_4$, a general $4$-D space. We then perform a transformation by another abelian subgroup $G_2$ over the partition function of ${\cal Q}$.

To detect the anomaly of $G_2$ on $X_4$ (which is a mixed $G_2$--$G_1$--gravity anomaly), we proceed as follows. We choose an element
$
u \in G_{2}
$
whose anomaly we wish to probe, and define the $5$D mapping torus \cite{Dai:1994kq,Garcia-Etxebarria:2018ajm}
\begin{eqnarray}
Y_{5}
=
\frac{X_{4}\times [0,1]}{(x,1)\sim (u\cdot x,0)}\,, 
\qquad
x \in (p, \mathcal{A}_{G_1}, \mathcal{B}_{G_2}) \, .
\end{eqnarray}
Here $p \in X_4$ is a spacetime point and $\mathcal{A}_{G_1}$ and $\mathcal{B}_{G_2}$ denote the background bundle data for $G_1$ and $G_2$, respectively. As we mentioned,  we turn on a nontrivial $2$-form curvature for $G_1$ on $X_4$, while the $G_2$ curvature on $X_4$ is set to zero; however, its holonomy around $S^{1}_{\tau}$ (constructed by identifying $\{0\}$ and $\{1\}$ in $[0,1]$) is nonvanishing. The twist
$
(x,1)\sim (u\cdot x,0)
$
is to be understood as the identification
$
(p, \mathcal{A}_{G_1}, \mathcal{B}_{G_2})(1)
\sim
(p, \mathcal{A}_{G_1}, u\cdot \mathcal{B}_{G_2})(0)\,.
$

The Dirac operator on $Y_5$ is given by
\begin{eqnarray}
\slashed D_{Y_5}&=&\gamma^5\left(\frac{\partial}{\partial\tau}+iB_\tau\right)+\slashed D_{X_4}\,,\quad
\slashed D_{X_4}=\left[\begin{array}{cc}0& L^\dagger\equiv{\hat\sigma}^a D_a\\ 
L\equiv\bar{\hat\sigma}^a D_a&0\,,
\end{array}\right],
\end{eqnarray}
or explicitly
\begin{eqnarray}\label{DIRACY5}
\slashed D_{Y_5}=\left[\begin{array}{cc} \left(\frac{\partial}{\partial \tau}+iB_\tau\right)I_2 &L^\dagger\\ 
L& - \left(\frac{\partial}{\partial \tau}+iB_\tau\right)I_2
\end{array}\right]\,,
\end{eqnarray}
which is simply the extension of $\slashed D_{X_4}$ to $5$-D using the $4$-D chiral operator $\quad \gamma^5=\mbox{diag}\left(I_2,-I_2\right)$. Here, $D_a$ is the Dirac operator coupling to both gravity and the background gauge field associated with the symmetry group $G_1$ (all $G_1$ group indices are suppressed in this section). Also, $B_\tau$ is the $G_2$ flat connection on $S^1_\tau$. We define $\hat\sigma^a$ and $\bar{\hat\sigma}^a$ as  $\hat\sigma^a = (1, i\vec{\hat\sigma})$ and $\bar{\hat\sigma}^a = (1, -i\vec{\hat\sigma})$, with $\vec{\hat\sigma}=(\hat\sigma_1,\hat\sigma_2,\hat\sigma_3)$ and $\hat\sigma_{1,2,3}$  the Pauli matrices; the Latin indices $a=1,2,3,4$ label components in the locally flat tangent space (orthonormal frame). 

According to the Dai-Freed theorem \cite{Dai:1994kq,Witten:2015aba,Witten:2019bou}, the anomaly of ${\cal Q}$ on  $X_4$ is the phase of the regularized partition function on $Y_5$, which is the $\eta$-invariant of the Dirac operator defined over $Y_5$:
\begin{eqnarray}\label{5DETa}
{\cal A}_{{\cal Q}}=e^{i2\pi \eta({Y_5})}\,, \quad\eta({Y_5})\equiv\frac{1}{2}\left(\mbox{lim}_{T\rightarrow 0}\sum_{\lambda_{(5)}\neq 0}\frac{\mbox{sign}(\lambda_{(5)})}{|\lambda|^{T}_{(5)}}+\mbox{dim ker}(i\slashed D_{Y_5})\right)\,,
\end{eqnarray}
where the sum is over the eigenvalues of the $5$-D Dirac operator $i\slashed D_{Y_5}$. One can easily calculate $\eta({Y_5})$, thanks to Theorem 1.7.1 in \cite{Gilkey2018}. This theorem was proved for compact $Y_5$, but we can easily adapt\footnote{We thank I. Garc\'{\i}a Etxebarria for illuminating discussion on this point.} it to open spaces of the type we are considering in our work $\mbox{EH}\times S^1_\tau/r$, and the twist $r$ is defined as above.

To this end, we first note that the operator $L \equiv \bar{\hat\sigma}^a D_a$ is not Hermitian and therefore lacks a real spectrum or a complete set of orthonormal eigenstates; indeed, $L$ maps left-handed fermions to right-handed ones. 
We instead define two Hermitian operators,
$
H_L \equiv L^\dagger L\,, H_R \equiv L L^\dagger\,.
$
These satisfy
$
H_L \phi_n(x) = \lambda_n^2\,\phi_n(x)\,, H_R \chi_n(x) = \lambda_n^2\,\chi_n(x)\,,
$
with identical nonzero eigenvalues $\lambda_n \neq 0$ and eigenfunctions related by $\chi_n = L\phi_n/\lambda_n$. 
We assume that, on the background $X_4$, the operator $H_R$ has no kernel (i.e.~$\hat\sigma^n D_n \bar\psi=0$, where $\bar\psi$ is the right-handed fermion, has no normalizable solution), while $H_L$ possesses a nontrivial kernel (meaning that the left-handed fermions admit zero modes), as occurs on the EH space. 
Thus, $H_L \phi_i^{(0)}=0$ for $i=1,2,\ldots,{\cal I}^{(X_4)}$. 
The set $\{\phi\}$, including the zero modes, forms a complete orthonormal basis.

Next, let $S^{1}_\tau$ have a unit length, and consider
$
i D_\tau \equiv i\left(\frac{d}{d\tau} + i B_{\tau}\right),
$
reminding that $B_{\tau}$ is the flat $U(1)$ connection whose holonomy is $e^{2\pi i \beta_\tau}$, and we take $\beta_\tau \in (0,1)$,  so that the spinor on  $S^1_\tau$ obeys the twisted boundary condition
$
\vartheta(\tau+1) =e^{-2\pi i\beta_\tau}\,\vartheta(\tau)\,, 0 < \beta_\tau < 1.
$
The spectrum of $iD_\tau $ (the solution of the eigenvalue problem $iD_\tau \vartheta=\omega\vartheta$) is then
$
\omega_{p} = 2\pi\,(p+\beta_\tau)\,, p\in\mathbb{Z},
$
with eigenvectors $\vartheta_p=e^{-i\omega_p\tau}$.

The eigenbasis
\begin{eqnarray}
{\cal W}_{n,p}=\left[\begin{array}{c}
\phi_n\\
\frac{L\phi_n}{\lambda_n}\end{array}
\right]\otimes \vartheta_p\,, \quad \lambda_n\neq 0
\end{eqnarray}
along with
\begin{eqnarray}
{\cal W}_{i,p}=\left[\begin{array}{c}
\phi_i^{(0)}\\
0\end{array}
\right]\otimes \vartheta_p\,, \quad i=1,2,.., {\cal I}^{(X_4)}
\end{eqnarray}
span the eigenspace of $\slashed D_{Y_5}$. In particular, for $\lambda_n\neq 0$, it is easily seen that
\begin{eqnarray}
i\slashed D_{Y_5}{\cal W}_{n,p}=\left[\begin{array}{cc}i\omega_p &i\lambda_n\\ i\lambda_n& -i\omega_p\end{array}\right]{\cal W}_{n,p}\,,
\end{eqnarray}
with eigenvalues of $i\slashed D_{Y_5}$ given by $\lambda_{(5)n,p}=\pm\sqrt{\omega_p^2+\lambda_n^2}$. These eigenvalues are distinct and opposite in sign, and thus, their contribution cancels out in the definition of the $\eta$-invariant of $\slashed D_{Y_5}$ given in (\ref{5DETa}). What remains is the contribution from the eigenvalues of ${\cal W}_{i,p}$. Given that there are ${\cal I}^{(X_4)}$ zero modes of $\slashed D_{X_4}$, the definition  (\ref{5DETa}) immediately yields:
\begin{eqnarray}
\eta({Y_5})={\cal I}^{(X_4)}\,\eta(i D_\tau)\,,
\end{eqnarray}
where  $\eta(iD_\tau)$ is the $\eta$-invariant of the $1$-D Dirac operator defined on the circle $S^1_\tau$. Simple calculations yield the result\footnote{
The spectrum of $i D_\tau $ is 
$
\omega_{n} = 2\pi\,(p+\beta_\tau)\,, p\in\mathbb{Z}.
$
The $\eta$-invariant is defined as
$
\eta(i D_\tau,s) =\frac{1}{2}\sum_{p\in\mathbb{Z}} 
\frac{\mathrm{sgn}(\omega_{p})}{|\omega_{p}|^{\,s}},
$
and it admits an analytic continuation to $s=0$. Evaluating at $s=0$ gives
$
\eta(iD_\tau) \equiv\eta(i D_\tau,0)
=
\frac{1}{2} -  \beta_\tau\,.
$
 The overall constant $\frac{1}{2}$ does not enter the anomaly calculations, so we can just drop it.
}
\begin{eqnarray}\label{mainanaomalyEQ}
{\cal A}_{{\cal Q}}=e^{-i2\pi \beta_\tau{\cal I}^{(X_4)}}\,.
\end{eqnarray}
When $G_2=U(1)$, then $\beta_\tau \in (0,1)$, while when $G_2=\mathbb Z_p$, then $\beta_\tau=\frac{k}{p}$, $k=1,2,..,p-1$.
 
 We must emphasize here that since the index ${\cal I}^{(X_4)}$ is a topological invariant (recall the discusion after Eq. (\ref{etavalrp3}), that the spectrum of the boundary Dirac operator defined on $\mathbb{RP}^3$ does not pass through $0$, and hence, the Dirac index on EH space does not change with smooth variation of the metric and gauge fields), no local counterterms can cancel the anomaly ${\cal A}_{{\cal Q}}$: the anomaly is robust under RG flow.

We can also follow a standard field-theoretic derivation to reproduce the anomaly expression~(\ref{mainanaomalyEQ}); see, e.g., \cite{Gibbons:1979kq} and Appendix C in \cite{Anber:2025vjo}.
Without loss of generality, we express the UV theory entirely in terms of left-handed Weyl fermions, collectively denoted by $\psi$, whose action takes the form
$
S_F = -\int_{X_4}\!\sqrt{g}\;\bar\psi\,\bar{\hat\sigma}^a D_a \psi\,,
$
where $g$ denotes the metric on $X_4$. 
We expand Weyl fermions $\psi$ and $\bar\psi$ in the basis $\phi_n$, $\chi_n$, and $\phi_i^{(0)}$ defined above:
\begin{eqnarray}\label{modeexpan}
\psi(x)
&=& \sum_{i=1}^{{\cal I}^{(X_4)}} \eta_i^{(0)}\,\phi_i^{(0)}(x)
   + \sum_{n>{\cal I}^{(X_4)}} \eta_n\,\phi_n(x)\,, 
   \qquad 
\bar\psi(x)
 = \sum_{n>{\cal I}^{(X_4)}} \bar\eta_n\,\chi_n(x)\,,
\end{eqnarray}
where $\eta_i^{(0)}, \eta_n,$ and $\bar\eta_n$ are Grassmann variables. 
We have omitted $\bar\eta_i^{(0)}$ in the expansion of $\bar\psi$ because $\hat\sigma^a D_a\bar\psi=0$ has no normalizable solutions. 
For non-compact spaces, the discrete sum over nonzero modes is replaced by an integral over a continuum of scattering states. 
In particular, for the EH space, the continuous spectrum of scattering states spans all values from $\lambda\to 0$ to $\lambda\to\infty$; see the analysis starting from Eq. (\ref{nonzeromodesanalysis}) in Appendix \ref{sec:fermions-EH}.

Substituting the mode expansion~(\ref{modeexpan}) into the partition function, and using the orthonormality of $\{\phi\}$, yields
\begin{eqnarray}\nonumber
&&Z[X_4,J,\bar J]
= \int \mathcal{D}\psi\,\mathcal{D}\bar\psi\;
   e^{-S_F + \int_{X_4}\!\sqrt g\,\big(J(x)\psi(x)+\bar J(x)\bar\psi(x)\big)}
   \;\longrightarrow\\\nonumber
&&\int \prod_{i=1}^{{\cal I}^{(X_4)}}\! d\eta_i^{(0)} 
   \prod_{n>{\cal I}^{(X_4)}}\! d\eta_n
   \prod_{n'>{\cal I}^{(X_4)}}\! d\bar\eta_{n'}\;\\
   &&\quad\times\exp\!\left[
   \sum_{n>{\cal I}^{(X_4)}}\!\lambda_n \bar\eta_n \eta_n
   +\sum_{i=1}^{{\cal I}^{(X_4)}} j_i^{(0)}\eta_i^{(0)}
   +\sum_{n>{\cal I}^{(X_4)}}\!\left(j_n \eta_n + \bar j_n \bar\eta_n\right)
   \right]\!.
\end{eqnarray}
Here, we have introduced classical sources $J(x)$ and $\bar J(x)$ and expanded them in the same basis:
$
J(x) = \sum_{i=1}^{{\cal I}^{(X_4)}} j_i^{(0)}\phi_i^{(0)}(x)
      + \sum_{n>{\cal I}^{(X_4)}} j_n\,\phi_n(x)\,,
\bar J(x) = \sum_{n>{\cal I}^{(X_4)}} \bar j_n\,\chi_n(x)\,.
$
Including sources is essential because, in their absence, the partition function vanishes due to the presence of fermionic zero modes.

Next, we examine the transformation properties of $Z[X_4,J,\bar J]$ under the symmetry $G_2$. 
Assuming that both the action and the fermion–source couplings are invariant under $G_2$, the Grassmann variables and sources transform as 
\begin{eqnarray}
\big(\eta_i^{(0)}, \eta_n, \bar j_n\big) 
  &\rightarrow& e^{-i2\pi\beta_\tau}\big(\eta_i^{(0)}, \eta_n, \bar j_n\big)\,, \quad
\big(\bar\eta_n, j_i^{(0)}, j_n\big)
  \rightarrow e^{\,i2\pi\beta_\tau}\big(\bar\eta_n, j_i^{(0)}, j_n\big)\,.
\end{eqnarray}
It follows immediately that all contributions to the measure from the nonzero-mode Grassmann variables $(\eta_n,\bar\eta_n)$ cancel under $G_2$—this remains true even for a continuous spectrum, as in the EH space, for all eigenvalues $0<\lambda<\infty$. 
The only surviving contribution comes from the zero-mode Grassmann variables $\eta_i^{(0)}$, which generate the anomalous phase
$
e^{-i2\pi\beta_\tau\,{\cal I}^{(X_4)}}\,,
$
a result that exactly matches (\ref{mainanaomalyEQ}) obtained via the mapping torus. 
A byproduct of this derivation is that one immediately sees that the condensate $\langle \psi^{{\cal I}^{(X_4)}} \rangle$ is nonvanishing, indicating that $G_2$ is broken\footnote{Here, we assume that the total number of the fermion zero modes is an even number so that the fermion number $(-1)^F$ is unbroken and the theory preserves its Lorentz invariance.  Also, we note that the breaking of $G_2$ here is a UV statement. In particular, for asymptotically free theories, the spectrum of the theory can change under the RG flow from the UV to the IR. The only thing that matters is that the UV anomaly must match the IR anomaly.}.

As an exercise, let us show that one recovers the standard quantum electrodynamics (QED) axial anomaly in the EH background. We consider a single Dirac fermion $\Psi$.
Classically, the theory possesses a global 
$U(1)_V \times U(1)_A$ symmetry. 
We decompose the Dirac spinor into two left-handed Weyl fermions, 
$\psi$ and $\tilde{\psi}$, whose charges under 
$U(1)_V \times U(1)_A$ are
$
Q(\psi) = (+1, +1), 
Q(\tilde{\psi}) = (-1, +1).
$
We now turn on a background $U(1)_V$ gauge field with flux
$
\int_{S^2} F_{(V)} = 2\pi \mathcal{C}\,,
\mathcal{C} \in \mathbb{Z}.
$
We then examine the fermionic partition function in this background while performing 
an axial rotation generated by $e^{i 2\pi \beta_A}$ of the $U(1)_A$ symmetry. 
Using Eqs.~(\ref{evenC}), (\ref{oddC}), (\ref{mainanaomalyEQ}), 
one finds the APS indices and anomalies:
\begin{align}\nonumber
\mathcal{I} &= p^2, 
& \mathcal{A} &= e^{i 4\pi \beta_A p^2}, 
& & \text{for } \mathcal{C} = 2p, \\[4pt]
\mathcal{I} &= p(p+1), 
& \mathcal{A} &= e^{i 4\pi \beta_A p(p+1)}, 
& & \text{for } \mathcal{C} = 2p + 1.
\end{align}
Even values of $\mathcal{C}$ correspond to trivial holonomy $(+1)$ 
at the boundary $\mathbb{RP}^3$, 
whereas odd values correspond to a holonomy of $(-1)$. 
Note that the minimal flux $\mathcal{C} = 1$ does not yield 
normalizable zero modes and thus contributes no anomaly. 
For $\mathcal{C} > 1$, zero modes appear (for every zero mode of $\psi$ there is a corresponding zero mode of $\tilde \psi$, and thus, the total number of zero modes is even in any background), inducing an axial anomaly.
In particular, when $\mathcal{C} = 2$, the $U(1)_A$ symmetry is broken\footnote{In fact, according to the new understanding of symmetries, $U_A$ becomes a noninvertible symmetry \cite{Choi:2022jqy,Cordova:2022ieu}. We do not discuss the realization of noninvertible symmetries in the EH background in this work, but we recognize that it is an important subject that we leave for a future study.} 
to its minimal discrete subgroup $\mathbb{Z}_2$, 
corresponding to the conserved fermion number in any Lorentz-invariant theory. We also note that the anomaly of $U(1)_B$ in the background of its own fluxes vanishes identically. We may also turn on fluxes along $U(1)_A$ and show that the anomaly of $U(1)_B$ vanishes in this background, which removes any obstruction to gauging $U(1)_B$, i.e., summing over $U(1)_B$ gauge bundles in the path integral\footnote{Although the two symmetries are present, this situation is 
different from the familiar vector/axial system in which the Bardeen
counterterm allows one to redistribute anomalies between two currents \cite{Bilal:2008qx}.
That mechanism requires both symmetries to admit independent background
gauge fields simultaneously.  In our case, only $G_1$ is equipped with a background field, while $G_2$ is
probed solely through the transformation $\beta_\tau$ (i.e., only via its holonomy) and has no independent
background gauge field with which a local counterterm could mix.}. 

The moral of this exercise is that the EH background
introduces no surprises in a theory as simple as QED:
it faithfully reproduces the expected $U(1)_A$ anomaly pattern.
However, there is a fundamental difference between the EH space
and closed manifolds, and this difference is tied to the
$\eta$-invariant contribution from $\mathbb{RP}^3$. To make this contrast explicit,
let us introduce $\theta$-angles $\theta_{R}$ and $\theta_{e}$
for the gravitational and electromagnetic sectors, respectively.
While $\theta_{e}$ multiplies the topological charge
$\int_{\scriptsize\mbox{EH}}F_{(V)}\wedge F_{(V)}/(8\pi^2)$,
the gravitational $\theta$-angle couples to the
Hirzebruch signature $\tau$ defined in (\ref{TINV}).
Recall that, on EH space, the second fundamental form $\Theta$
vanishes, and likewise the signature $\eta$-invariant.
Thus, the QED partition function with the $\theta$-angles switched on is
\begin{eqnarray}\nonumber
Z[\theta_{R},\theta_e]=\int D[\Psi]D[\bar\Psi]\exp\left[-i\frac{\theta_R}{24\pi^2}\int_{\scriptsize \mbox{EH}}\mbox{tr}R\wedge R+i\frac{\theta_e}{8\pi^2}\int_{\scriptsize\mbox{EH}}F_{(V)}\wedge F_{(V)}\right]e^{-S_{\scriptsize\mbox{QED}}}\,.\\
\end{eqnarray}
It is well known that on closed manifolds one can always rotate away
one $\theta$-angle for each anomalous symmetry in the theory.
Since $U(1)_A$ is anomalous, on a closed manifold one of the
$\theta$-angles, either the electromagnetic or the gravitational one,
can be removed by an axial rotation. Let us now ask whether the same is
true on the non-compact EH space. Under an axial transformation,
the measure transforms as
\begin{eqnarray}\nonumber
&& D[\Psi]D[\bar\Psi]\rightarrow  D[\Psi]D[\bar\Psi]\\
&&\times \exp\left\{i4\pi\beta_A\left[\frac{1}{24\cdot 8\pi^2}\int_{\scriptsize \mbox{EH}} \mbox{tr}R\wedge R+\frac{1}{8\pi^2}\int_{\scriptsize\mbox{EH}}F_{(V)}\wedge F_{(V)}-\eta(\mathbb{RP}^3)\right]\right\}\,.  
\end{eqnarray}
Thus, choosing $4\pi\beta_A=-\theta_e$ eliminates the bulk term
$\int_{\scriptsize\mbox{EH}}F_{(V)}\wedge F_{(V)}$.
Nevertheless, a residual phase $e^{i\theta_e\eta(\mathbb{RP}^3)}$
remains: the boundary $\eta$-invariant term (of the Dirac operator) still retains information\footnote{A follow-up question is whether the residual $\theta_e$ is physically observable, which we leave for a future investigation. This question might also be relevant to the Standard Model in the background of EH instantons, a subject that we discussed in \cite{Anber:2025gvb} in the context of gauging the SM $1$-form symmetry.}
about $\theta_e$. The same behaviour persists if we try to eliminate $\theta_{R}$ instead. 

The fact that an axial rotation can be used to remove the bulk $\theta_e$ term but leaves
behind a residual phase proportional to the $\eta$-invariant does not imply that the
anomaly is $3$-dimensional in nature. Rather, this is precisely what one expects from
the Atiyah--Patodi--Singer formulation of the four-dimensional chiral anomaly on a
manifold with boundary (and remember that the boundary lives at asymptotic infinity). In this setting, the anomaly functional naturally decomposes into
a bulk index density and a boundary $\eta$-invariant,
so that the effect of an anomalous axial rotation on the fermion measure necessarily
involves both bulk and boundary contributions. In particular, we are not introducing any
independent $3$-D degrees of freedom on the asymptotic $\mathbb{RP}^3$
boundary; the residual $\theta_e$-dependence is entirely a property of the
$4$-dimensional fermion determinant in this background. 

The feature we saw for the $\theta$-angle terms will persist when we turn to anomaly
matching with composite degrees of freedom, as we explain next.
In the following, however, we will set all $\theta$-angles to zero.

\subsection{The anomaly on EH space}
\label{anomaliesonEHspace}
EH space is simply connected and spin, with $H^2(\mathrm{EH},\mathbb Z)\cong \mathbb Z$ generated by the exceptional $2$-sphere of self-intersection $2$. For any integer charge $q_{\Psi}\in\mathbb Z$, there exists a globally defined line bundle $L^{q_\Psi}$ (associated to a gauge field), so Dirac fermions can be defined as sections of $S\otimes L^{q_\Psi}$, where $S$ is the spin bundle. Although the bulk cohomology is torsion-free, the boundary $\partial \mathrm{EH}\simeq\mathbb{RP}^3$ has torsion $H_1(\mathbb{RP}^3,\mathbb Z)\cong \mathbb Z_2$, implying that the bulk flux seen by the fermion mod $2$ determines the $\mathbb Z_2$ holonomy at infinity. Thus, EH space carries distinct topological information compared to other $4$-manifold, e.g., $S^4$, $S^2\times S^2$, $\mathbb T^4$, and $\mathbb{CP}^2$,  systematically used to detect anomalies of $0$- and $1$-form symmetries (see Appendix \ref{anomaliesonT4CP2} for a review of the anomalies on $\mathbb T^4$ and $\mathbb{CP}^2$). Therefore, we expect that the EH space can detect anomalies that cannot be seen in these spaces. Indeed, we shall show this is the case.

The anomalies studied in this work are of the ’t~Hooft type.  Matching ’t~Hooft anomalies on all backgrounds compatible with a given global symmetry places strong constraints on the possible IR dynamics of theories that flow to strong coupling. One logical possibility is that the theory preserves its global symmetries along the renormalisation group flow, in which case the IR spectrum must contain massless composite states that exactly reproduce the UV anomalies (we remain ignorant about the possibility that an IR emergent symmetry-preserving TQFT could match the anomalies).

To understand how the anomaly on the EH space differs from the ones identified on $S^4$, $S^2\times S^2$, $\mathbb T^4$, and $\mathbb{CP}^2$,  consider an $SU(N)$ gauge theory with global symmetry containing $U(1)_1 \times U(1)_2$ and a nonabelian flavour symmetry.  Assume that the IR vacuum preserves the full global symmetry.  The IR spectrum must then include composite operators that reproduce the complete set of UV ’t~Hooft anomalies.

To probe the anomalies, we introduce background fluxes for the color, flavour, and $U(1)_1$ symmetries, and apply a $U(1)_2$ transformation to the partition function via $e^{-i2\pi\beta_\tau}$.  On $S^4$, let $Q_c$, $Q_f$, $Q_{u_1}$, and $Q_G$ denote, respectively, the color, flavour, $U(1)_1$, and gravitational topological charges\footnote{Technically, there is no $U(1)$ instanton number on $S^4$. However, manifolds with non-trivial second cohomology group, e.g., $S^2\times S^2$, admit a non-zero $U(1)$ instanton number, and thus, we shall keep $Q_{u_1}$ as a bookkeeping device to compare anomalies on general manifolds.}.  The ratio of the UV and IR anomaly phases then takes the form
\begin{eqnarray}
\frac{{\cal A}_{\text{UV}}}{{\cal A}_{\text{IR}}}
= \frac{e^{i2\pi\beta_\tau{\cal I}_{UV}}}{e^{i2\pi\beta_\tau{\cal I}_{IR}}}=
\frac{
  e^{\,i2\pi\beta_\tau\left(0Q_c+ \ell_f^{(\text{UV})} Q_f + \ell_{u_1}^{(\text{UV})} Q_{u_1}+\ell_G^{(\text{UV})} Q_G \right)}
}{
  e^{\,i2\pi\beta_\tau\left(0Q_c+ \ell_f^{(\text{IR})} Q_f + \ell_{u_1}^{(\text{IR})} Q_{u_1}+\ell_G^{(\text{IR})} Q_G \right)}
}\,.
\end{eqnarray}
Here, the coefficients $\ell$ are group-theoretic integers determined by the matter content in the UV and IR. The color charge $Q_c$ is multiplied by zero because we assume $U(1)_2$ remains a good symmetry in the color background and the IR composites are color singlets. Since we are dealing with continuous (perturbative) symmetries, the anomalies must match exactly, implying \cite{Anber:2019nze}
\begin{eqnarray}
\ell_f^{(\text{UV})} = \ell_f^{(\text{IR})}, \qquad
\ell_{u_1}^{(\text{UV})} = \ell_{u_1}^{(\text{IR})}, \qquad
\ell_G^{(\text{UV})} = \ell_G^{(\text{IR})}.
\end{eqnarray}

Now, consider placing the same theory on $\mathbb{T}^4$ (or $S^2\times S^2$) and turning on fractional fluxes (e.g., $PSU(N)$ fluxes) that induce fractional topological charges. Provided such fluxes are compatible with the background geometry and matter content, the continuous nature of the symmetry, namely $U(1)_2$, ensures that the anomaly matching between the UV and IR sectors continues to hold automatically, since the $\ell$’s are fixed group-theoretic constants. 

However, this straightforward correspondence no longer holds on the EH space. According to the APS index theorem, there is now an additional contribution to the index (\ref{APSINDEXTHEO}, \ref{indexmastermix}), the $\eta$-invariant contribution $\text{sgn}(\text{Hol})/8$, which was absent in the perturbative anomalies. Consequently, the set of IR composites that matched the perturbative anomalies on $S^4$ is not guaranteed \emph{a priori} to reproduce this extra contribution on the EH background.

If $U(1)_2$ is replaced by a discrete symmetry, such as $\mathbb{Z}_q$, the composites that reproduce the perturbative anomalies on $S^4$ are no longer guaranteed to match the anomalies on $\mathbb{T}^4$ or $S^2 \times S^2$. This is especially relevant when fractional fluxes are present—for example, those associated with $1$-form symmetries—because anomaly matching between the UV and IR is then only required modulo $q$. Consequently, additional constraints are imposed on the spectrum of IR composites \cite{Anber:2019nze,Anber:2020gig}. Placing the theory on the EH space strengthens these constraints further, as its nontrivial topology introduces new, independent anomaly conditions that must also be satisfied.

In certain cases, the anomaly on the EH space admits a direct geometric interpretation via the capping construction described in Section~\ref{Gluing construction}. Suppose that in the UV there exists a manifold $\mathrm{EH}'$, equipped with a $U(1)'$ flux ${\cal C}'$, such that the glued space ${\cal M} = \mathrm{EH} \cup \mathrm{EH}'$ preserves the Dirac index of all fermions. If no analogous capping can be realized in the IR—using the same $U(1)'$ flux—such that the Dirac index of the composites is preserved, then the theory may exhibit a new anomaly.

If the composites fail to match the anomaly on the EH background, either additional gapless degrees of freedom must appear, the composites must couple to extra topological sectors, or the theory breaks its symmetry spontaneously.

\subsection{RG flow and varying the bolt size}
\label{Conclusion}

The EH metric introduces a geometric length scale $a$, set by the size of the bolt $2$-sphere. It is convenient to compare this to the dynamical strong scale $\Lambda$ of the gauge theory. Throughout, it is important to emphasize that the renormalized gauge coupling $g(\mu)$ is a function of the momentum scale $\mu$, not of the coordinate position. 

Modes that are localized near the bolt and vary over distances of order $a$ have typical momenta
\begin{equation}
  \mu_{\rm bolt} \sim 1/a \, .
\end{equation}
By contrast, long-wavelength excitations that probe the asymptotically flat region are associated with much smaller momenta $\mu \ll 1/a$. With this in mind, we distinguish two parametric regimes.

\paragraph{Regime $a\Lambda \ll 1$ (bolt scale in the UV).}

When $a\Lambda \ll 1$, the scale set by the bolt lies deep in the ultraviolet of the gauge theory,
\begin{equation}
  a\Lambda \ll 1 \quad \Longleftrightarrow \quad \mu_{\rm bolt} \sim 1/a \gg \Lambda \, ,
\end{equation}
and asymptotic freedom implies $g(1/a) \ll 1$. In this regime, the fermion modes whose profile is tied to the bolt --- in particular, the normalizable zero modes of the microscopic Dirac operator --- probe a weakly coupled regime of the gauge dynamics. Their structure can therefore be analyzed semiclassically using the explicit Dirac operator on the EH background. As we saw in the previous section, the quantum numbers of these microscopic zero modes reproduce the APS index ${\cal I}^{\rm (EH)}$ and hence the  't~Hooft anomaly in the UV is
\begin{equation}
  {\cal A}_{\rm UV} = e^{-i 2\pi \beta_\tau\, {\cal I}^{\rm (EH)}} \, .
\end{equation}

\paragraph{Regime $a\Lambda \gg 1$ (bolt scale in the IR).}

In the opposite limit $a\Lambda \gg 1$ the curvature scale $1/a$ is far below the strong scale $\Lambda$,
\begin{equation}
  a\Lambda \gg 1 \quad \Longleftrightarrow \quad \mu_{\rm bolt} \sim 1/a \ll \Lambda \, .
\end{equation}
Now the entire EH geometry lies deep in the infrared of the gauge theory: any mode whose wavelength is comparable to the size of the bolt probes momenta well below the scale at which the microscopic quarks and gluons become strongly coupled. If the IR spectrum contains massless composite fermions,\footnote{More generally, one can consider Goldstone fields, (strongly-coupled) conformal field theory, or an IR TQFT; we focus on composite fermions for concreteness.} then it is natural in this regime to introduce an \emph{effective} Dirac operator of these composite degrees of freedom. Placing this IR Dirac operator on the EH background, and imposing the same boundary conditions we used in the UV, one can again ask for normalizable zero modes of the composites.

Because the APS index is an RG-invariant quantity, the index of the IR Dirac operator must coincide with the microscopic EH index,
\begin{equation}
  {\cal I}_{\rm IR}^{\rm (EH)} = {\cal I}_{\rm UV}^{\rm (EH)} \equiv {\cal I}^{\rm (EH)} \, ,
\end{equation}
so that the composite zero modes reproduce the same anomaly phase
\begin{equation}
  {\cal A}_{\rm IR} = e^{-i 2\pi \beta_\tau\, I_{\rm IR}^{\rm (EH)}} = e^{-i 2\pi \beta_\tau\, I^{\rm (EH)}} = {\cal A}_{\rm UV} \, .
\end{equation}
In other words, in the regime $a\Lambda\gg 1$, the EH anomaly can be computed entirely within the IR effective description.

\paragraph{No double counting of anomaly-carrying states.}

The preceding discussion might suggest that, for intermediate values of $a\Lambda$, one could simultaneously talk about microscopic zero modes and composite zero modes on the same EH background, leading to an apparent ``double counting'' of anomaly-carrying states: one set localized near the bolt in the UV, and another set of IR composites propagating on the full manifold. This is not the correct interpretation.

There is a single topological index ${\cal I}^{\rm (EH)}$, determined by the APS index and the background data $(g_{\rm EH},A_{G_1})$, and a single anomaly phase ${\cal A}=e^{-i2\pi\beta_\tau {\cal I}^{\rm (EH)}}$. The microscopic description and the IR effective description are two different ways of packaging the \emph{same} anomaly-carrying degrees of freedom at different RG scales. In the regime $a\Lambda\ll 1$, it is appropriate to use the microscopic fermions at scale $\mu\sim 1/a$; in the regime $a\Lambda\gg 1$, it is appropriate to use the composite fermions at the same geometric scale. 

Equivalently, one can view ${\cal I}^{\rm (EH)}$ as a property of the full RG trajectory of the theory in the given background: it is computed reliably in the UV using semiclassical zero modes when $a\Lambda\ll 1$, and it must be matched by whatever IR phase the theory flows to when $a\Lambda\gg 1$. The EH anomaly, therefore, provides a nontrivial constraint on the allowed IR dynamics, but it does not imply the existence of two distinct sets of zero modes at the same time.

\paragraph{Are there boundary degrees of freedom on $\mathbb{RP}^3$?} We assume that there are no independent $3$-D degrees of freedom dynamically generated on the asymptotic $\mathbb{RP}^3$ boundary in the IR. The boundary is merely the spatial infinity of a complete noncompact space, so any bulk normalizable excitation decays at large radius. Combined with our UV definition (plus trivial boundary theory), this leads us to the working assumption that the EH anomaly must be saturated by $4$-D bulk IR physics---such as composite fermions, chiral Lagrangian, or TQFT. 

\subsection{Hilbert-space interpretation}

The EH anomaly arises from placing a theory on a non-compact ALE space with bulk flux and boundary holonomy. This can be naturally understood as a {\em relative anomaly} associated with that class of bulk-boundary backgrounds, though we will not attempt a full cohomological classification of the corresponding relative group.

To clarify this point, we now present a construction that gives a Hilbert–space interpretation of the EH anomaly. A related construction was developed in \cite{tHooft:1988wxy,Anber:2025gvb}, where the EH geometry appears as a saddle in the path integral of semi-classical Euclidean gravity. In the present work, however, our focus is on the anomaly itself, and we will not revisit the detailed interpretation of the states, referring instead to \cite{tHooft:1988wxy,Anber:2025gvb} for those aspects.
Let $r$, the radial coordinate on EH space, be the Euclidean ``time'' $\tau$, and truncate the EH space between two radii,
$
  \tau_{\rm in} < \tau_{\rm out},
$
to obtain a compact region ${\cal M}_{\rm EH}$ with two boundary components:
\begin{itemize}
  \item an inner boundary $\Sigma_{\rm in}$, e.g.\ a small $S^3/\mathbb Z_2$  surrounding the bolt region, and
  \item an outer boundary $\Sigma_{\rm out} \cong \mathbb{RP}^3$ at large radius, representing the asymptotic region.
\end{itemize}
Quantizing the theory on a spatial slice $\Sigma$ gives a Hilbert space $\mathcal{H}_\Sigma$. The Euclidean path integral over ${\cal M}_{\rm EH}$ with a fixed $G_1$ background $A_{G_1}$ defines a Euclidean time-evolution operator
\begin{equation}
  U_{\rm EH}(A_{G_1}) : \mathcal{H}_{\Sigma_{\rm in}} \longrightarrow \mathcal{H}_{\mathbb{RP}^3}.
\end{equation}
In a Hamiltonian language, one may think of
\begin{equation}
  U_{\rm EH}(A_{G_1})=
  \mathcal{T} \exp\!\left(-\int_{\tau_{\rm in}}^{\tau_{\rm out}} d\tau\,
  H(\tau; A_{G_1})\right),
\end{equation}
where $H(\tau;A_{G_1})$ is the Euclidean Hamiltonian in the given background and $\mathcal{T}$ denotes Euclidean time ordering. Choosing a reference ``initial'' state
\begin{equation}
  |0_{\rm in}\rangle \in \mathcal{H}_{\Sigma_{\rm in}},
\end{equation}
 with appropriate boundary conditions\footnote{The boundary conditions on states are discussed in \cite{tHooft:1988wxy,Anber:2025gvb}, which amounts to applying an antipodal identification, thanks to the modding by $\mathbb Z_2$. Under such identification, the prepared states are entangled states between two halves of the space. These details, however, are not vital for the anomaly we are discussing.}, the path integral on ${\cal M}_{\rm EH}$ prepares a state on $\mathbb{RP}^3$:
\begin{equation}
  |\Psi_{\rm EH}(A_{G_1})\rangle \;:=\; U_{\rm EH}(A_{G_1})\,|0_{\rm in}\rangle
  \;\in\; \mathcal{H}_{\mathbb{RP}^3}.
\end{equation}
Thus, the geometry of EH with background $A_{G_1}$ and a choice of initial condition produces a specific state in the Hilbert space of the $4$-D theory quantized on $\mathbb{RP}^3$.

To connect to the usual language of a ``partition function on EH space'', let $|a_{G_1}\rangle$ be a basis vector in $\mathcal{H}_{\mathbb{RP}^3}$ labeled by a configuration $a_{G_1}$ of the $G_1$ background restricted to the boundary (e.g. $a_{G_1}$ is the holonomy of $A_{G_1}$ seen at the boundary). The wavefunctional of the EH-prepared state is then
\begin{equation}
  \Psi_{\rm EH}[a_{G_1}]
  \;=\;
  \langle a_{G_1} | \Psi_{\rm EH}(A_{G_1})\rangle
  \;=\;
  \langle a_{G_1} | U_{\rm EH}(A_{G_1}) |0_{\rm in}\rangle.
  \label{eq:wavefunctional}
\end{equation}
This $\Psi_{\rm EH}[a_{G_1}]$ is precisely what one informally calls ``the partition function of the theory on EH space in the background $A_{G_1}$ with asymptotic data $a_{G_1}$''. The Euclidean time evolution is carried by $U_{\rm EH}$; the resulting object is a particular matrix element between an initial state on $\Sigma_{\rm in}$ and a boundary configuration on $\mathbb{RP}^3$.

Now let $u \in G_2$ be a global symmetry transformation. Quantization on $\mathbb{RP}^3$, then  a unitary operator $  U_{G_2}(u) $ acting on a state defined on $\mathbb{RP}^3$ yields:
\begin{equation}
  U_{G_2}(u) : \mathcal{H}_{\mathbb{RP}^3} \to \mathcal{H}_{\mathbb{RP}^3}.
\end{equation}
In the presence of an 't~Hooft anomaly, the state $|\Psi_{\rm EH}(A_{G_1})\rangle$ transforms as
\begin{equation}
  U_{G_2}(u)\,|\Psi_{\rm EH}(A_{G_1})\rangle
  \;=\;
  \mathcal{A}_{{\cal Q}}\,|\Psi_{\rm EH}(A_{G_1})\rangle,
  \label{eq:state_phase}
\end{equation}
for some phase $  \mathcal{A}_{{\cal Q}} \in U(1)$. In terms of the wavefunctional, one has
\begin{equation}
  \Psi_{\rm EH}[a_{G_1}] \;\longmapsto\;
  \Psi'_{\rm EH}[a_{G_1}]
  \;=\;
  \langle a_{G_1}|U_{G_2}(u)|\Psi_{\rm EH}(A_{G_1})\rangle
  \;=\;
    \mathcal{A}_{{\cal Q}}\,\Psi_{\rm EH}[a_{G_1}],
\end{equation}
so the ``partition function on EH space'' in that background acquires the phase $  \mathcal{A}_{{\cal Q}}$ under the $G_2$ transformation. This is the Hilbert space version of the statement that anomalies show up as phases of the partition function on the EH background.

If the theory has both a UV description (e.g.\ Yang--Mills plus fermions) and an IR description (e.g.\ composites, Goldstones, or a TQFT), then on the EH background they define two states
\begin{equation}
  |\Psi_{\rm EH}^{\rm UV}(A_{G_1})\rangle,
  \quad
  |\Psi_{\rm EH}^{\rm IR}(A_{G_1})\rangle
  \;\in\; \mathcal{H}_{\mathbb{RP}^3}.
\end{equation}
't~Hooft anomaly matching requires that these states transform with the same projective phase under $G_2$:
\begin{align}\nonumber
  U_{G_2}(u)\,|\Psi_{\rm EH}^{\rm UV}(A_{G_1})\rangle
  &=   \mathcal{A}_{{\cal Q}}\,|\Psi_{\rm EH}^{\rm UV}(A_{G_1})\rangle,\\
  U_{G_2}(u)\,|\Psi_{\rm EH}^{\rm IR}(A_{G_1})\rangle
  &=   \mathcal{A}_{{\cal Q}}\,|\Psi_{\rm EH}^{\rm IR}(A_{G_1})\rangle.
\end{align}
A mismatch in $ \mathcal{A}_{{\cal Q}}$ between the UV and IR computations signals that the IR description fails to reproduce the full anomaly and must be supplemented by additional/different gapless modes or an extra TQFT. In this way, the EH background provides a refined anomaly probe: it prepares a specific state in $\mathcal{H}_{\mathbb{RP}^3}$ via Euclidean time evolution, and the anomaly is precisely the phase with which this state transforms under the symmetry $G_2$.

In the following sections, we analyze the phases of several gauge theories, including both vector-like and chiral models. We employ the EH anomaly, along with anomaly constraints on $S^4$, $\mathbb{T}^4$, and $\mathbb{CP}^2$, to narrow down the possible infrared dynamics. We also provide worked examples of the theoretical construction outlined above.

\section{Applications I: Vector-like gauge theories}
\label{ApplicationsI}

We consider $SU(N_c)$ gauge theory with $N_f$ Dirac fermions in a color representation ${\cal R}$ of $N$-ality $n_c$.  
The gauge group that faithfully acts on the fermions is
\begin{equation}
\frac{SU(N_c)}{\mathbb Z_p}, 
\qquad p=\gcd(N_c,n_c).
\end{equation}
Hence, the fermions are charged under the $\mathbb Z_{N_c/p}$ subgroup of the center of $SU(N_c)$.
After quotienting by redundancies, the global symmetry group is
\begin{equation}\label{globalG}
G_{\text{global}}
=\frac{SU(N_f)_L \times SU(N_f)_R \times U(1)_B
\times \mathbb Z^\chi_{\,2\,N_f\,T_{{\cal R}}}}
{\mathbb Z_{N_c/p}\times \mathbb Z_{N_f}\times \mathbb Z_2^{\rm F}}\times \mathbb Z_p^{(1)}.
\end{equation}
Here, $U(1)_B$ is the baryon-number symmetry acting on the fermions, while $T_{\cal R}$ denotes the Dynkin index of the representation ${\cal R}$ (normalized as $T_{\Box}=1$ in the defining representation).  We assume that the chiral symmetry $\mathbb Z_{2\,N_fT_{{\cal R}}}^\chi$ is a genuine symmetry of the theory; thus, it cannot be absorbed in the continuous part of $G_{\scriptsize\mbox{global}}$ (this can be checked on a case-by-case basis). The discreet group $\Z_2^{\rm F}$ in the denominator denotes fermion number, $\Z_{N_c/p}$ is in the center of $SU(N_c)$, while $\Z_{N_f}$ is in the center of the contnous flavour symmetries $SU(N_f)_L \times SU(N_f)_R$.
In addition, a $\mathbb Z_p^{(1)}$ $1$-form center symmetry acts on Wilson loops whenever $\gcd(N_c,n_c)=p>1$.  

In the following, we shall think of a Dirac fermion $\Psi$ as two left-handed Weyl fermions: $\psi$ and $\tilde\psi$. We summarize their gauge and global quantum numbers in the table below:
\begin{equation}\label{charges1}
\begin{array}{c|ccccc}
 & SU(N_c) & SU(N_f)_L & SU(N_f)_R & U(1)_B & \mathbb{Z}^{\chi}_{2N_f T_{\cal R}} \\ \hline
\psi        & {\cal R} & \Box & 1 & 1 & 1 \\
\tilde\psi  & \overline{\cal R} & 1 & \overline{\Box} & -1 & 1
\end{array}
\end{equation}
To find anomalies, we  put $\psi$ and $\tilde\psi$ in various spaces, $\mathbb T^4$, $\mathbb{CP}^2$, and EH, in the background of gauge fields along the Cartan subalgebra of $SU(N_c)$ as well as $U(1)_B$.

\subsection{Chiral symmetry breaking in single-flavour theories}

  We consider asymptotically free theories that possess genuine discrete chiral 
symmetries, and examine their partition functions in the background of EH spaces with fluxes localized at the bolt. 
When ${\cal R}$ is a complex representation, we shall consider a single Dirac flavour, while when ${\cal R}$ is real (${\cal R}=\overline{\cal R}$), then we shall consider a single Weyl flavour\footnote{Recall that a theory with a single Weyl fermion in a complex representation has a gauge anomaly, and thus, it must be avoided by considering a Dirac fermion instead.}. In this latter case, the theory does not have $U(1)_B$ symmetry, and thus, we should only use the background gauge field along the Cartan generators of $SU(N_c)$ to consistently define the theory in the EH background. 

We seek to answer the following question. If a theory exhibits chiral symmetry breaking in the infrared, what is the minimal pattern of chiral breaking required to match the ’t Hooft anomaly of the ultraviolet theory? Equivalently, how strongly does the anomaly constrain the degree of chiral symmetry breaking in the IR?

Suppose a given theory admits ${\cal I}_{\cal R}$ fermion zero modes and enjoys a 
$\mathbb{Z}_{2T_{\cal R}}^\chi$ chiral symmetry. Then the action of 
$\mathbb{Z}_{2T_{\cal R}}^\chi$ on the partition function is
\begin{eqnarray}\label{anomalyphase}
Z[\mathrm{EH}] 
\xrightarrow{\;\mathbb{Z}_{2T_{\cal R}}^\chi\;} 
e^{i\frac{2\pi {\cal I}_{\cal R}}{T_{\cal R}}}\,Z[\mathrm{EH}] \,.
\end{eqnarray}
This phase is valued in $\mathbb Z_{\scriptsize T_{\cal R}/\mbox{gcd}(T_{\cal R},{\cal I}_{\cal R})}$, and the most constraining anomaly results when $\mbox{gcd}(T_{\cal R},{\cal I}_{\cal R})=1$, which yields a phase in the proper $\mathbb Z_{T_{\cal R}}$. The phase must be reproduced by the infrared dynamics, thereby forcing the chiral symmetry to fully break down to the $(-1)^F$ fermion number. 

Recall that the background fluxes are designated by two free parameters $m_{N_c}, {\cal C}\in \mathbb Z$, as in (\ref{ASUNCBULK}, \ref{abelian background needed}). Then, the corresponding integral entries ${\cal C}_a$ in $A_t$, see (\ref{finalAt}), are dependent on $m_{N_c}, {\cal C}$ as well as the specific representation ${\cal R}$, and they can be found using the method of Verma basis.  In the following, we shall set $m_{N_c}=1$ for all complex representations, while the value of $\cal{C}$ will be chosen to yield the most constraining phase in (\ref{anomalyphase}). When ${\cal R}$ is real, then there is no $U(1)_B$ and we need to select $m_{N_c}$ such that the fermions are rendered well-defined in the background gauge field of $SU(N_c)$. The number of the zero modes can be determined from the original APS-index formula (\ref{indexmastermix}) after using the Verma basis. Alternatively, we can use Eq. (\ref{masterMformula}) to determine the number of zero modes, provided that the gluing method introduced in Section \ref{Gluing construction} can be applied. We shall see that there are situations where one cannot find $\mbox{EH}'$ that can be smoothly glued to EH while preserving the number of the zero modes in the original space.

In Table \ref{breaking pattern}, we list all the single-flavour vector-like $SU(N_c)$ asymptotically-free theories (this is determined from the $1$-loop $\beta$-function (\ref{beta function})) with $3\leq N_c\leq 9$. The Dynkin indices and dimension of representation are directly obtained from expressions  (\ref{TRGR})-(\ref{beta function}). In each case, we compute the traditional anomaly $\mathbb Z_{2T_{\cal R}}^\chi [U(1)_B]^2$ (we use the letter B to denote this anomaly in Table \ref{breaking pattern}), the Baryon-Color (BC) anomaly on spin and non-spin manifolds (see below), as well as the anomaly on the EH space.

\begin{table}[htbp]
\centering
\resizebox{\textwidth}{!}{%
\begin{minipage}{0.48\textwidth}
\centering
\scriptsize
\tabcolsep=0.11cm
\begin{tabular}{|c|c|c|c|c|c|c||c|}
\hline
$G$ &$\cal R$ & $T_{\cal R}$ & $\dim{\cal R}$ & B &  $\mathbb T^4$  & $\mathbb {CP}^2$ & EH \\\hline
$SU(3)$& $(2,0)$ & $5$ & $6$ & $\mathbb  Z_5$ & $\mathbb Z_5$ &  $\mathbb Z_5$ &  $\mathbb Z_5$  \\\hline
&$(3,0)$ & $15$ & $10$ & $\mathbb  Z_3$ & $\mathbb Z_3$ & $\mathbb Z_3$ & \colorbox{green}{$\mathbb Z_{15}$}\\\hline
&$(1,1)$ & $6$ & $8$ & --& $\mathbb Z_3$ &-- & $\mathbb Z_3$\\\hline
&$(2,1)$ & $20$ & $15$ & $\mathbb  Z_4$ & $\mathbb Z_4$ & $\mathbb Z_4$ & \colorbox{green}{$\mathbb{Z}_{20}$}\\\hline
\hline\hline
$SU(4)$&$(2,0,0)$ & $6$ & $10$ & $\mathbb  Z_3$ & $\mathbb Z_6$ & $\mathbb Z_6$ & $\mathbb Z_6$ \\\hline
&$(3,0,0)$ & $21$ & $20$ & $\mathbb  Z_{21}$ & $\mathbb Z_{21}$ & $\mathbb Z_{21}$ & $\mathbb Z_{21}$ \\\hline
&$(0,1,0)$ & $2$ & $6$ & -- & $1$ & --  & $1$\\\hline
&$(0,2,0)$ & $16$ & $20$ & -- & $\mathbb Z_4$ & -- & \colorbox{green}{$\mathbb Z_8$}\\\hline
&$(1,0,1)$ & $8$ & $15$ & -- & $\mathbb Z_{4}$ &-- & $\mathbb Z_4$ \\\hline
&$(1,1,0)$ & $13$ & $20$ & $\mathbb  Z_{13}$ &  $\mathbb Z_{13}$ & $\mathbb Z_{13}$& $\mathbb  Z_{13}$ \\\hline
&$(2,0,1)$ & $33$ & $36$ & $\mathbb  Z_{11}$ &  $\mathbb Z_{11}$ & $\mathbb Z_{11}$ & $\mathbb Z_{11}$\\\hline
\hline\hline
$SU(5)$ & $(2,0,0,0)$ & $7$ & $15$ & $\mathbb Z_7$ & $\mathbb Z_7$ & $\mathbb Z_7$ & $\mathbb Z_7$\\\hline
& $(0,1,0,0)$ & $3$ & $10$ & $\mathbb Z_3$ & $\mathbb Z_3$ &  $\mathbb Z_3$ & $\mathbb Z_3$\\\hline
& $(1,0,0,1)$ & $10$ & $24$ & -- & $\mathbb Z_5$ &-- & $\mathbb Z_5$\\\hline
& $(1,1,0,0)$ & $22$ & $40$  & $\mathbb Z_{11}$ & $\mathbb Z_{11}$ & $\mathbb Z_{11}$& $\mathbb Z_{11}$\\\hline
& $(1,0,1,0)$ & $24$ & $45$ & $\mathbb Z_8$ & $\mathbb Z_{8}$ &$\mathbb Z_{8}$ & $\mathbb Z_8$\\\hline
\end{tabular}
\end{minipage}
\hfill
\begin{minipage}{0.48\textwidth}
\centering
\scriptsize
\tabcolsep=0.11cm
\begin{tabular}{|c|c|c|c|c|c|c||c|}
\hline
$G$ &$\cal R$ & $T_{\cal R}$ & $\dim {\cal R}$ & B & $\mathbb T^4$ & $\mathbb {CP}^2$& EH\\\hline
$SU(6)$ & $(2,0,0,0,0)$ & $8$ & $21$ & $\mathbb Z_8$ & $\mathbb Z_8$ & $\mathbb Z_8$ & $\mathbb Z_8$\\\hline
& $(0,1,0,0,0)$ & $4$ & $15$ & $\mathbb Z_4$ & $\mathbb Z_4$ &  $\mathbb Z_4$ & $\mathbb Z_4$\\\hline
& $(0,0,1,0,0)$ & $6$ & $20$ & -- & $\mathbb Z_3$ & --& $\mathbb Z_3$\\\hline
& $(1,0,0,0,1)$ & $12$ & $35$ &-- & $\mathbb Z_{6}$ &-- & $\mathbb Z_{6}$  \\\hline
& $(1,1,0,0,0)$ & $33$ & $70$ & $\mathbb Z_{33}$ & $\mathbb Z_{33}$ & $\mathbb Z_{33}$& $\mathbb Z_{33}$\\\hline
\hline\hline
$SU(7)$ & $(2,0,0,0,0,0)$ & $9$ & $28$ & $\mathbb Z_9$ & $\mathbb Z_9$ &  $\mathbb Z_9$ & $\mathbb Z_9$\\\hline
& $(0,1,0,0,0,0)$ & $5$ & $21$ & $\mathbb Z_5$ & $\mathbb Z_5$ & $\mathbb Z_5$ & $\mathbb Z_5$\\\hline
 & $(1,0,0,0,0,1)$ & $14$ & $48$ & -- & $\mathbb Z_7$ &--  & $\mathbb Z_7$\\\hline
& $(0,0,1,0,0,0)$ & $10$ & $35$ & $\mathbb Z_2$ & $\mathbb Z_2$ &$\mathbb Z_2$ & $\mathbb Z_2$\\\hline
\hline\hline
$SU(8)$ & $(2,0,0,0,0,0,0)$ & $10$ & $36$ & $\mathbb Z_5$ & $\mathbb Z_{10}$  & $\mathbb Z_{10}$ & $\mathbb Z_{10}$\\\hline
& $(0,1,0,0,0,0,0)$ & $6$ & $28$ & $\mathbb Z_3$ & $\mathbb Z_6$  &  $\mathbb Z_6$  & $\mathbb Z_6$ \\\hline
& $(0,0,0,1,0,0,0)$ & $20$ & $70$ & $\mathbb Z_2$ & $\mathbb Z_4$ &-- & {$\mathbb Z_4$}\\\hline
& $(1,0,0,0,0,0,1)$ & $16$ & $63$ & --& $\mathbb Z_{8}$ & --& $\mathbb Z_8$ \\\hline
& $(0,0,1,0,0,0,0)$ & $15$ & $56$ & $\mathbb Z_{15}$ & $\mathbb Z_{15}$  & $\mathbb Z_{15}$ & {$\mathbb Z_{15}$}\\\hline
\hline\hline
$SU(9)$ & $(2,0,0,0,0,0,0,0)$ & $11$ & $45$ & $\mathbb Z_{11}$ & $\mathbb Z_{11}$ &  $\mathbb Z_{11}$& $\mathbb Z_{11}$\\\hline
& $(0,1,0,0,0,0,0,0)$ & $7$ & $36$ & $\mathbb Z_{7}$ & $\mathbb Z_7$ & $\mathbb Z_7$ & $\mathbb Z_7$\\\hline
& $(1,0,0,0,0,0,0,1)$ & $18$ & $80$ & -- & $\mathbb Z_9$ & -- &  $\mathbb Z_9$\\\hline
\end{tabular}
\end{minipage}
}
\caption{\label{breaking pattern}We consider the asymptotically free representations of the gauge groups $SU(3)$ through $SU(9)$. Each representation is specified by its Dynkin labels, ${\cal R} = (n_1, n_2, \ldots, n_{N_c-1}) \equiv \sum_{a=1}^{N_c-1} n_a\, \bm{w}_a$, where $\bm{w}_a$ denote the fundamental weights. A representation is said to be \emph{real} if its Dynkin labels satisfy $(n_1, n_2, \ldots, n_{N_c-1}) = (n_{N_c-1}, n_{N_c-2}, \ldots, n_1)$. For example, the representations $(1,1)$, $(1,0,1)$, $(0,1,0)$, and $(0,2,0)$ are all real. In such cases, special care is required because the $U(1)_B$ baryon number symmetry is enhanced to an $SU(2)_f$ flavor symmetry. To avoid this additional complication, we consider a single Weyl fermion whenever the representation is real. The discrete chiral symmetry then becomes $\mathbb{Z}_{T_{\cal R}}^{\chi}$, and both the baryon number symmetry and the mixed anomaly $\mathbb{Z}_{2T_{\cal R}}^{\chi}[U(1)_B]^2$ vanish. We explicitly exclude the defining (fundamental) representation, $(1,0,0,\ldots,0)$, since gauge theories with fundamentals do not possess genuine discrete chiral symmetries. We list the phases associated with the $\mathbb{Z}_{2T_{\cal R}}^{\chi}[U(1)_B]^2$ anomaly (which we denote by B, noting that it coincides with the $\mathbb{Z}_{2T_{\cal R}}^{\chi}$--gravity anomaly), the BC anomaly on $\mathbb T^4$, the BC anomaly on $\mathbb{CP}^2$, and the anomaly on the EH space. The cases where the EH anomaly is the strongest are highlighted in green. 
}
\end{table}

The $\mathbb{Z}_{2T_{\mathcal{R}}}^\chi [U(1)_B]^2$ anomaly is simply characterized by the phase $e^{i 2 \pi \frac{\dim \mathcal{R}}{T_{\mathcal{R}}}}$, which is valued in the subgroup $\mathbb{Z}_{T_{\mathcal{R}}/\mathrm{gcd}(T_{\mathcal{R}}, \dim \mathcal{R})}$.  In addition to this mixed chiral--baryon anomaly, there also exists a mixed $\mathbb{Z}_{2T_{\mathcal{R}}}^{d\chi}-\mathrm{gravity}$ anomaly. The most restrictive contribution to the latter arises when the anomaly is evaluated on a non-spin manifold.
Fermions are ill-defined on non-spin manifolds such as $\mathbb{CP}^2$, and in order to make the fermions well-defined, one must introduce a background $U(1)_B$ monopole of charge $1/2$. The resulting fractional baryon flux combines with the fractional flux of the gravitational $\mathbb{CP}^2$ instanton, yielding an integer Dirac index equal to $1$ for each color component of the Weyl fermion; see \cite{Anber:2020gig} for details. Consequently, one again obtains the anomalous phase $e^{i 2\pi \frac{\dim \mathcal{R}}{T_{\mathcal{R}}}}$, which coincides precisely with the phase of the $\mathbb{Z}_{2T_{\mathcal{R}}}^\chi [U(1)_B]^2$ anomaly.

The BC anomaly is obtained by turning on background fluxes in the quotient group $SU(N_c) \times U(1)_B/\mathbb{Z}_p$, where $p = \mathrm{gcd}(N_c, n_c)$, exactly like we do in the EH case. The BC anomaly, however, is detected on manifolds such as $\mathbb{T}^4$, $S^2 \times S^2$, or $\mathbb{CP}^2$; see Appendix \ref{anomaliesonT4CP2} for a succinct review of the topology of $\mathbb {T} ^4$ and $\mathbb{CP}^2$ and the BC anomalies on these manifolds\footnote{In principle, one could refer to anomalies on the EH background as BC anomalies. However, doing so would require continual specification of the underlying manifold. Since the EH anomaly represents a distinct and previously unexplored case, we will refer to it simply as the \emph{EH anomaly} in order to clearly distinguish it from BC anomalies computed on compact manifolds such as $\mathbb{T}^4$ or $\mathbb{CP}^2$.
}.  
This type of anomalies and their applications have been studied extensively in \cite{Anber:2019nze,Anber:2020xfk,Anber:2021iip,Anber:2021lzb,Anber:2023urd,Anber:2023yuh,Nakajima:2022jxg,Kang:2024elv}. In particular, it was shown in \cite{Anber:2021lzb} that the BC anomaly\footnote{\label{BCANOMALY}The BC anomaly is given by the phase $e^{i 2 \pi \frac{\mathcal{I}_{\mathcal{R}}}{T_{\mathcal{R}}}}$, where  $\mathcal{I}_{\mathcal{R}}$ is the Dirac index.

 On $\mathbb{T}^4$ (or equivalently $S^2\times S^2$) the index is $ \mathcal{I}_{\mathcal{R}} = T_{\mathcal{R}} Q_c + \dim \mathcal{R} \, Q_B$, and $Q_c$ and $Q_B$ are the topological charges of the fractional fluxes in $SU(N_c)$ and $U(1)_B$, given by $Q_c = m_{12} m_{34} (1 - \frac{1}{N_c})$ and $Q_B = (n_1 + m_{12}\frac{n_c}{N_c})(n_2 + m_{34}\frac{n_c}{N_c})$, where we have excited the center fluxes along the Cartan generators of $SU(N_c)$; see \cite{Anber:2019nze} and Appendix \ref{anomaliesonT4CP2} for details. Here, $m_{12}$ and $m_{34}$ specify the amount of $\mathbb Z_{N_c}$ center flux turned on in the $12$ and $34$ planes, respectively, while $n_1, n_2$ are arbitrary integers.
  
  On $\mathbb{CP}^2$ the index is $ \mathcal{I}_{\mathcal{R}} = T_{\mathcal{R}} Q_c + \dim \mathcal{R} \, \left(Q_B+Q_G\right)$, and $Q_c$, $Q_B$ are in this case are given by $Q_c=\frac{m^2}{2}\left(1-\frac{1}{N_c}\right)$, $Q_B=\frac{1}{2}\left(\frac{1}{2}+m\frac{n_c}{N_c}+n\right)^2$, and $Q_G=-\frac{1}{8}$ is the topological gravitational charge; see  \cite{Anber:2020gig} and Appendix \ref{anomaliesonT4CP2} for details. Here, we used $U(1)_B$ to turn on an additional half-integer $U(1)$ flux (the extra factor of $\frac{1}{2}$ inside the parentheses in $Q_B$) to render the fermions globally well-defined on $\mathbb{CP}^2$. The integer $m$ specifies the amount of $\mathbb Z_{N_c}$ center flux turned on the $2$-cycle  $\mathbb{CP}^1\subset \mathbb{CP}^2$, while $n$ is an arbitrary integer.
 } 
 is stronger than or equal to the $\mathbb{Z}_{2T_{\mathcal{R}}}^\chi [U(1)_B]^2$ anomaly. In addition, one can also turn on BC fluxes on nonspin manifolds, e.g.,  $\mathbb{CP}^2$ \cite{Anber:2020gig}. In this case, the fermions are rendered globally well-defined by turning on a flux in the $U(1)_B$ direction. This anomaly is as strong as the BC anomaly on $\mathbb T^4$ for all single-flavour vector-like asymptotically-free gauge theories. These observations are evident from Table~\ref{breaking pattern}. 

In the majority of cases, the EH anomaly is as strong as the BC anomaly. In exceptional instances---highlighted in green in Table~\ref{breaking pattern}---the EH anomaly provides a strictly stronger constraint. 

It is instructive to examine one of the exceptional cases in detail and understand why it is stronger than the BC anomaly. Consider a Dirac fermion in the representation ${\cal R} = (3,0)$, which has $N$-ality $n_c=0$. By setting $m_{(3)} = 1$ and ${\cal C} = 0$, and working in the Verma basis, one readily finds that the total background gauge field $A_t$ in~(\ref{finalAt}) takes the form
\begin{eqnarray}
A_t = \mathrm{diag}_{\;\dim(3,0)=10}\left(2,1,1,0,0,0,-1,-1,-1,-1\right)\frac{\sigma_z a^2}{r^2} \,.
\end{eqnarray}
This background field induces the following holonomy (see (\ref{holrp3})) at the boundary $\mathbb{RP}^3$:
\begin{eqnarray}
\mathrm{Hol}(A_t) = \mathrm{diag}_{\;\dim(3,0)=10}\left(1,-1,-1,1,1,1,-1,-1,-1,-1\right) \,,
\end{eqnarray}
with signature $\mathrm{sgn}(\mathrm{Hol}) = 4 - 6 = -2$. 
Using either the explicit counting formulae~(\ref{evenC}, \ref{oddC}, \ref{explicit}) or, equivalently, the APS index in~(\ref{indexmastermix}), we obtain
${\cal I}^{EH}_{(3,0)} = 1$. Consequently, the anomalous phase on the EH background is
$
e^{i 2\pi \frac{{\cal I}_{(3,0)}}{T_{(3,0)}}} = e^{i \frac{2\pi}{15}} \,.
$
Using any value of the flux ${\cal C}\in \mathbb Z$ always yields fermionic zero modes in the EH background. Consequently, there is no choice of auxiliary flux ${\cal C}'$ on a second copy $\mathrm{EH}'$ that can be glued to the original space to form ${\cal M} = \mathrm{EH}\cup \mathrm{EH}' \cong S^2 \times S^2$ while preserving the number of zero modes. Were such a choice possible, the anomaly on EH space would necessarily coincide with the anomaly on a compact spin manifold—such as $\mathbb T^4$ or $S^2 \times S^2$—on which the BC anomaly is computed.

Let us also compute the BC anomaly on $\mathbb T^4$ in this case. Using the data summarized in Footnote~\ref{BCANOMALY}, and specifying 
$m_{12} = m_{34} = 1$ with $n_1 = n_2 = n_c = 0$, we obtain 
${\cal I}_{(3,0)}^{BC} = 10$ zero modes on $\mathbb{T}^4$. Consequently, the BC anomaly induces a phase 
$e^{i 2\pi \frac{2}{3}}$, which is weaker than the corresponding EH anomaly. This calculation highlights a key point 
already emphasized in the introduction: the topological charge carried by the $U(1)$ fluxes localized at the EH bolt 
is more finely quantized, leading to smaller Dirac indices and, therefore, a more stringent anomaly.  

Given that the number of fermionic zero modes on EH space in the representation ${\cal R} = (3,0)$ of $SU(3)$ is equal to $2$, and assuming that the anomaly is saturated through chiral symmetry breaking\footnote{The theory has a $\mathbb Z_3^{(1)}$ $1$-form center symmetry. It is assumed that the theory confines in the IR, preserving its center symmetry. Also, $U(1)_B$ symmetry cannot be broken in vector-like theories \cite{Vafa:1983tf}.}, it follows that a gauge-invariant bilinear condensate must form. In this minimal scenario, a condensate of the form $\langle \tilde\psi \psi \rangle \sim \Lambda^3$, where $\Lambda$ denotes the strong coupling scale, maximally breaks the chiral symmetry.

Let us also comment on the $(0,2,0)$ representation of $SU(4)$. Because the representation is real, we consider a single Weyl fermion, and there is no conserved baryon number symmetry. The fermion zero modes are induced upon turning on an $SU(N_c=4)$ background gauge field of the form
\begin{eqnarray}\label{special}
A_t=A_{SU(N_c=4)} =\left(\bm H_{(0,2,0)}\cdot \bm \nu_{a}+{\cal C}\right) \frac{\sigma_z a^2}{r^2}\,.
\end{eqnarray}
The minimal number of zero modes, equal to $6$, arises for ${\cal C} = \pm 1$, while for ${\cal C} = 0$ no normalizable zero modes exist. Although the latter might appear suitable for constructing a capping geometry $\mathrm{EH}'$, the two configurations cannot in fact be glued smoothly: backgrounds with ${\cal C}=\pm 1$ and ${\cal C}=0$ have inequivalent boundary holonomies, which constitute a topological obstruction.
This obstruction explains why the anomaly present on EH space is not reproduced on the compact manifolds $\mathbb T^4$ or $S^2\times S^2$ considered here, indicating that it is not captured by these conventional BC anomaly probes.

\subsection*{Anomaly matching via symmetry breaking}

The most natural mechanism for matching the EH anomaly is through spontaneous symmetry breaking. Suppose that a theory with a discrete chiral symmetry $\mathbb Z_{2n}^{\chi}$ has an EH anomaly phase $e^{2\pi i/n}$. Matching this anomaly by symmetry breaking then requires
$
\mathbb Z_{2n}^{\chi}\rightarrow \mathbb Z_2^{\rm F}\,,
$
or equivalently $\mathbb Z_n\to 1$, so that the low-energy theory contains multiple vacua separated by domain walls. In this subsection, we briefly review how such a symmetry-breaking pattern can account for the anomaly.

To describe the broken phase, we label the vacua by an order parameter
$
\phi(x)\in \mathbb Z_n.
$
A domain wall is a codimension-one defect across which $\phi$ jumps by a nontrivial element of $\mathbb Z_n$. A convenient topological description of this jump is obtained by introducing a $\mathbb Z_n$-valued cochain (or background field) $a$ in the bulk, allowed to fail to be closed precisely on the wall. In cochain language,
\begin{equation}
\delta a \;=\; \mathrm{PD}(W)\qquad (\mathrm{mod}\;n),
\label{eq:delta-a}
\end{equation}
where $\delta$ is the coboundary operator, $W\subset M_4$ is the wall worldvolume, and $\mathrm{PD}(W)$ is its Poincar\'e dual. This relation expresses the fact that the $\mathbb Z_n$ field $a$ has a singularity, or discontinuity, localized on the wall.

The EH anomaly associated with the $\mathbb Z_n$ symmetry can be represented by a bulk topological term of the schematic form
\begin{equation}
S_{\mathrm{anom}} \;\sim\; \frac{2\pi i}{n}\int_{M_4} a \smile \nu(A),
\label{eq:bulk-anom}
\end{equation}
where $A$ denotes the relevant background fields on the EH space, $\nu(A)$ is an integral $4$-cocycle constructed from them, and $\smile$ is the cup product. In the presence of a domain wall, $a$ is no longer closed, and using \eqref{eq:delta-a} one can perform the usual descent. Since $\nu(A)$ is closed, the bulk term localizes schematically as
$
\int_{M_4} a \smile \nu(A)
\;\leadsto\;
\int_W \omega_3(A),
$
where $\omega_3(A)$ is a $3$-cochain on $W$ whose coboundary reproduces $\nu(A)$ upon pullback, in analogy with the continuum relation $d\,\omega_3(A)=\nu(A)$. The anomaly functional therefore reduces to a wall term,
\begin{equation}
S_{\mathrm{wall}} \;\sim\; \frac{2\pi i}{n}\int_W \omega_3(A).
\label{eq:wall-response}
\end{equation}
This is the anomaly inflow mechanism: the domain wall must support a $(2+1)$-dimensional theory whose anomalous variation precisely cancels the inflow from the bulk.

\subsection{Composites in theories with fermions in the $2$-index antisymmetric representation}

In the next example, we consider the possibility of 't Hooft anomaly matching via composites, taking fermions in the $2$-index antisymmetric representation as an example.

The representation ${\cal R} = (0,1,0,\ldots,0)\equiv \tiny\yng(1,1)$ has dimension and Dynkin index
$
\dim{\cal R} = \frac{N_c(N_c-1)}{2}\,, T_{\cal R} = N_c - 2\,,
$
and therefore the theory possesses a chiral symmetry $\mathbb{Z}_{2(N_c-2)}^{\chi}$.  
To probe this symmetry on EH space, we turn on the background fields by setting $m_{N_c}=1$ and ${\cal C}=1$ in (\ref{ASUNCBULK}) and (\ref{abelian background needed}):
\begin{eqnarray}\nonumber\label{2indexback}
A_{SU(N_c)} \;&=&\; \, \mbox{diag}_{N_c}\left(1-\frac{1}{N_c},-\frac{1}{N_c},...,-\frac{1}{N_c}\right) \frac{\sigma_z\, a^2}{r^2}\,,\\
A_{U(1)_B} &=&\left(\frac{2}{N_c} + 1\right) \frac{\sigma_z\, a^2}{r^2}\,,
\end{eqnarray}
noting that $A_{SU(N_c)} $ is expressed using the defining representation. 
 One may then show, by induction (using the Verma basis in the $2$-index antisymmetric representation), that the signature of holonomy  is given by
\begin{equation}
\mathrm{sgn}(\mathrm{Hol}) = 2(N_c-1) - \dim{\cal R}
= \frac{(N_c-1)(4-N_c)}{2}\,.
\end{equation}
Substituting this result together with (\ref{SUNU1BTC}) into (\ref{indexmastermix}), we obtain
\begin{equation}
{\cal I}_{(0,1,0,\ldots,0)}^{(\rm EH)} = N_c - 1\,,
\end{equation}
which yields the anomalous phase
$
\exp\!\left(i\frac{2\pi}{N_c-2}\right)
$
 in (\ref{anomalyphase}). The index obtained here exactly matches the result (\ref{index2ind}) obtained using the gluing method discussed in Section \ref{Gluing construction}.

The same anomalous phase can also be derived from the BC anomaly, either on $\mathbb{T}^4$ or on $\mathbb{CP}^2$. Using the data summarized in Footnote~\ref{BCANOMALY}, and choosing $m_{12} = m_{34} = m = 1$, $n_c = 2$, and $n_1 = n_2 = n = 0$, the BC anomaly yields the phase $ \exp\!\left(i\,\frac{2\pi}{N_c - 2}\right)$, which agrees with the result obtained on the EH background. Moreover, the conventional mixed anomaly $\mathbb{Z}_{2(N_c-2)}^\chi \!\left[U(1)_B\right]^2$ produces the same phase $ \exp\!\left(i\,\frac{2\pi}{N_c - 2}\right)$ whenever $N_c$ is not divisible by $4$.

In the deep infrared, all anomalies can be matched by the bilinear fermion condensate 
$
\langle \tilde{\psi}\psi \rangle \neq 0\,,
$
which breaks $\mathbb{Z}_{2(N_c-2)}^{\chi}$ down to the fermion number symmetry $\mathbb{Z}_2^F$, thereby saturating the anomaly.

When $N_c$ is odd, or more generally when $N_c \not\equiv 0 \pmod{4}$, the theory also admits gauge-singlet composite fermions. In principle, one could attempt to match the anomalies using a set of such massless composites. The nontrivial question is whether a spectrum of composite fermions (without the help of any TQFT) can simultaneously match all anomalies.

\underline{\textbf{$N_c$ odd, $N_c\geq 5$}}. 
Consider a set of gauge-invariant massless composite fermions, 
$
({\cal O} \sim \psi^{N_c}, \ \tilde{\cal O} \sim \tilde{\psi}^{N_c}),
$
with multiplicity ${\cal N}$. Their charges under $U(1)_B$ are $(N_c, -N_c)$, while under the chiral symmetry $\mathbb{Z}_{2(N_c-2)}^\chi$ they have charges  $(N_c, N_c)$.
We first determine the value of ${\cal N}$ required to match the BC anomaly on $\mathbb T^4$. Using the information from Footnote~\ref{BCANOMALY}, and setting $m_{12}=m_{34}=1$, $n_c = 2$, and $n_1 = n_2 = 0$, the Dirac index of a single composite ${\cal O}$ is
$
{\cal I}_{\cal O}^{BC} = N_c^2 \left(\frac{2}{N_c}\right)^2 = 4.
$
In brief, since ${\cal O}$ is a gauge singlet, it does not couple to the $SU(N_c)$ center flux. However, it does couple to the $U(1)_B$ flux with charge $N_c$, leading to the index shown above. Consequently, the infrared contribution of the ${\cal N}$ composites to the BC anomaly is
$
\exp\!\left(i\, 2\pi \frac{4{\cal N}N_c}{\,N_c-2\,}\right)=\exp\!\left(i\, 2\pi \frac{8{\cal N}}{\,N_c-2\,}\right).
$
To match the UV value of the BC anomaly, we must therefore require
\begin{equation}\label{BCcond}
8{\cal N}= 1 \quad (\mathrm{mod}\ N_c - 2)\,\quad (\mbox{BC anomaly on}\,\, \mathbb T^4)\,.
\end{equation}
One may verify numerically that the required value of ${\cal N}$ is independent of the specific background data $(m_{12}, m_{34}, n_1, n_2)$.\footnote{One may also check that the mixed anomaly $\mathbb{Z}_{2(N_c-2)}^\chi[U(1)_B]^2$  is consistently matched in the IR whenever ${\cal N}$ satisfies $8{\cal N}= 1$ (mod $N_c - 2$).}

We can repeat the computation of the BC anomaly for the composites on $\mathbb{CP}^2$. 
Using the expressions given in Footnote~\ref{BCANOMALY} and setting $m=1$, $n=0$, we obtain the condition
\begin{equation}\label{BCcondCP2}
\mathcal{N} N_c \, \frac{(N_c + 3)(N_c + 5)}{8} 
\equiv 1 \pmod{N_c - 2},
\qquad \text{(BC anomaly on } \mathbb{CP}^2\text{)}.
\end{equation}
A numerical analysis shows that for all values of $N_c \geq 7$, there exists no multiplicity 
$\mathcal{N}$ that simultaneously satisfies both conditions~(\ref{BCcond}) and~(\ref{BCcondCP2}). 
However, the BC anomaly constraints still allow the composites as a viable possibility for the case $N_c = 5$.

We now turn to the infrared matching of the EH anomaly. The composite ${\cal O}$ is neutral under the $SU(N_c)$ background field $A_{SU(N_c)}$ in (\ref{2indexback}), but it does couple to the $U(1)_B$ background. Since ${\cal O}$ carries baryon charge $N_c$, its covariant derivative couples to the effective background field
\begin{equation}
N_c\left(\frac{2}{N_c} + 1\right)\frac{\sigma_z\, a^2}{r^2}
= \left(2 + N_c\right)\frac{\sigma_z\, a^2}{r^2}.
\end{equation}
Using the zero-mode counting formulas in (\ref{evenC}), (\ref{oddC}), and (\ref{explicit}), we obtain
\begin{equation}
{\cal I}^{(\rm EH)}_{\cal O} = \frac{(N_c + 1)(N_c + 3)}{4},
\end{equation}
so that the infrared contribution of the ${\cal N}$ composites to the EH anomaly is
\begin{equation}
\exp\!\left(i\,2\pi\,\frac{{\cal N}N_c(N_c+1)(N_c+3)}{4(N_c - 2)}\right).
\end{equation}
Matching this with the ultraviolet EH anomaly requires
\begin{equation}\label{EHcond}
\frac{{\cal N}N_c(N_c+1)(N_c+3)}{4} = 1
\quad (\mathrm{mod}\ N_c - 2) \qquad \text{(EH anomaly)}.
\end{equation}

The anomaly~(\ref{EHcond}) on EH space is distinct from the anomalies on $\mathbb T^4$ or $S^2 \times S^2$. At first sight, this may seem counterintuitive, since the fermion zero modes of the UV theory can be reproduced by capping the original EH space without changing their total number. The key point, however, is that the same UV cap $\mathrm{EH}'$ cannot be consistently used once the IR degrees of freedom are composites.
In the UV, the capping space $\mathrm{EH}'$ carries background gauge fields specified by (\ref{2indexfluxes}), with $C' = -1$, corresponding to our choice ${\cal C} = 1$. In the IR, the composite states instead experience a $U(1)$ flux of $(-N)$ on the cap $\mathrm{EH}'$, rather than $(-1)$. This larger flux necessarily induces additional fermionic zero modes, altering the anomaly matching condition.
This explains why composite fermions that successfully reproduce the anomaly on $\mathbb{T}^4$ or $S^2 \times S^2$ generally fail to match the anomaly on EH space.

Now, we are in a position to use the various anomalies to constrain the IR dynamics. 

Without the use of the BC anomaly condition on $\mathbb{CP}^2$, one can verify numerically that for every odd value of $N_c \geq 5$, there exists no choice of $\mathcal{N}$ that simultaneously satisfies both conditions~(\ref{BCcond}) and~(\ref{EHcond}). 
We thus conclude that the set of composites
$
(\mathcal{O} \sim \psi^{N_c}, \tilde{\mathcal{O}} \sim \tilde{\psi}^{N_c})
$
fails to reproduce the complete set of anomaly matching conditions and must therefore be amended with an extra sector. 
In this sense, the EH anomaly imposes a stronger constraint on the composite spectrum than the BC anomaly on non-spin manifolds (recall that the BC anomalies on $\mathbb T^4$ and $\mathbb{CP}^2$ exclude composites starting from $N_c \geq 7$).

We can generalize the previous exercise to include higher composites of the form\\ $\left(\psi^{(2k+1)N_c},\tilde\psi^{(2k+1)N_c} \right)$ with multiplicities ${\cal N}_k\geq 0$, where $k=0,1,2,..$. The anomaly matching conditions are given in Table \ref{tab:anomalies}.
\begin{table}[t]
\centering
\renewcommand{\arraystretch}{1.6}
\begin{tabular}{|c|l|}
\hline
\textbf{Space} & \textbf{Anomaly Condition} \\ \hline
$\mathbb{T}^4$ & 
$\displaystyle 
8 \sum_{k=0}^{} \mathcal{N}_k (2k+1)^3 
\qquad 
$ \\ \hline
$\mathbb{CP}^2$ &
$\displaystyle 
\sum_{k=0}^{} 
\mathcal{N}_k N_c (2k+1) 
\frac{(3 + N_c + 2k(4 + N_c))(5 + N_c + 2k(4 + N_c))}{8} 
\qquad 
$ \\ \hline
EH &
$\displaystyle 
\sum_{k=0}^{} 
\mathcal{N}_k N_c (2k+1)
\Biggl(2k+1 + \frac{N_c(2k+1)-1}{2}\Biggr)
\Biggl(2k+1 + \frac{N_c(2k+1)+1}{2}\Biggr) 
\qquad
$ \\ \hline
\end{tabular}
\caption{Anomaly matching conditions for the composite spectrum on different four-manifolds. 
All conditions must give $1$ mod $N_c-2$.}
\label{tab:anomalies}
\end{table}
The values of $\{{\cal N}_k\}$ are bounded by the $a$-theorem as
\begin{eqnarray}
\underbrace{\frac{7}{2}\sum_{k=0}{\cal N}_k}_{\scriptsize\mbox{composites}}\leq \underbrace{2(N_c^2-1)}_{\scriptsize\mbox{gluons}}+\underbrace{\frac{7}{4}N_c(N_c-1)}_{\scriptsize\mbox{UV fermions}}\,.
\end{eqnarray}

For $N_c = 5$, the IR EH anomaly evaluates to $0 \pmod{3}$ for all choices of $\mathcal{N}_k$, 
demonstrating that the composites of this theory cannot match the UV anomaly $1 \pmod{3}$.  
Our numerical analysis further shows that for $N_c = 7,9,11$, the three composite multiplicities, 
${\cal N}_0$, $\mathcal{N}_1$, $\mathcal{N}_2$, are not enough to satisfy all anomaly matching conditions. 
For larger values of $N_c$ such as $N_c = 13, 15, \dots$, at least three multiplicities, 
$\mathcal{N}_0, \mathcal{N}_1, \mathcal{N}_2$, are needed to match all anomalies; etc.

\subsection{More on anomalies of the $SU(5)$ $2$-index antisymmetric theory}
\label{More on the EH anomaly in the 2-index antisymmetric theory}

In this work, we demonstrated that there exist anomalies that cannot be detected on conventional closed manifolds such as $S^4$, $S^2\times S^2$, $\mathbb T^4$, and $\mathbb{CP}^2$, but can be detected on EH spaces. We do not claim to have proven that the anomalies detected on EH space cannot be seen on an arbitrary closed $4$-manifold. This remains an important gap in the present construction: one would like to classify precisely the anomalies detected by EH space, for example, by determining the corresponding bordism invariants. What we can do at present is show, by explicit examples, that in certain theories the constraints imposed by the EH anomalies cannot be obtained from the conventional closed manifolds listed above. In some instances, the EH anomalies cannot be seen on any closed manifold.  In this section, we illustrate this last point with a simple example.

In the previous section, we observed that the $SU(N_c=5)$ theory with fermions in the two-index antisymmetric representation is quite distinct from the theories with $N_c\neq 5$, in that the EH anomaly cannot be matched by any number of composite states in the IR, provided that the IR is not supplemented by an additional TQFT sector. A natural question is then whether there exists any anomaly detectable on a closed manifold that already rules out these composites as viable IR candidates.

To address this question, recall that for a QFT with global symmetry $G$, defined on a closed $4$-dimensional manifold, the anomaly is classified by the bordism group
$
\Omega_{5}^{\mathcal S}(BG),
$
where $BG$ is the classifying space of $G$ and $\mathcal S$ specifies the relevant tangential structure (such as $\mathrm{Spin}$, $\mathrm{Spin}^c$, or $\mathrm{Pin}^{\pm}$) \cite{Dai:1994kq,Witten:2019bou}. The bordism group $\Omega_{5}^{\mathrm{Spin}}(BG)$ classifies the closed $5$-D spaces with spin structure.  Here, we are interested in the $\mathrm{Spin}$ case, since we eventually want to check whether the anomaly detected on EH space, which is spin, can also be seen using a bordism group with the same tangential structure. For the $SU(5)$ gauge theory with a single Dirac fermion in the two-index antisymmetric representation, the faithful global symmetry group is (see Eq.~(\ref{globalG}))
\begin{eqnarray}
G=\frac{SU(5)\times U(1)_B\times \mathbb Z_{6}^\chi}{\mathbb Z_5\times \mathbb Z_2^{\rm F}}
\cong U(5)\times \mathbb Z_3^{\chi}\,.
\end{eqnarray}
Using
$
B\!\left(U(5)\times \mathbb{Z}_3^{\chi}\right)\simeq BU(5)\times B\mathbb{Z}_3^{\chi},
$
together with the Atiyah--Hirzebruch spectral sequence, one finds (see Appendix \ref{app:bordismU5Z3})
\begin{eqnarray}\label{bordsom5spinu5}
\Omega_5^{\mathrm{Spin}}\!\left(B\!\left(U(5)\times \mathbb{Z}_3^{\chi}\right)\right)
\cong \mathbb Z_9\oplus \mathbb Z_3\oplus \mathbb Z_3\oplus \mathbb Z_3.
\end{eqnarray}

One can understand this result physically by identifying the generators of the summands. The $\mathbb Z_9$ factor is generated by the $5$-dimensional lens space $L^3(3)\equiv S^5/\mathbb Z_3$, and the corresponding anomaly is a genuine nonperturbative anomaly, not accounted for in our previous analysis. The next two $\mathbb Z_3$ factors are mixed anomalies involving the $\mathbb Z_3^\chi$ background and the $U(5)$ characteristic classes; more precisely, they are detected by the bordism invariants associated with the classes $c_1^2\, a_\chi$ and $c_2\, a_\chi$, where $a_\chi\in H^1(B\mathbb Z_3^\chi;\mathbb Z_3)$. Geometrically, these can be represented by product manifolds of the form $M_4\times S^1$, where $L^1(3)\equiv S^1/\mathbb Z_3\simeq S^1$ carries the $\mathbb Z_3^\chi$ background and $M_4$ is equipped with a suitable $U(5)$ bundle. In particular, one may use $S^2\times S^2\times S^1$ or $S^4\times S^1$ with appropriate bundle data to isolate the two independent $\mathbb Z_3$ classes. The anomaly detected on $S^4\times S^1$ is determined by the manifold and the choice of $U(5)$ background bundle. With a bundle for which $\int_{S^4} c_2\neq 0$ and $\int_{S^4} c_1^2=0$, the corresponding bordism invariant detects the mixed class $c_2 a_\chi$, which is the bordism representative of the perturbative $[SU(5)]^2\mathbb Z_3^\chi$ anomaly.\footnote{For a $U(5)$ background connection $A$ with curvature $F=dA+A\wedge A$, the first and second Chern classes are given by
$
c_1=\frac{1}{2\pi}\,\mathrm{tr}\,F\,,
c_2=\frac{1}{8\pi^2}\Bigl(\mathrm{tr}(F\wedge F)-\mathrm{tr}F\wedge \mathrm{tr}F\Bigr),
$
so that
$
c_1^2=\left(\frac{1}{2\pi}\,\mathrm{tr}\,F\right)^2
$.} In the present theory, this anomaly vanishes, since $\mathbb Z_3^\chi$ is a genuine symmetry in the $SU(5)$ color background. By contrast, on $S^2\times S^2\times S^1$, with a suitable $U(5)$ bundle for which $\int_{S^2\times S^2} c_1^2\neq 0$, one detects the mixed class $c_1^2 a_\chi$, which corresponds to the baryon-color mixed anomaly discussed above. The generator of the last $\mathbb Z_3$ factor in (\ref{bordsom5spinu5}) is the $K3$ surface, and the anomaly detected on $K3$ is the mixed gravitational-$\mathbb Z_3^\chi$ anomaly.

For general $\mathbb Z_n$ with $n$ odd, the anomaly on the lens space $L^5(n)=S^5/\mathbb Z_n$ is encoded in the  $\eta$-invariant \cite{Gilkey2018,Garcia-Etxebarria:2018ajm}:
\begin{eqnarray}
\eta\!\left(L^5(n)\right)=\frac{1}{n}\sum_{\omega\neq 1}\left(\omega^s-1\right)\left(\frac{\sqrt{\omega}}{\omega-1}\right)^3\,,
\end{eqnarray}
where $s$ is the $\mathbb Z_n$ charge of the fermion and $\omega$ is an $n$th root of unity. The square root is chosen so that $(\sqrt{\omega})^n=1$. Using $n=3$, $\dim {\cal R}=10$ and recalling that there are two Weyl fermions, $\psi$ and $\tilde\psi$, contributing to the anomaly in the UV, one finds
\begin{eqnarray}
e^{2\pi i\,\eta_{\rm UV}}=e^{2\pi i\frac{2}{9}}\,.
\end{eqnarray}
In the IR, one may postulate ${\cal N}$ copies of the composite operators ${\cal O}\sim \psi^5$ and $\tilde{\cal O}\sim (\tilde\psi)^5$. These composites match all bordism anomalies classified by $\Omega_5^{\rm Spin}(B(U(5)\times \mathbb Z_3^\chi))$, including the non-perturbative one, provided that
\begin{eqnarray}
{\cal N}=2+3p \quad \mbox{and}\quad  p+1=0\; (\mbox{mod}\, 3)\,.
\label{the first condition from L3}
\end{eqnarray}
Indeed, anomalies detectable on ordinary closed spin $5$-manifolds with background $B(U(5)\times \mathbb Z_3^\chi)$ are classified by this bordism group, and matching its generators is sufficient to match all such anomalies.

Yet, as we demonstrated in the previous section, no IR composites can match the EH anomaly in the $SU(5)$ theory with $2$-index antisymmetric fermions. This shows that the EH anomaly is not captured by the ordinary closed-manifold bordism invariants that we have analyzed for this theory. It remains an open question how to classify the EH anomaly. We speculate that such anomalies may be related to relative bordism.

\subsection*{What happens at asymptotic infinity?}

In our analysis, the QFT on EH space is defined with trivial asymptotic boundary conditions at infinity, and all anomaly constraints are derived within this definition of the theory. One may nevertheless ask whether relaxing this assumption, while preserving the full global symmetry, could allow the composite operators to match the EH anomaly in the IR.

A first possibility is to supplement the IR theory by an invertible boundary phase localized on the asymptotic $\mathbb{RP}^3$. A possible contribution of this kind could come from an invertible fermionic phase of APS $\eta$-invariant type. On $\mathbb{RP}^3$, such phases yield eighth roots of unity, $e^{i2\pi k/8}$ (these phases can be easily seen from our earlier treatment, and they are related to the fact that $H_1(\mathbb{RP}^3, \mathbb Z)=\mathbb Z_2$). Since the UV EH anomaly gives the phase $e^{i2\pi/3}$, no boundary phase of this purely gravitational $\eta$-invariant type can reproduce the required anomaly (recall that the IR composites do not yield any nontrivial phase).

A second possibility is that the asymptotic boundary supports a noninvertible topological sector (e.g., anyons) whose partition function contributes the phase $e^{i2\pi/3}$. We do not know any good physical reason for such boundary topological order to emerge in the present $SU(5)$ theory, so we regard this possibility as speculative. Another possibility, which is not excluded by our analysis, is that the IR contains an additional symmetry-preserving $4$-D TQFT sector that matches the EH anomaly in the presence of the composites. We do not address this possibility here, and it would be interesting to investigate it in future work.

\section{Applications II: Bars-Yankielowicz models}
\label{ApplicationII}

In this section, we consider a class of chiral gauge theories with fermions in the fundamental and two-index symmetric or antisymmetric representations. These are known as Bars–Yankielowicz (BY) models \cite{Bars:1981se}. We focus on two specific types of these theories, denoted by $\{S, N_c, p\}$ and $\{A, N_c, p\}$, where $S$ and $A$ indicate the symmetric and antisymmetric representations, respectively.  

\underline{\bf In the first class}, the $\{S, N_c, p\}$ model, the gauge group is $SU(N_c)$, and the matter content consists of a single left-handed Weyl fermion $\psi$ in the two-index symmetric representation, $p$ left-handed Weyl fermions $\xi$ in the fundamental representation, and $4+N_c+p$ left-handed Weyl fermions $\eta$ in the anti-fundamental representation. Assuming $p\geq 1$, the theory possesses the global symmetry
\begin{equation}
G_{\scriptsize\mbox{Global}} = SU(4+N_c+p) \times SU(p) \times U(1)_{\psi\eta} \times U(1)_{\psi\chi},
\end{equation}
up to a possible quotient by a discrete subgroup. The charges of the fermions under $G_{\scriptsize\mbox{Global}}$ are given in the table below, and we emphasize that in the following we restrict our attention to the case where either $N_c$ or $p$ (or both) are odd and that $\mbox{gcd}(N_c+p+4,p)=1$.  
\begin{equation}\label{charges1BYAS}
\begin{array}{c|ccccc}
 & SU(N_c) & SU(4+N_c+p) & SU(p) & U(1)_{\psi\eta} &U(1)_{\psi\chi} \\ \hline
\psi        & \scriptsize\,\yng(2)\,& 1& 1 & N_c+4+p & p \\
\eta  & \overline{\scriptsize\,\yng(1)} & \scriptsize\,\yng(1)& 1 & -(N_c+2) & 0\\
\xi &\scriptsize\,\yng(1) & 1 & \scriptsize\,\yng(1) &0 &-(N_c+2)
\end{array}
\end{equation}

A well-known set of composite operators was shown in \cite{Bars:1981se}  to match the conventional 0-form 't Hooft anomalies. These composites are
$
B_1 \sim \psi \eta \eta,  B_2 \sim \bar{\psi} \bar{\eta} \xi, B_3 \sim \psi \bar{\xi} \bar{\xi},
$
with their charges under the global symmetry group $G_{\rm Global}$ summarized in the following Table.
\begin{equation}\label{charges1CBYAS}
\begin{array}{c|ccccc}
 & SU(N_c) & SU(4+N_c+p) & SU(p) & U(1)_{\psi\eta} & U(1)_{\psi\chi} \\ \hline
B_1 & 1 & \scriptsize\yng(1,1) & 1 & -N_c + p & p \\
B_2 & 1 & \overline{\scriptsize\yng(1)} & \scriptsize\yng(1) & -(p+2) & -(N_c+p+2) \\
B_3 & 1 & 1 & \overline{\scriptsize\yng(2)} & N_c + 4 + p & 2N_c + 4 + p
\end{array}
\end{equation}
If we turn on background fluxes in the color, flavor, and $U(1)$ directions (so-called CFU backgrounds), then the set of composites $B_1, B_2, B_3$ must also match the corresponding mixed anomalies of the type $U(1)_{\psi\eta}[\rm CFU]$ or $U(1)_{\psi\chi}[\rm CFU]$ on $\mathbb T^4$. This was proven in \cite{Anber:2019nze} and emphasized in Section \ref{anomaliesonEHspace}: any set of composites that satisfies the traditional 't Hooft anomalies for continuous symmetries must automatically match $U(1)[\rm CFU]$ anomalies as well.

The situation changes when the theory is placed on the EH space. In this case, we turn on a {\em unit} background flux along the $U(1)_{\psi\eta}$ direction, localized at the bolt of the EH space. Such a flux induces fermion zero modes (determined by the charge of the fermion under $U(1)_{\psi\eta}$), and if the partition function acquires a phase under a $U(1)_{\psi\chi}$ transformation, an anomaly arises. For the proposed set of composites to serve as a viable infrared realization of the symmetries, they must reproduce this same anomaly. 
However, we will show that, in certain situations, the composites fail to match the anomaly on the EH background. This discrepancy originates from the APS index~(\ref{indexmastermix}): while the composites reproduce all contributions to the index except for the final term, $\mathrm{sgn}(\mathrm{Hol})$, this additional term represents a new contribution that is not necessarily matched by the composites. This mismatch occurs whenever either $p$ or $N_c$ is odd.

The anomaly takes the form 
$e^{i\beta_\tau \sum Q_{\psi\chi} {\cal I}^{(\rm EH)}}$, 
where ${\cal I}^{(\rm EH)}$ denotes the number of fermion zero modes in the $U(1)_{\psi\eta}$ flux background, $Q_{\psi\chi}$ is the charge under $U(1)_{\psi\chi}$, $\beta_\tau$ is the corresponding $U(1)_{\psi\chi}$ phase, and the sum runs over all fermion species. 
As discussed earlier, all terms in the APS index formula~(\ref{indexmastermix}) are matched between the UV and IR theories, except for the final term, which contributes an anomalous phase of the form 
$\exp[i{\cal A}] \equiv \exp[i\frac{\beta_\tau}{8} \sum {\cal N} Q_{\psi\chi} (-1)^{Q_{\psi\eta}}]$, 
where $(-1)^{Q_{\psi\eta}}$ is the holonomy of the $U(1)_{\psi\eta}$ background field seen by the fermion, and ${\cal N}$ is the multiplicity of each fermion species, determined from the group representations of the fermion under $SU(N_c)$, $SU(4+N_c+p)$, and $SU(p)$.  
A direct computation yields, in the UV, 
\begin{eqnarray}
8{\cal A}_{UV}&=&\frac{p (N_c + 1) N_c}{2} (-1)^{N_c + p + 4} - (N_c + 2)N_c p\,,
\end{eqnarray}
and in the IR
\begin{eqnarray}\nonumber
8{\cal A}_{IR}&=&  \frac{(N_c + 4 + p) (N_c + 3 + p)}{2}p(-1)^{-N_c + p} -  p (N_c + 4 + p) (N_c + p +  2)(-1)^{p + 2} \\
&&+  \frac{p (p + 1)}{2} (2 N_c + p + 4)(-1)^{N_c + p + 4}\,.
\end{eqnarray}
It is straightforward to verify that ${\cal A}_{\mathrm{UV}} = {\cal A}_{\mathrm{IR}}$ when both $N_c$ and $p$ are even, while this equality fails whenever either $N_c$ or $p$ is odd. Hence, for odd $N_c$ or $p$, the proposed composites cannot, by themselves (without the aid of an additional TQFT), match the EH  anomaly\footnote{The authors of \cite{Bolognesi:2021yni} identified a {\em singular} vortex that would exclude the composites when both $N$ and $p$ are even. This is a different conclusion from what we have here, where we turn on a flux of $U(1)_{\psi\eta}$ and apply a global transformation by $U(1)_{\psi\chi}$. This is only one example, and it will be left for the future to check whether turning on fluxes of other groups can also exclude the composites for even values of $N_c$ and $p$.}.

\underline{\bf In the second class}, the $\{A, N_c, p\}$ model, the gauge group is $SU(N_c)$, and the matter content consists of a single left-handed Weyl fermion $\chi$ in the two-index antisymmetric representation, $p$ left-handed Weyl fermions $\xi$ in the fundamental representation, and $N_c - 4 + p$ left-handed Weyl fermions $\eta$ in the anti-fundamental representation. Assuming $p \geq 1$ and $N_c - 4 + p \geq 1$, the theory exhibits the global symmetry
\begin{equation}
G_{\scriptsize\mbox{Global}} = SU(N_c - 4 + p) \times SU(p) \times U(1)_{\chi\eta} \times U(1)_{\chi\xi},
\end{equation}
up to a possible quotient by a discrete subgroup. The charges of the fermions under $G_{\scriptsize\mbox{Global}}$ are listed in the table below. In what follows, we restrict attention to the case where either $N_c$ or $p$ (or both) are odd, and where $\gcd(N_c - 4 + p,\, p) = 1$.
\begin{equation}\label{charges1BYAS}
\begin{array}{c|ccccc}
 & SU(N_c) & SU(N_c-4+p) & SU(p) & U(1)_{\chi\eta} & U(1)_{\chi\xi} \\ \hline
\chi & \scriptsize\yng(1,1)\, & 1 & 1 & N_c - 4 + p & p \\
\eta & \overline{\scriptsize\,\yng(1)} & \scriptsize\,\yng(1) & 1 & -(N_c - 2) & 0 \\
\xi  & \scriptsize\,\yng(1) & 1 & \scriptsize\,\yng(1) & 0 & -(N_c - 2)
\end{array}
\end{equation}
The IR composite operators and their charges are given by:
\begin{equation}\label{charges1CBYAS}
\begin{array}{c|ccccc}
 & SU(N_c) & SU(N_c-4+p) & SU(p) & U(1)_{\chi\eta} & U(1)_{\chi\xi} \\ \hline
B_1 & 1 & \scriptsize\,\yng(2)\, & 1 & -N_c + p & p \\
B_2 & 1 & \overline{\scriptsize\yng(1)} & \scriptsize\yng(1) & -(p - 2) & -(N_c + p - 2) \\
B_3 & 1 & 1 & \overline{\scriptsize\yng(1,1)} & N_c - 4 + p & 2N_c - 4 + p
\end{array}
\end{equation}

We now turn on a background for $U(1)_{\chi\eta}$ and study the transformation of the partition function under $U(1)_{\chi\xi}$. A direct computation of the $\eta$-invariant contribution to the anomaly yields, in the UV,
\begin{eqnarray}
8{\cal A}_{UV} &=& \frac{p (N_c - 1) N_c}{2} (-1)^{N_c - 4 + p} - (N_c - 2) N_c\, p \,,
\end{eqnarray}
and in the IR,
\begin{eqnarray}\nonumber
8{\cal A}_{IR} &=& \frac{(N_c - 4 + p)(N_c - 3 + p)}{2}\, p\, (-1)^{-N_c + p}
- p (N_c - 4 + p)(N_c + p - 2)(-1)^{-(p - 2)} \\
&& + \frac{p (p - 1)}{2} (2N_c - 4 + p)\, (-1)^{N_c - 4 + p}\,.
\end{eqnarray}
It is then straightforward to verify that for odd $N_c$ or $p$, the proposed composites fail to reproduce the EH anomaly.

\section{Future directions}
\label{Future directions}

In this work, we have shown that asymptotically locally Euclidean (ALE) spaces can diagnose quantum anomalies that are invisible on conventional closed $4$-manifolds such as $S^4$, $\mathbb T^4$, $S^2 \times S^2$, or $\mathbb{CP}^2$. Using the Eguchi--Hanson (EH) space as the simplest nontrivial ALE background, we uncovered a novel EH anomaly that imposes additional consistency conditions on quantum field theories beyond those inferred from standard anomaly probes. Our analysis suggests that ALE geometries may provide a genuinely new window into the space of admissible UV theories and their possible IR realizations. Building on this work, we highlight several directions for future investigation:

\begin{enumerate}

\item \textbf{Systematic exploration of gauge theories.}

In this paper, we focused on a restricted set of gauge theories to illustrate the appearance and consequences of the EH anomaly. A natural next step is to undertake a systematic survey of gauge theories, with particular emphasis on chiral gauge theories, see, e.g., \cite{Bolognesi:2022beq,Bolognesi:2021yni,Bolognesi:2020mpe,Bolognesi:2017pek,Csaki:2021xhi,Csaki:2021aqv,Anber:2021iip,Anber:2023yuh,Smith:2021vbf,Sheu:2022odl}, that are otherwise consistent on familiar backgrounds. In addition, one seeks to better understand whether the notation of complementarity between the Higgs and confined phases persists in the background of ALE spaces. Supersymmetric theories, in particular Seiberg dualities \cite{Seiberg:1994pq} and s-confinement \cite{Csaki:1996sm}, are also worth exploring in the EH background.

\item \textbf{IR TQFTs as anomaly absorbers on ALE backgrounds.}

For many theories, one expects that a nontrivial infrared topological quantum field theory (TQFT) can contribute to anomaly matching by carrying (part of) the anomaly that cannot be realized by gapless degrees of freedom alone. An immediate concrete task is to analyze explicit examples in which a candidate IR description built solely from composite fields fails the EH anomaly constraint, and to ask whether tensoring it with an appropriately chosen TQFT restores consistency. This requires constructing IR TQFTs whose partition functions and anomaly inflow properties on EH space (and more general ALE backgrounds) precisely compensate the deficit of the would-be IR composites. 

\item \textbf{Nonabelian backgrounds and Kronheimer--Nakajima classification.}

Our analysis has primarily focused on turning on background fluxes along Cartan subalgebras, i.e.\ effectively abelian backgrounds embedded in nonabelian gauge groups. A more complete picture requires studying genuinely nonabelian configurations on EH and on general ALE spaces. In this regard, an important input is the systematic construction of Yang--Mills instantons on ALE spaces carried out by Kronheimer and by Kronheimer and Nakajima, who classified $SU(N)$ instantons and abelian gauge bundles on all ALE spaces ~\cite{Kronheimer:1989zs,PeterBKronheimer:1990zmj}. Incorporating these nonabelian instantons into the anomaly analysis would allow one to probe a richer set of backgrounds, potentially revealing new classes of anomalies or strengthening the constraints on allowed matter representations. It would also be interesting to understand how the space of nonabelian instantons gives rise to discrete flux sectors, and how the boundary holonomy data, which are central to the EH anomaly, are encoded.

\item \textbf{Symmetry breaking and Wess--Zumino--Witten-type effective actions on ALE spaces.}

It is generally assumed that spontaneous symmetry breaking is one of the standard mechanisms by which anomalies are matched: the anomaly is then encoded in topological terms in the effective action of the Goldstone modes, often of Wess--Zumino--Witten (WZW) type. However, it remains to be understood in detail how this mechanism operates in the presence of EH or more general ALE backgrounds. In particular, it would be highly desirable to construct WZW-like Lagrangians whose actions on EH space reproduce the required anomaly inflow and thereby match the EH anomaly. This entails, for example, understanding how the nontrivial topology and asymptotics of ALE spaces affect the quantization conditions and homotopy arguments that underlie WZW terms, and determining whether new types of topological terms (absent on compact manifolds) become relevant. A satisfactory construction of such effective actions would provide a geometric realization of anomaly matching via symmetry breaking in the ALE setting and could offer novel insights into the interplay between geometry, topology, and IR effective field theory.

\item \textbf{How to classify anomalies on ALE spaces?}

 It is still unclear to us whether the anomaly revealed on
EH space can already be detected on some (currently unknown to us) closed $4$-manifold. In the $SU(5)$ theory with $2$-index antisymmetric fermions discussed in Section \ref{More on the EH anomaly in the 2-index antisymmetric theory}, we showed that this is not the case. 
If this turns out to be the case for some other theories, the EH anomaly would simply correspond to an element
of the standard spin bordism group $\Omega_{5}^{\mathrm{Spin}}(BG)$, perhaps expressed in an
unusual basis. If not (as in Section \ref{More on the EH anomaly in the 2-index antisymmetric theory}) it remains an open --- and in our view interesting --- question to identify
the appropriate generalized cohomology theory (for example, a suitable bordism group for ALE
spaces with fixed asymptotics) that classifies such anomalies.

\item \textbf{The role of ALE spaces beyond anomalies.}

Beyond their appearance in 't~Hooft anomaly constraints, ALE spaces also arise naturally in semiclassical gravity. It has long been known that these geometries furnish saddle points of the Einstein action \cite{Gibbons:1978tef,Gibbons:1979xm}, and it is therefore natural to expect them to contribute to gravitational path integrals relevant for quantum processes \cite{Perry:1978fd,Hawking:1978ghb,Smilga:1983ib,tHooft:1988wxy}. Recently, their significance---especially that of the EH space---was highlighted in the context of gauging the Standard Model $\mathbb Z_6^{(1)}$ $1$-form symmetry \cite{Anber:2025gvb}. In addition, \cite{Guevara:2025psg} introduced a novel class of near-extremal black holes in a spacetime of signature $(+1,+1,-1,-1)$ whose exterior develops an infinite near-horizon throat governed by the EH instanton. This provides a concrete setting in which ALE saddles may control quantum-gravitational effects associated with black holes. The methods developed here are well-suited to further probe and clarify these recent directions.

\end{enumerate}

{\bf \flushleft{Acknowledgments:}} We would like to thank I.~Garc\'{\i}a Etxebarria and T.~Sulejmanpasic for many illuminating discussions, and I.~Garc\'{\i}a Etxebarria and E.~Poppitz for helpful comments on the manuscript. Any errors are solely the responsibility of the author. This work was supported by the STFC under grant ST/X000591/1.

\appendix

\section{Lie algebras and conventions}
\label{app:LieAlgebra}

In this appendix, we summarize the basic definitions and conventions for Lie algebras used throughout this work. Standard reviews include~\cite{Slansky:1981yr,Georgi:1999wka}. 

\subsection*{Definitions}
The Lie algebra $\mathfrak g$ of a group $G$ consists of generators $\{t^a\}$, with $a=1,2,\ldots,\dim G$, satisfying
\begin{align}\nonumber
[t^a,t^b] &= i f_{abc}\, t^c, \label{eq:Lie-algebra}\\[4pt]
[t^a,[t^b,t^c]] + \text{cyclic perms} &= 0, 
\end{align}
where $f_{abc}$ are the structure constants.  
If the $t^a$ are chosen Hermitian, then $f_{abc}\in\mathbb R$.  
Equivalently, one may define $\mathfrak g$ as the algebra generated by $[t^a,t^b]=if_{abc}t^c$ with $f_{abc}$ totally antisymmetric.

\subsection*{Cartan--Weyl basis}

A Cartan subalgebra consists of $r$ mutually commuting generators $H^i$,
\begin{equation}
[H^i,H^j]=0, \qquad i,j=1,\ldots,r,
\end{equation}
where $r=\mathrm{rank}(G)$.  
The remaining generators can be organized into raising and lowering operators $E_{\boldsymbol\beta}$, $E_{-\boldsymbol\beta}=E_{\boldsymbol\beta}^\dagger$, associated with the roots $\boldsymbol\beta=(\beta^1,\ldots,\beta^r)\in \mathbb R^r$. They satisfy
\begin{align}\nonumber
[H^i,E_{\boldsymbol\beta}] &= \beta^i E_{\boldsymbol\beta}, \\\nonumber
[E_{\boldsymbol\beta},E_{-\boldsymbol\beta}] &= \beta_i H^i, \\
[E_{\boldsymbol\beta},E_{\boldsymbol\gamma}] &= \mathcal N_{\boldsymbol\beta,\boldsymbol\gamma}\,E_{\boldsymbol\beta+\boldsymbol\gamma}.
\label{eq:Cartan-algebra}
\end{align}
There are $\dim G-r$ roots, split evenly into positive and negative roots.  
Accordingly, there are $(\dim G-r)/2$ raising operators $E_{\boldsymbol\beta}$ (positive roots) and $(\dim G-r)/2$ lowering operators $E_{-\boldsymbol\beta}$.  
The constants $\mathcal N_{\boldsymbol\beta,\boldsymbol\gamma}$ can be fixed but will not be needed here.

\subsection*{Weights, simple roots, and co-roots}

In a representation ${\cal R}$, the Cartan generators act diagonally:
\begin{equation}
H^i|\mu,{\cal R}\rangle = \mu_i |\mu,{\cal R}\rangle,
\end{equation}
where $\bm\mu=(\mu_1,\ldots,\mu_r)$ is a weight.  
The weights of the adjoint representation are precisely the roots $\{\bm\beta\}$.  

A weight is called \emph{positive} if its first nonzero component is positive.  
The \emph{simple roots} $\{\bm\alpha_a\}$ are those positive roots which cannot be written as sums of other positive roots. The $su(N_c)$ Lie algebra has rank $r=N_c-1$, and there are $N_c-1$ simple roots.  

The affine root is defined by
\begin{equation}
\bm\alpha_0 = -\sum_{a=1}^r k_a \bm\alpha_a,
\end{equation}
where $k_a$ are the Kac labels. For $su(N_c)$, all $k_a=1$.  

The co-roots are defined as
\begin{equation}
\bm\alpha^* \equiv \frac{2}{\bm\alpha^2}\,\bm\alpha.
\end{equation}
For simply-laced algebras such as $su(N_c)$, we normalize $\bm\alpha^2=2$ so that $\bm\alpha^*=\bm\alpha$.  
The affine co-root takes the form
\begin{equation}
\bm\alpha_{r+1}^* = -\sum_{a=1}^r k_a^* \bm\alpha_a^*,
\end{equation}
with dual Kac labels $k_a^*$. For $su(N_c)$, $k_a^*=1$ for all $a$.

\subsection*{Fundamental weights and Dynkin labels}

The fundamental weights $\bm w_a$ are defined by
\begin{equation}
\bm w_a \cdot \bm\alpha_b^* = \delta_{ab}, \qquad a,b=1,\ldots,r.
\end{equation}
The highest weight of a representation ${\cal R}$ is
\begin{equation}
\bm\mu_h = \sum_{a=1}^r m_a \bm w_a,
\end{equation}
where the nonnegative integers $\{m_a\}$ are the Dynkin labels.  
All other weights are obtained by applying lowering operators $E_{-\bm\alpha}$.  
Thus, we denote a representation by its Dynkin labels,
\begin{equation}
{\cal R} = (m_1,m_2,\ldots,m_r).
\end{equation}

The conjugate of $(m_1,m_2,\ldots,m_r)$ is $(m_r,\ldots,m_2,m_1)$.  
A representation is real if it equals its conjugate. For example, the adjoint representation $(1,0,\ldots,0,1)$ of $su(N_c)$ is real.  
Some important examples are:
\begin{center}
\begin{tabular}{|c|c|}
\hline
Representation & Dynkin labels \\
\hline
Fundamental ($F$) & $(1,0,\ldots,0)$ \\
Anti-fundamental ($\overline F$) & $(0,\ldots,0,1)$ \\
Adjoint (adj) & $(1,0,\ldots,0,1)$ \\
$n$-index symmetric & $(n,0,\ldots,0)$ \\
Two-index antisymmetric & $(0,1,0,\ldots,0)$ \\
\hline
\end{tabular}
\end{center}

The weights of the defining representation of $su(N_c)$ are
\begin{equation}
\bm\nu_a = \bm w_1 - \sum_{b=1}^{a-1}\bm\alpha_b, \qquad a=1,\ldots,N_c.
\end{equation}

\subsection*{Young tableau}

The symmetry of a tensor in representation $(m_1,m_2,\ldots,m_r)$ is encoded in its Young tableau. For example, the $(3,3,3)$ representation of $su(4)$ corresponds to
\begin{equation}
\ytableausetup{centertableaux}
\begin{ytableau}
{} & {} & {} & {} & {} & {} & {} & {} & {} \\
{} & {} & {} & {} & {} & {} \\
{} & {} & {}
\end{ytableau}.
\end{equation}

The $N$-ality of a representation ${\cal R}$ of $su(N_c)$ is defined as the sum of the boxes in a Young tableau modulo $N_c$. In terms of the Dynkin labels, it is given by
\begin{eqnarray}
n_c=\sum_{a=1}^{N_c-1} am_a\,\mbox{Mod}\, N_c\,.
\end{eqnarray}

\subsection*{Casimir and trace operators, representation dimensions, and the $\beta$-function}
\label{app:Casimir}

The quadratic Casimir of a representation ${\cal R}$ is defined by
\begin{equation}
t^a_{\cal R} t^a_{\cal R} = C_2({\cal R})\,I_{\dim{\cal R}},
\end{equation}
and in particular, the adjoint Casimir is denoted $C_2(G)$.  

The trace normalization $T_{\cal R}$ is defined by
\begin{equation}
\mathrm{tr}\,(t^a_{\cal R}t^b_{\cal R}) = T_{\cal R}\,\delta^{ab}.
\end{equation}
From these definitions, it follows that
\begin{equation}\label{TRGR}
T_{\cal R}\, \dim G = C_2({\cal R})\, \dim{\cal R},
\end{equation}
where $\dim{\cal R}$ is the dimension of ${\cal R}$.  This relation can be used to find $T_{\cal R}$ knowing $C_2({\cal R})$ and $\dim{\cal R}$, which we now turn to.

For a representation with Dynkin labels $(a_1,\ldots,a_{N_c-1})$, the quadratic Casimir is~\cite{White:1992aa}
\begin{align}
C_2({\cal R}) &= \frac{1}{N_c}\sum_{m=1}^{N_c-1}\Big[ N_c(N_c-m) m a_m + m(N_c-m) a_m^2 \notag \\
&\quad + \sum_{n=0}^{m-1} 2n(N_c-m) a_n a_m \Big],
\end{align}
and the dimension is
\begin{equation}
\dim{\cal R} = \prod_{p=1}^{N_c-1}\left\{ \frac{1}{p!}\prod_{q=p}^{N_c-1}\left[\sum_{r=q-p+1}^q (1+a_r)\right]\right\}.
\end{equation}

In particular,
\begin{equation}
C_2(G)=2N_c, \qquad \dim G=N_c^2-1,
\end{equation}
and
\begin{eqnarray}
T_{\Box}=1
\end{eqnarray}
for the defining (fundamental) representation.

The $1$-loop $\beta$ function is given by
\begin{eqnarray}
\beta(g)=-\beta_0\frac{g^3}{(4\pi)^2}\,,\quad
\beta_0=\frac{11}{6}C_2(G)-\sum_{\cal R}\frac{1}{3}T_{\cal R}n_{\cal R}~,
\label{beta function}
\end{eqnarray} 
where $n_{\cal R}$ is the number of the Weyl flavors in representation ${\cal R}$. 

\section{Constructing weights using Verma basis}
\label{secverma-bases}

\begin{table}[t]
\centering
\renewcommand{\arraystretch}{1.2}
\begin{tabular}{|c|l|}
\hline
$su(2)$ & $0\leq a_1 \leq m_1$ \\
        & $0\leq a_2 \leq m_2+a_1$ \\
\hline
$su(3)$ & $0\leq a_3 \leq \min(m_2,a_2)$ \\\hline
  $su(4)$      & $0\leq a_4 \leq m_3+a_2$ \\
        & $0\leq a_5 \leq \min(m_3+a_3,a_4)$ \\
 & $0\leq a_6 \leq \min(m_3,a_5)$ \\
\hline
$\vdots$ & $\vdots$ \\
\hline
$su(N_c)$ & $0\leq a_{N-N_c+2}\leq m_{N_c-1}+a_{N-2N_c+4}$ \\
          & $0\leq a_{N-N_c+3}\leq \min(m_{N_c-1}+a_{N-2N_c+5},a_{N-N_c+2})$ \\
           & $0\leq a_{N-N_c+4}\leq \min(m_{N_c-1}+a_{N-2N_c+6},a_{N-N_c+3})$ \\
          & $\cdots$ \\
          & $0\leq a_{N}\leq \min(m_{N_c-1},a_{N-1})$ \\
\hline
\end{tabular}
\caption{Verma basis inequalities~\cite{:/content/aip/journal/jmp/27/3/10.1063/1.527222}.}
\label{verma-basis-table}
\end{table}

The weights of any representation of $su(N_c)$ can be constructed systematically using the Verma basis \cite{:/content/aip/journal/jmp/27/3/10.1063/1.527222}. 
For a representation ${\cal R}$, denoted by its Dynkin labels
\begin{equation}
{\cal R} = (m_1,m_2,\ldots,m_{N_c-1}),
\end{equation}
the generic basis vectors take the form
\begin{align}
\Big[&\,(E_{-\bm \alpha_1})^{a_N} (E_{-\bm \alpha_2})^{a_{N-1}} \cdots (E_{-\bm \alpha_{N_c-1}})^{a_{N-N_c+2}} \Big]
\Big[ (E_{-\bm \alpha_1})^{a_{N-N_c+1}} \cdots (E_{-\bm \alpha_{N_c-2}})^{a_{N-2N_c+4}} \Big] \notag \\
&\hspace{3cm}\cdots 
\Big[(E_{-\bm \alpha_1})^{a_3}(E_{-\bm \alpha_2})^{a_2}\Big] 
(E_{-\bm \alpha_1})^{a_1} |{\cal R}\rangle ,
\label{eq:verma-basis}
\end{align}
where $N=N_c(N_c-1)/2$, $\{E_{-\bm \alpha_a}\}$ $(a=1,2,\ldots,N_c-1)$ are the simple-root generators, and $\{a_i\}$ are nonnegative integers constrained by inequalities given in Table~\ref{verma-basis-table} (see~\cite{:/content/aip/journal/jmp/27/3/10.1063/1.527222}).

\paragraph{Example: $su(3)$.}  
For the algebra $su(3)$, a representation $(m_1,m_2)$ has basis vectors of the form
\begin{equation}
(E_{-\bm \alpha_1})^{a_3}(E_{-\bm \alpha_2})^{a_2}(E_{-\bm \alpha_1})^{a_1}|(m_1,m_2)\rangle ,
\end{equation}
with the constraints
\begin{equation}
0 \leq a_1 \leq m_1, 
\qquad 
0 \leq a_2 \leq m_2+a_1, 
\qquad 
0 \leq a_3 \leq \min(m_2,a_2).
\end{equation}

As a concrete case, consider the adjoint representation $G=(1,1)$. The corresponding Verma basis reads
\begin{align}
\{\, & |(1,1)\rangle,\ 
E_{-\bm \alpha_1}|(1,1)\rangle,\ 
E_{-\bm \alpha_2}|(1,1)\rangle,\ 
E_{-\bm \alpha_2}E_{-\bm \alpha_1}|(1,1)\rangle,\ 
E_{-\bm \alpha_1}E_{-\bm \alpha_2}|(1,1)\rangle, \notag \\
& (E_{-\bm \alpha_2})^2 E_{-\bm \alpha_1}|(1,1)\rangle,\ 
E_{-\bm \alpha_1}E_{-\bm \alpha_2}E_{-\bm \alpha_1}|(1,1)\rangle,\ 
E_{-\bm \alpha_1}(E_{-\bm \alpha_2})^2E_{-\bm \alpha_1}|(1,1)\rangle
\,\}.
\label{eq:adjoint-verma}
\end{align}
The simple roots  are
\begin{equation}
\bm \alpha_1=\left(\tfrac{1}{\sqrt{2}}, \tfrac{\sqrt{3}}{\sqrt{2}}\right), 
\qquad 
\bm \alpha_2=\left(\tfrac{1}{\sqrt{2}}, -\tfrac{\sqrt{3}}{\sqrt{2}}\right),
\end{equation}
and the fundamental weights are
\begin{equation}
\bm w_1=\left(\tfrac{1}{\sqrt{2}}, \tfrac{1}{\sqrt{6}}\right), 
\qquad 
\bm w_2=\left(\tfrac{1}{\sqrt{2}}, -\tfrac{1}{\sqrt{6}}\right).
\end{equation}
The highest weight state is given by
\begin{equation}
|(m_1,m_2)\rangle \;\equiv\; m_1 \bm w_1 + m_2 \bm w_2,
\end{equation}
and the remaining weights are obtained by successive subtraction of simple roots $\bm \alpha_1$ and/or $\bm \alpha_2$ according to the Verma basis~\eqref{eq:adjoint-verma}.

\section{Fermionic zero modes in the EH geometry with $U(1)$ flux}
\label{sec:fermions-EH}

In this appendix, we analyze how fermions behave in the EH gravitational background with a localized $U(1)$ flux supported at the bolt. The objective is to identify the spectrum of fermionic zero modes by studying the Dirac operator in this curved space. Before we begin, it is helpful to list the key geometric quantities required to solve the Dirac equation in the EH background. We use the definitions and conventions of  \cite{Eguchi:1980jx}. First, the vielbeins are given by
\begin{eqnarray}\label{velbspher}
e^a=(e^0,e^1,e^2,e^3)=\left(f(r)dr,r\sigma_x,r\sigma_y,rf^{-1}(r)\sigma_z\right)\,,
\end{eqnarray}
and we adopt the convention $e^0\wedge e^1\wedge e^2\wedge e^3$ is positive. Also,  the spin connection $1$-forms are
\begin{eqnarray}\label{omegaEH}
\omega_{10} = \omega_{23} = \frac{\sigma_x}{f}\,, \quad \omega_{20} = \omega_{31} = \frac{\sigma_y}{f}\,, \quad \omega_{30} = \omega_{12} = (2 - f^{-2}) \sigma_z\,,
\end{eqnarray}
and the corresponding curvature $2$-forms:
\begin{eqnarray} \label{CT2EH} \nonumber
R_{10} &=& R_{23} = -\frac{2a^4}{r^6}(e^1 \wedge e^0 + e^2 \wedge e^3)\,, \\\nonumber
R_{20} &=& R_{31} = -\frac{2a^4}{r^6}(e^2 \wedge e^0 + e^3 \wedge e^1)\,, \\
R_{30} &=& R_{12} = \frac{4a^4}{r^6}(e^3 \wedge e^0 + e^1 \wedge e^2)\,.
\end{eqnarray}
Remarkably, in the standard EH coordinates given by the metric (\ref{EH_metric}), we find that the spin connection $1$-forms and curvature $2$-forms satisfy the anti-self-duality condition, for example $\tilde{R}_{12} = -R_{12}$, where $\tilde R_{ab}\equiv \frac{1}{2}\epsilon_{abcd}R_{cd}$ (and we use $\epsilon_{0123}=1$), and similarly for the others.

The starting point is the Dirac equation
\begin{equation}\label{fullDirac}
\slashed{D}\Psi = 0\,, \qquad 
\slashed{D} = \gamma^a E_a^{\;\mu}\Bigl(\partial_\mu + \tfrac{1}{2}\Omega_\mu + i A_\mu\Bigr)\,,
\end{equation}
where the $U(1)$ gauge field is
\begin{equation}
A_{} \;=\; {\cal C}\, \frac{\sigma_z\, a^2}{r^2},\quad {\cal C}\in \mathbb Z
\end{equation}
 $\Omega_\mu$ denotes the spin connection contribution,
\begin{equation}
\Omega_\mu = \omega_{ab\mu}\,\Sigma^{ab}, \qquad \Sigma^{ab}=\tfrac{1}{4}[\gamma^a,\gamma^b]\,,
\end{equation}
and $\gamma^a$, $a=0,1,2,3$, satisfies the Clifford algebra $\{\gamma^a,\gamma^b\}=2\delta^{ab}$.

The inverse vielbein $E_a^{\;\mu}$ is obtained from $E_a^{\;\mu} e^b_{\;\mu}=\delta^b_a$. Explicitly,
\begin{align}\nonumber
E_0^{\mu} &= \Bigl(\tfrac{1}{f},0,0,0\Bigr), &
E_1^{\mu} &= \Bigl(0, \tfrac{2\sin\psi}{r}, -\tfrac{2\cos\psi}{r\sin\theta}, \tfrac{2\cos\psi}{r}\cot\theta\Bigr),\\
E_2^{\mu} &= \Bigl(0, -\tfrac{2\cos\psi}{r}, -\tfrac{2\sin\psi}{r\sin\theta}, \tfrac{2\sin\psi}{r}\cot\theta\Bigr), &
E_3^{\mu} &= \Bigl(0,0,0,\tfrac{2f}{r}\Bigr).
\end{align}
Inserting this into $\slashed{D}$ gives
\begin{align}
\slashed{D} &= \gamma^0 \tfrac{1}{f}D_r 
+ \biggl[\gamma^1 \tfrac{2\sin\psi}{r} - \gamma^2 \tfrac{2\cos\psi}{r}\biggr]D_\theta \notag\\
&\quad + \biggl[-\gamma^1 \tfrac{2\cos\psi}{r\sin\theta} - \gamma^2 \tfrac{2\sin\psi}{r\sin\theta}\biggr]D_\varphi \notag\\
&\quad + \biggl[\gamma^1 \tfrac{2\cos\psi}{r}\cot\theta + \gamma^2 \tfrac{2\sin\psi}{r}\cot\theta + \gamma^3\tfrac{2f}{r}\biggr]D_\psi\,,
\end{align}
and the individual covariant derivatives take the form
\begin{align}
D_r &= \partial_r\,, \qquad 
D_\psi = \partial_\psi + \tfrac{1}{2}(\Sigma^{30}+\Sigma^{12})(2-f^{-2}) + i\,\tfrac{\mathcal{C}a^2}{2r^2}\,, \notag\\
D_\theta &= \partial_\theta + \tfrac{ (\Sigma^{10}+\Sigma^{23})\sin\psi - (\Sigma^{20}+\Sigma^{31})\cos\psi }{2f}\,, \notag\\
D_\varphi &= \partial_\varphi - \tfrac{\sin\theta}{2f}\Bigl[(\Sigma^{10}+\Sigma^{23})\cos\psi + (\Sigma^{20}+\Sigma^{31})\sin\psi\Bigr] \notag\\
&\qquad\qquad + \tfrac{1}{2}(\Sigma^{30}+\Sigma^{12})(2-f^{-2})\cos\theta 
+ i \tfrac{\mathcal{C}a^2}{2r^2}\cos\theta.
\end{align}

For the gamma matrices, we use the chiral representation,
\begin{equation}
\gamma^a = \begin{pmatrix} 0 & \hat\sigma^a \\ \bar{\hat\sigma}^a & 0 \end{pmatrix}, 
\end{equation}
with $\hat\sigma^a=(1,i\vec{\hat\sigma})$ and $\bar{\hat\sigma}^a=(1,-i\vec{\hat\sigma})$.  
Because EH space is self-dual, only one chirality of fermions couples to the spin connection. In particular,
\begin{equation}
\Sigma^{10}+\Sigma^{23}=i\begin{pmatrix}\hat\sigma_x&0\\0&0\end{pmatrix},\quad
\Sigma^{20}+\Sigma^{31}=i\begin{pmatrix}\hat\sigma_y&0\\0&0\end{pmatrix},\quad
\Sigma^{30}+\Sigma^{12}=i\begin{pmatrix}\hat\sigma_z&0\\0&0\end{pmatrix}.
\end{equation}

To streamline the angular dependence, we introduce the operators
\begin{align}
\hat{\mathcal{J}}_x &= i\Bigl(\sin\psi\,\partial_\theta - \cos\psi\csc\theta\,\partial_\varphi + \cos\psi\cot\theta\,\partial_\psi\Bigr),\notag\\
\hat{\mathcal{J}}_y &= i\Bigl(-\cos\psi\,\partial_\theta - \sin\psi\csc\theta\,\partial_\varphi + \sin\psi\cot\theta\,\partial_\psi\Bigr),\notag\\
\hat{\mathcal{J}}_z &= -i\,\partial_\psi,
\end{align}
which form an $SU(2)$ algebra $[\hat{\mathcal{J}}_x,\hat{\mathcal{J}}_y]=i\hat{\mathcal{J}}_z$ and cyclic permutations.

In this language, the Dirac equation becomes block-diagonal,
\begin{equation}
\slashed{D}\Psi = 
\begin{pmatrix}
0 & \mathcal{A}\\
\mathcal{B} & 0
\end{pmatrix}\Psi=0,
\end{equation}
with
\begin{align}\label{andb}
\mathcal{A} &= \tfrac{1}{f}\partial_r I_2 - \tfrac{a^2\mathcal{C}f}{r^3}\hat\sigma_z 
+ \tfrac{2}{r}\bigl(\hat{\mathcal{J}}_x\hat\sigma_x+\hat{\mathcal{J}}_y\hat\sigma_y - f\hat{\mathcal{J}}_z\hat\sigma_z\bigr), \notag\\
\mathcal{B} &= \Bigl(\tfrac{1}{f}\partial_r + \tfrac{2f}{r}+\tfrac{1}{fr}\Bigr)I_2 
+ \tfrac{a^2\mathcal{C}f}{r^3}\hat\sigma_z 
- \tfrac{2}{r}\bigl(\hat{\mathcal{J}}_x\hat\sigma_x+\hat{\mathcal{J}}_y\hat\sigma_y - f\hat{\mathcal{J}}_z\hat\sigma_z\bigr).
\end{align}

We write the Dirac spinor as 
\begin{equation}
\Psi =
\begin{bmatrix}
\lambda_\alpha \\[4pt]
\bar\chi^{\dot\alpha}
\end{bmatrix},
\end{equation}
which leads to the coupled equations
\begin{align}
\Bigg[ \frac{1}{f}\frac{\partial}{\partial r}I_2 
- \frac{a^2{\cal C} f}{r^3}\hat\sigma_z 
+ \frac{2}{r}\Big(\hat {\cal J}_x\hat\sigma_x
+\hat {\cal J}_y\hat\sigma_y
-f\hat {\cal J}_z\hat\sigma_z\Big)\Bigg]\bar\chi &= 0,\nonumber  \\[6pt]
\Bigg[ \Big(\frac{1}{f}\frac{\partial}{\partial r}
+\frac{2f}{r}+\frac{1}{fr}\Big)I_2 
+ \frac{a^2{\cal C} f}{r^3}\hat\sigma_z 
- \frac{2}{r}\Big(\hat {\cal J}_x\hat\sigma_x
+\hat {\cal J}_y\hat\sigma_y
-f\hat {\cal J}_z\hat\sigma_z\Big)\Bigg]\lambda &= 0. \label{eq:lambda-eq}
\end{align}

The $\bar\chi$ equation does not lead to normalizable zero modes. Thus, we focus on the $\lambda$-equation. After straightforward manipulations, it reduces to
\begin{align}
\Bigg[ \frac{1}{f}\frac{\partial}{\partial r} 
+ \frac{2f}{r} + \frac{1}{fr} 
+ \frac{a^2{\cal C}f}{r^3} 
+ \frac{2f}{r}\hat{\cal J}_z\Bigg]\lambda_1
- \frac{2}{r}\hat{\cal J}_-\lambda_2 &= 0, \label{eq:lambda1} \\[6pt]
-\frac{2}{r}\hat{\cal J}_+\lambda_1
+ \Bigg[ \frac{1}{f}\frac{\partial}{\partial r} 
+ \frac{2f}{r} + \frac{1}{fr} 
- \frac{a^2{\cal C}f}{r^3} 
- \frac{2f}{r}\hat{\cal J}_z\Bigg]\lambda_2 &= 0. \label{eq:lambda2}
\end{align}
Here $\hat{\cal J}_{\pm} = \hat{\cal J}_x \pm i\hat{\cal J}_y$.

The solutions can be factorized into radial and angular parts,
\begin{equation}
\lambda_1= g_1(r)\,|j_1,m_1',m_1\rangle, 
\qquad 
\lambda_2 = g_2(r)\,|j_2,m_2',m_2\rangle,
\label{eq:lambda-decomp}
\end{equation}
where $|j,m',m\rangle$ are Wigner $D$-matrix states \cite{Sakurai:2011zz}:
\begin{align}
|j,m',m\rangle 
&\equiv D^j_{m',m}(\theta,\varphi,\psi) 
= d^j_{m',m}(\theta)\, e^{im\varphi} e^{im'\psi}, 
\qquad j = 0,\tfrac12,1,\ldots,\quad |m|,|m'|\leq j, \\[4pt]
d^j_{m',m}(\theta) 
&= \sum_k (-1)^{k-m+m'}
\frac{\sqrt{(j+m)!(j-m)!(j+m')!(j-m')!}}
     {(j+m-k)!\,k!\,(j-k-m')!\,(k-m+m')!} \nonumber \\[2pt]
&\quad \times \left(\cos\frac{\theta}{2}\right)^{2j-2k+m-m'}
        \left(\sin\frac{\theta}{2}\right)^{2k-m+m'}.
\end{align}
The action of $\hat{\cal J}_z$ and $\hat{\cal J}_\pm$ is standard:
\begin{align}\nonumber
\hat{\cal J}_z|j,m',m\rangle &= m'|j,m',m\rangle, \\
\hat{\cal J}_\pm|j,m',m\rangle &= \sqrt{j(j+1)-m'(m'\pm1)}\,|j,m'\!\pm\!1,m\rangle.
\end{align}
The states $\{|j,m',m\rangle\}$ form a complete orthonormal basis with
\begin{equation}
\int_0^{2\pi}\! d\psi \int_0^\pi \! d\theta\,\sin\theta 
\int_0^{2\pi}\! d\varphi \,
\langle j_1,m_1',m_1 | j_2,m_2',m_2\rangle
= \frac{8\pi^2}{2j_1+1}\,\delta_{j_1j_2}\delta_{m_1'm_2'}\delta_{m_1m_2}.
\end{equation}

Substituting \eqref{eq:lambda-decomp} into \eqref{eq:lambda1}--\eqref{eq:lambda2} gives
\begin{align}
\Bigg[ \frac{1}{f}\frac{\partial}{\partial r} 
+ \frac{2f}{r} + \frac{1}{fr} 
+ \frac{a^2{\cal C}f}{r^3} 
+ \frac{2fm_1'}{r}\Bigg]g_1(r)\,|j_1,m_1',m_1\rangle \nonumber \\
-\frac{2}{r}\sqrt{j_2(j_2+1)-m_2'(m_2'-1)}\, g_2(r)\,|j_2,m_2'-1,m_2\rangle &= 0,
\label{eq:radial1}
\end{align}
and
\begin{align}
-\frac{2}{r}\sqrt{j_1(j_1+1)-m_1'(m_1'+1)}\, g_1(r)\,|j_1,m_1'+1,m_1\rangle \nonumber \\
+ \Bigg[ \frac{1}{f}\frac{\partial}{\partial r} 
+ \frac{2f}{r} + \frac{1}{fr} 
- \frac{a^2{\cal C}f}{r^3} 
- \frac{2fm_2'}{r}\Bigg]g_2(r)\,|j_2,m_2',m_2\rangle &= 0.
\label{eq:radial2}
\end{align}

A careful analysis shows that consistent solutions exist only when
\begin{equation}
\lambda_1(r,\Omega) = g_1(r)\,|j,m'=j,m\rangle, 
\qquad 
\lambda_2(r,\Omega) = g_2(r)\,|j,m'=-j,m\rangle, 
\qquad |m|\leq j.
\end{equation}
Using $\hat{\cal J}_-|j,-j,m\rangle=0$ and $\hat{\cal J}_+|j,j,m\rangle=0$, the system simplifies to
\begin{align}
\Bigg[ \frac{1}{f}\frac{\partial}{\partial r}
+ \frac{2f}{r} + \frac{1}{fr} 
+ \frac{a^2{\cal C}f}{r^3} 
+ \frac{2fj}{r}\Bigg] g_1(r) &= 0,\nonumber \\
\Bigg[ \frac{1}{f}\frac{\partial}{\partial r}
+ \frac{2f}{r} + \frac{1}{fr} 
- \frac{a^2{\cal C}f}{r^3} 
+ \frac{2fj}{r}\Bigg] g_2(r) &= 0. \label{eq:g2}
\end{align}

The norm is defined as
\begin{equation}
\int |\Psi|^2\, r^3 dr\wedge \sigma_x\wedge\sigma_y\wedge\sigma_z < \infty.
\label{eq:norm}
\end{equation}
For ${\cal C}>0$, only $g_2$ admits normalizable solutions; for ${\cal C}<0$, only $g_1$ does. If ${\cal C}=0$, no normalizable zero modes exist.  

When ${\cal C}>0$, the normalizable solutions of \eqref{eq:g2} exist for ${\cal C}>2j$ and are given by
\begin{equation}
g_2(r) = \frac{1}{r}(r^2-a^2)^{\frac{{\cal C}-2j-2}{4}}
(r^2+a^2)^{-\frac{{\cal C}+2j+2}{4}},
\qquad j \geq 0, \quad {\cal C}>2j.
\end{equation}
Near $r=a$ one has $g_2^2 \sim (r-a)^{({\cal C}-2j-2)/2}$, so the norm \eqref{eq:norm} converges if ${\cal C}>2j$. At infinity, $r^3g_2^2 \sim r^{-4j-3}$, which is convergent for all $j\geq 0$.

Under gravity and gauge fields, parallel transporting the spinor along the contractible path $\ell_{r \approx a}$ (connecting $(r \approx a,\theta,\varphi,\psi)$ and $(r \approx a,\theta,\varphi,\psi+2\pi)$) gives 
\begin{align}\label{PARALLTA2}
\Psi(r \approx a,\theta,\varphi, \psi+2\pi) &= e^{i\pi{\cal C}} \Psi(r \approx a,\theta,\varphi, \psi)\,.
\end{align}
On the other hand, from the solution of the zero modes, we can directly calculate $\Psi$ at $\psi+2\pi$:
\begin{align}\label{aretheysame}\nonumber
\Psi(r,\theta,\varphi,\psi) &= \left[\begin{array}{c}
0\\
g_2(r)e^{-ij\psi}e^{im\varphi}d_{m',m}^j(\theta)\\
0\\
0
\end{array}\right] \longrightarrow\\
\Psi(r,\theta,\varphi,\psi+2\pi)&= e^{-i2\pi j}\left[\begin{array}{c}
0\\
g_2(r)e^{-ij\psi}e^{im\varphi}d_{m',m}^j(\theta)\\
0\\
0
\end{array}\right]\,.
\end{align}
The fermion zero modes are well-defined on EH space, i.e., they are unambiguously defined on a contractible path, if and only if the r.h.s. of (\ref{PARALLTA2}) and (\ref{aretheysame}) are identical.

If ${\cal C}=2p$ is even, the holonomy at the bolt is trivial, restricting $j$ to integers. The number of zero modes is
\begin{equation}
{\cal I}_{U(1)} = \sum_{j=0}^{p-1} (2j+1) = p^2.
\end{equation}
If ${\cal C}=2p+1$ is odd, the holonomy contributes a $(-1)$ phase, restricting $j$ to half-integers. The number of zero modes is
\begin{equation}
{\cal I}_{U(1)} = \sum_{j=1/2}^{p-1/2} (2j+1) = p(p+1).
\end{equation}
These results agree with the Atiyah–Patodi–Singer index theorem.

In the remainder of this appendix, we examine the behavior of the nonzero modes of the Dirac operator, focusing on the $s$-wave sector (the mode with angular momentum $j=0$) as the simplest illustrative example. We are interested in solving
\begin{equation}\label{nonzeromodesanalysis}
\slashed{D}\Psi = -i\omega\,\Psi 
\quad \Longleftrightarrow \quad
\begin{bmatrix}
0 & \mathcal{A} \\[4pt]
\mathcal{B} & 0
\end{bmatrix}
\begin{bmatrix}
\lambda_\alpha \\[4pt]
\bar\chi^{\dot\alpha}
\end{bmatrix}
=
-i\omega
\begin{bmatrix}
\lambda_\alpha \\[4pt]
\bar\chi^{\dot\alpha}
\end{bmatrix},
\end{equation}
where $\mathcal{A}$ and $\mathcal{B}$ are defined in~\eqref{andb}. Since our goal is only to understand the qualitative behavior of the nonzero modes rather than construct the full solution, it suffices to study a single component, say $\bar\chi^{\dot\alpha}$. The equation of motion of this component is
\begin{equation}
\mathcal{B}\mathcal{A}\,\bar\chi^{\dot\alpha} \;=\; -\omega^2\,\bar\chi^{\dot\alpha}.
\end{equation}
For the lowest angular-momentum mode, we take
$
\bar\chi^{\dot\alpha}(r,\Omega)
=
\mathcal{G}^{\dot\alpha}(r)\,
\big| j=0,\,m'=0,\,m=0\big\rangle,
$
and for concreteness we consider the equation obeyed by $\mathcal{G}^{\dot 1}(r)$:
\begin{equation}
\left(
\frac{1}{f}\frac{d}{dr}
+ \frac{2f}{r}
+ \frac{1}{fr}
+ \frac{a^2 \mathcal{C} f}{r^3}
\right)
\left(
\frac{1}{f}\frac{d}{dr}
- \frac{a^2\mathcal{C} f}{r^3}
\right)
\mathcal{G}^{\dot 1}(r)
= -\omega^2 \mathcal{G}^{\dot 1}(r).
\end{equation}

For long wavelengths, $|\omega|\ll a^{-1}$,  the relevant region is $r\sim |\omega|^{-1}\gg a$, so that $f-1=\mathcal{O}((a/r)^4)=\mathcal{O}((a\omega)^4)$ and the $U(1)$ term is of order $a^{2}|\omega|^{3}$, both parametrically suppressed relative to $\omega^{2}$.
Under this approximation, the radial equation is solved by
\begin{equation}
\mathcal{G}^{\dot 1}(r) \;\sim\; \frac{H_1(r|\omega|)}{r},
\end{equation}
where $H_1(x)$ is the Hankel function of the first or second kind.

For short wavelengths, $|\omega|\gg a^{-1}$, the modes see a locally flat $\mathbb R^4$ and thus they are just more violently oscillatory scattering states. 

This analysis shows the expected behavior of the non-zero modes. Since the EH space is noncompact and asymptotically locally Euclidean, the Dirac operator approaches the flat-space operator at large $r$, and its nonzero eigenfunctions become asymptotically oscillatory; hence, apart from the finitely many normalizable zero modes, the spectrum is continuous for all $\omega>0$ (with the continuum beginning at $\omega\rightarrow 0$ and extending to infinity).

\section{Anomalies on  $\mathbb T^4$ and $\mathbb {CP}^2$}
\label{anomaliesonT4CP2}

On a general $4$-D manifold $X_4$, topological gauge sectors of an abelian field
strength $F$ are classified by its second cohomology class
$
 \frac{F}{2\pi} \in H^2(X_4;\mathbb{Z}).
$
Thus the allowed fluxes form a lattice $\Lambda = H^2(X_4;\mathbb{Z})$
equipped with the bilinear pairing
$
 (x,y)=\int_{X_4} x\wedge y
$. 

The simplest case of $X_4$ is $S^4$, which has $H^2(S^4; \mathbb Z)=0$ and therefore carries no
nontrivial $2$-form flux sectors;  $S^4$ can be used to detect $0$-form and perturbative anomalies.   

On the contrast, for $X_4=\mathbb{T}^4$ one has
$
H^2(\mathbb{T}^4;\mathbb{Z}) \cong \mathbb{Z}^6,
$
and the intersection pairing is \emph{even and unimodular}.
 Unimodular: the intersection matrix has determinant $\pm 1$;
  even: every vector has integer even pairing
  $(x,x)\in 2\mathbb{Z}$.
Consequently\footnote{This is easily seen on $\mathbb T^4$ (with unit periods) via the use of ${\cal K}=m_{12}dx^1\wedge dx^2+m_{13}dx^1\wedge dx^3+m_{14}dx^1\wedge dx^4+m_{23}dx^2\wedge dx^3+m_{24}dx^2\wedge dx^4+m_{34}dx^3\wedge dx^4$, for  $m_{\mu\nu}\in \mathbb Z$ and $\mu,\nu=1,2,3,4$. Then, direct calculations give $\int_{\mathbb{T}^4} {\cal K}\wedge {\cal K} =\frac{1}{4}\epsilon_{\mu\nu\alpha\beta}m_{\mu\nu}m_{\alpha\beta}=2m_{12}m_{34}+...\in 2\mathbb Z$. The intersection matrix is given by
\begin{equation}
Q \;=\;
\begin{pmatrix}
0 & 0 & 0 & 0 & 0 & 1\\
0 & 0 & 0 & 0 & -1 & 0\\
0 & 0 & 0 & 1 & 0 & 0\\
0 & 0 & 1 & 0 & 0 & 0\\
0 & -1 & 0 & 0 & 0 & 0\\
1 & 0 & 0 & 0 & 0 & 0
\end{pmatrix}\,,
\end{equation}
with determinant $\mbox{Det}\, Q=-1$. Notice that all the intersections of the $2$-cycles $\mathbb T^2$ are mutual and there are no self-intersections.
}, for an abelian bundle ${\cal K}$ that satisfies $\int_{\mathbb T^2\subset \mathbb T^4} {\cal K}=1$, one finds
$
\int_{\mathbb{T}^4} {\cal K}\wedge {\cal K} \in 2\mathbb{Z}
$. Thus, for a $U(1)$ gauge bundle, $F_{(1)}=2\pi{\cal K}$, the $U(1)$ topological charge is $(1/8\pi^2)\int_{\mathbb T^4}F_{(1)}\wedge F_{(1)}\in \mathbb Z$.  

For a nonabelian $\mathrm{SU}(N)$ gauge theory the topological charge $(1/8\pi^2)\int_{\mathbb{T}^4} \mathrm{tr} F_{(N)}\wedge F_{(N)}\in \frac{\mathbb Z}{N}$, i.e., can be fractional.  This does not contradict even unimodularity: fractional instantons arise not from $H^2(\mathbb{T}^4;\mathbb{Z})$ (the cohomological flux lattice), but from nontrivial bundle topology valued in the center $\mathbb Z_N$ of the gauge group.  More precisely, $PSU(N)$ bundles on $\mathbb{T}^4$ are classified by $H^2(\mathbb{T}^4, \mathbb Z_N) \cong (\mathbb Z_N)^6$, and a nonzero class in $H^2(\mathbb{T}^4,\mathbb Z_N)$ produces a 't~Hooft flux which shifts the instanton number by a multiple of $1/N$.  Equivalently, these fractional topological charges encode the coupling of the theory to a background $1$-form $\mathbb Z_N^{(1)}$ center-symmetry gauge field. As $\mathbb T^4$ admits both fractional and integral topological charges, it serves as a probe of anomalies of both $0$-form and  $1$-form symmetries. This is precisely the mechanism by which one diagnoses what is dubbed in \cite{Anber:2019nze} as the baryon–colour–flavour (BCF) anomalies: they are generalized 't Hooft anomalies involving both $0$-form and $1$-form symmetries.

Next, we come to $\mathbb{CP}^2$.
The second cohomology group of $\mathbb{CP}^2$ is $H^2(\mathbb{CP}^2;\mathbb{Z}) = \mathbb{Z}$, and the space contains a unique $2$-cycle, the projective line $\mathbb{CP}^1 \subset \mathbb{CP}^2$. Let ${\cal K}$ be a closed $2$-form normalized such that $\int_{\mathbb{CP}^1} {\cal K} = 1$. Then the intersection form satisfies $\int_{\mathbb{CP}^2} {\cal K} \wedge {\cal K} = 1$, so the self-intersection number of $\mathbb{CP}^2$ is $+1$, and the intersection pairing on $H^2(\mathbb{CP}^2;\mathbb{Z})$ is odd and unimodular.
For a $U(1)$ field strength $F_{(1)} = 2\pi {\cal K}$ with $\int_{\mathbb{CP}^1} (F_{(1)}/2\pi) = m \in \mathbb{Z}$, the topological charge is $(1/8\pi^2) \int_{\mathbb{CP}^2} F_{(1)} \wedge F_{(1)} = m^2/2$, which is generally half--integral. The factor of $1/2$ arises from the unit self-intersection number on $\mathbb{CP}^2$ (in contrast to $\mathbb{T}^4$, whose intersection form is even).

The manifold $\mathbb{CP}^2$ is not spin, since the second Stiefel-Whitney class $w_2(T\mathbb{CP}^2) \neq 0$, so there is an obstruction to lifting the $SO(4)$ tangent bundle to $\mathrm{Spin}(4)$. As a result, bare fermions are not globally well-defined. Indeed, in the absence of background gauge fields, the gravitational contribution to the Dirac index is ${\cal I}_{\mathbb{CP}^2} = 1/(24 \cdot 8\pi^2) \int_{\mathbb{CP}^2} \mathrm{tr}\,R \wedge R = -1/8$, a fractional value.
To define fermions consistently on $\mathbb{CP}^2$, one must instead work with a $\mathrm{spin}^c$ structure and turn on a $U(1)$ flux obeying $(1/2\pi) \int_{\mathbb{CP}^1} F_{(1)} = 1/2 + \mathbb{Z}$. This corresponds to using the $\mathrm{spin}^c$ bundle $(\mathrm{Spin}(4) \times U(1))/\mathbb{Z}_2$. For such half-integral flux, the gauge contribution to $(1/8\pi^2) \int F_{(1)} \wedge F_{(1)}$ equals $(1/2) (1/2 + \mathbb{Z})^2$, which combines with the gravitational term to produce an integer Dirac index; see \cite{Anber:2020gig} for more details. 

More generally, $H^2(\mathbb{CP}^2;\mathbb{Z}_N) \cong \mathbb{Z}_N$, so $\mathbb{CP}^2$ supports discrete 't~Hooft fluxes. For an $SU(N)/\mathbb{Z}_N$ bundle or background for a $\mathbb{Z}_N^{(1)}$ $1$-form symmetry, one finds\\ $(1/8\pi^2) \int_{\mathbb{CP}^2} \mathrm{tr}\,F_{(N)} \wedge F_{(N)} \in (1/2)(1/N + \mathbb{Z})$, again reflecting the geometric factor of $1/2$ coming from the odd intersection form.

These refined geometric and topological features distinguish $\mathbb{CP}^2$ from $\mathbb{T}^4$ and make it a powerful background for diagnosing anomalies that are invisible on $\mathbb{T}^4$ \cite{Anber:2020gig,Brennan:2023vsa}. For example, turning on BCF fluxes on $\mathbb{CP}^2$ in~\cite{Anber:2020gig} was used to constrain and rule out candidate IR phases in strongly-coupled gauge theories.

\section{Computation of \texorpdfstring{$\Omega_{5}^{\mathrm{Spin}}(B(U(5)\times \mathbb Z_{3}))$}{Omega5Spin(B(U(5)xZ3))}}
\label{app:bordismU5Z3}

In this appendix, we compute\footnote{We would like to thank I. Garc\'{\i}a Etxebarria for enomours help with this computation.}
\begin{equation}
\Omega_{5}^{\mathrm{Spin}}\!\bigl(B(U(5)\times \mathbb Z_{3})\bigr)\,,
\end{equation}
where $BG$ is the classifying space of group $G$,
using the Atiyah--Hirzebruch spectral sequence (AHSS)
\begin{equation}
E^{2}_{p,q}
=
H_{p}\!\left(B(U(5)\times \mathbb Z_{3});\Omega^{\mathrm{Spin}}_{q}(\mathrm{pt})\right)
\Longrightarrow
\Omega^{\mathrm{Spin}}_{p+q}\!\bigl(B(U(5)\times \mathbb Z_{3})\bigr).
\label{eq:AHSSspinU5Z3}
\end{equation}
Our goal is to determine all contributions in total degree $p+q=5$. Here, we follow the procedure explained in \cite{McCleary1985UsersGT,Garcia-Etxebarria:2018ajm,Davighi:2019rcd,Davighi:2023mzg}.

\section*{1. Low-dimensional spin bordism groups}

In low degrees, the spin bordism groups of a point are
\begin{eqnarray}\nonumber
\Omega^{\mathrm{Spin}}_{0}(\mathrm{pt})&=&\mathbb Z,\qquad
\Omega^{\mathrm{Spin}}_{1}(\mathrm{pt})=\mathbb Z_{2},\qquad
\Omega^{\mathrm{Spin}}_{2}(\mathrm{pt})=\mathbb Z_{2},\qquad
\Omega^{\mathrm{Spin}}_{3}(\mathrm{pt})=0,\\
\Omega^{\mathrm{Spin}}_{4}(\mathrm{pt})&=&\mathbb Z.
\label{eq:spinbordismlow}
\end{eqnarray}
Therefore, for total degree $5$, the only potentially relevant $E^{2}$-terms are
\begin{equation}
E^2_{5,0},\qquad E^2_{4,1},\qquad E^2_{3,2},\qquad E^2_{1,4}.
\end{equation}

\section*{2. Homology of the classifying space}

Since
\begin{equation}
B(U(5)\times \mathbb Z_{3})\simeq BU(5)\times B\mathbb Z_{3},
\label{eq:classifyingspacefactorization}
\end{equation}
we compute the homology of the two factors separately.

For $BU(5)$ one has
\begin{equation}
H_{*}(BU(5);\mathbb Z)=\{\mathbb Z,0,\mathbb Z,0,\mathbb Z^{2},0,\mathbb Z^{3},\ldots\},
\label{eq:BU5homology}
\end{equation}
so in particular
\begin{equation}
H_{2}(BU(5);\mathbb Z)\cong \mathbb Z,\qquad
H_{4}(BU(5);\mathbb Z)\cong \mathbb Z^{2},\qquad
H_{5}(BU(5);\mathbb Z)=0.
\end{equation}
For $B\mathbb Z_{3}$,
\begin{equation}
H_{*}(B\mathbb Z_{3};\mathbb Z)=\{\mathbb Z,\mathbb Z_{3},0,\mathbb Z_{3},0,\mathbb Z_{3},\ldots\},
\label{eq:BZ3homology}
\end{equation}
so
\begin{equation}
H_{1}(B\mathbb Z_{3};\mathbb Z)\cong \mathbb Z_{3},\qquad
H_{3}(B\mathbb Z_{3};\mathbb Z)\cong \mathbb Z_{3},\qquad
H_{5}(B\mathbb Z_{3};\mathbb Z)\cong \mathbb Z_{3}.
\end{equation}

Applying the K\"unneth theorem, and noting that there are no relevant Tor terms because the homology of $BU(5)$ is free abelian in these degrees, we obtain
\begin{align}
H_{5}(BU(5)\times B\mathbb Z_{3};\mathbb Z)
&\cong
H_{0}(BU(5);\mathbb Z)\otimes H_{5}(B\mathbb Z_{3};\mathbb Z)
\nonumber\\
&\quad\oplus
H_{2}(BU(5);\mathbb Z)\otimes H_{3}(B\mathbb Z_{3};\mathbb Z)
\nonumber\\
&\quad\oplus
H_{4}(BU(5);\mathbb Z)\otimes H_{1}(B\mathbb Z_{3};\mathbb Z)
\nonumber\\
&\cong
\mathbb Z_{3}\oplus \mathbb Z_{3}\oplus \mathbb Z_{3}^{2}
\nonumber\\
&\cong
\mathbb Z_{3}^{4}.
\label{eq:H5product}
\end{align}
Similarly,
\begin{equation}
H_{1}(BU(5)\times B\mathbb Z_{3};\mathbb Z)\cong \mathbb Z_{3}.
\label{eq:H1product}
\end{equation}

\section*{3. Mod-2 homology and the relevant \texorpdfstring{$E^2$}{E2}-page entries}

Since $3$ is odd, the positive-degree homology of $B\mathbb Z_{3}$ vanishes with $\mathbb Z_{2}$ coefficients:
\begin{equation}
H_{i}(B\mathbb Z_{3};\mathbb Z_{2})=0
\qquad\text{for }i>0.
\label{eq:BZ3mod2}
\end{equation}
Hence all mod-$2$ contributions come entirely from the $BU(5)$ factor. In particular,
\begin{equation}
H_{4}(BU(5)\times B\mathbb Z_{3};\mathbb Z_{2})
\cong
H_{4}(BU(5);\mathbb Z_{2})
\cong
\mathbb Z_{2}^{2},
\label{eq:H4mod2}
\end{equation}
and
\begin{equation}
H_{3}(BU(5)\times B\mathbb Z_{3};\mathbb Z_{2})=0.
\label{eq:H3mod2}
\end{equation}

Therefore the only potentially nonzero $E^{2}$-terms with $p+q=5$ are
\begin{align}
E^{2}_{5,0}
&=
H_{5}\!\left(B(U(5)\times \mathbb Z_{3});\mathbb Z\right)
\cong
\mathbb Z_{3}^{4},
\nonumber\\
E^{2}_{4,1}
&=
H_{4}\!\left(B(U(5)\times \mathbb Z_{3});\mathbb Z_{2}\right)
\cong
\mathbb Z_{2}^{2},
\nonumber\\
E^{2}_{3,2}
&=
H_{3}\!\left(B(U(5)\times \mathbb Z_{3});\mathbb Z_{2}\right)
=
0,
\nonumber\\
E^{2}_{1,4}
&=
H_{1}\!\left(B(U(5)\times \mathbb Z_{3});\mathbb Z\right)
\cong
\mathbb Z_{3}.
\label{eq:relevantE2terms}
\end{align}

\section*{4. The \texorpdfstring{$d_{2}$}{d2} differential and Steenrod square \texorpdfstring{$Sq^{2}$}{Sq2}}

The only possible $2$-primary contribution in total degree $5$ comes from
\begin{equation}
E^{2}_{4,1}\cong \mathbb Z_{2}^{2}.
\end{equation}
To determine whether this term survives, we study the $d_{2}$-differentials
\begin{equation}
d_{2}:E^{2}_{6,0}\longrightarrow E^{2}_{4,1},
\qquad
d_{2}:E^{2}_{4,1}\longrightarrow E^{2}_{2,2}.
\end{equation}
For spin bordism, the homological differential $d_{2}$ is dual to the cohomological Steenrod square $Sq^{2}$.

The mod-$2$ cohomology ring of $BU(5)$ is
\begin{equation}
H^{*}(BU(5);\mathbb Z_{2})
=
\mathbb Z_{2}[\bar c_{1},\bar c_{2},\bar c_{3},\bar c_{4},\bar c_{5}],
\label{eq:BU5cohomology}
\end{equation}
where $\bar c_i$ denotes the mod-$2$ reduction of the Chern class $c_i$. In low degrees we have
\begin{equation}
H^{2}(BU(5);\mathbb Z_{2})=\langle \bar c_{1}\rangle,
\qquad
H^{4}(BU(5);\mathbb Z_{2})=\langle \bar c_{1}^{2},\bar c_{2}\rangle.
\end{equation}

\subsection*{Outgoing differential from \texorpdfstring{$E^{2}_{4,1}$}{E241}}

The outgoing differential
\begin{equation}
d_{2}:E^{2}_{4,1}\longrightarrow E^{2}_{2,2}
\end{equation}
is dual to
\begin{equation}
Sq^{2}:H^{2}(BU(5);\mathbb Z_{2})\longrightarrow H^{4}(BU(5);\mathbb Z_{2}).
\end{equation}
Since $\bar c_{1}$ has degree $2$, one has
\begin{equation}
Sq^{2}(\bar c_{1})=\bar c_{1}^{2}\neq 0.
\label{eq:Sq2c1}
\end{equation}
Therefore this $Sq^{2}$ map has rank $1$, and hence the dual differential
\begin{equation}
d_{2}:E^{2}_{4,1}\to E^{2}_{2,2}
\end{equation}
also has rank $1$. Since $E^{2}_{4,1}\cong \mathbb Z_{2}^{2}$, it follows that
\begin{equation}
\dim_{\mathbb Z_{2}}\ker\!\left(d_{2}:E^{2}_{4,1}\to E^{2}_{2,2}\right)=1.
\label{eq:kerdim1}
\end{equation}

\subsection*{Incoming differential to \texorpdfstring{$E^{2}_{4,1}$}{E241}}

Next consider
\begin{equation}
d_{2}:E^{2}_{6,0}\longrightarrow E^{2}_{4,1},
\end{equation}
which is dual to
\begin{equation}
Sq^{2}:H^{4}(BU(5);\mathbb Z_{2})\longrightarrow H^{6}(BU(5);\mathbb Z_{2}).
\end{equation}
Using the Cartan formula together with the standard action of Steenrod squares on the mod-$2$ Chern classes, one finds
\begin{equation}
Sq^{2}(\bar c_{1}^{2})=0,
\qquad
Sq^{2}(\bar c_{2})=\bar c_{1}\bar c_{2}+\bar c_{3}\neq 0.
\label{eq:Sq2c2}
\end{equation}
Hence this map also has rank $1$, so
\begin{equation}
\dim_{\mathbb Z_{2}}\operatorname{im}\!\left(d_{2}:E^{2}_{6,0}\to E^{2}_{4,1}\right)=1.
\label{eq:imdim1}
\end{equation}

Since the image of the incoming differential is always contained in the kernel of the outgoing differential, and both have dimension $1$, they must coincide:
\begin{equation}
\operatorname{im}\!\left(d_{2}:E^{2}_{6,0}\to E^{2}_{4,1}\right)
=
\ker\!\left(d_{2}:E^{2}_{4,1}\to E^{2}_{2,2}\right).
\end{equation}
Therefore
\begin{equation}
E^{3}_{4,1}=0.
\label{eq:E41zero}
\end{equation}
No later differential can revive this term, and so
\begin{equation}
E^{\infty}_{4,1}=0.
\end{equation}
Thus no $2$-torsion survives in total degree $5$.

\section*{5. Surviving \texorpdfstring{$3$}{3}-primary pieces and the extension problem}

The surviving graded pieces in total degree $5$ are therefore
\begin{equation}
E^{\infty}_{5,0}\cong \mathbb Z_{3}^{4},
\qquad
E^{\infty}_{1,4}\cong \mathbb Z_{3}.
\label{eq:survivinggradedpieces}
\end{equation}
At the level of the associated graded, this gives five copies of $\mathbb Z_{3}$. The only remaining issue is the extension problem.

To fix the extension, note that there is a canonical inclusion
\begin{equation}
i:B\mathbb Z_{3}\hookrightarrow BU(5)\times B\mathbb Z_{3},
\end{equation}
obtained by sending the $BU(5)$ factor to the basepoint, and a projection
\begin{equation}
p:BU(5)\times B\mathbb Z_{3}\to B\mathbb Z_{3},
\end{equation}
with
\begin{equation}
p\circ i=\mathrm{id}_{B\mathbb Z_{3}}.
\end{equation}
Therefore
\begin{equation}
\Omega_{5}^{\mathrm{Spin}}(B\mathbb Z_{3})
\end{equation}
is a direct summand of
\begin{equation}
\Omega_{5}^{\mathrm{Spin}}(BU(5)\times B\mathbb Z_{3}).
\end{equation}
Using the standard result
\begin{equation}
\Omega_{5}^{\mathrm{Spin}}(B\mathbb Z_{3})\cong \mathbb Z_{9},
\label{eq:BZ3extension}
\end{equation}
we conclude that the bordism group is
\begin{equation}
\Omega_{5}^{\mathrm{Spin}}\!\bigl(B(U(5)\times \mathbb Z_{3})\bigr)
\cong
\mathbb Z_{9}\oplus \mathbb Z_{3}\oplus \mathbb Z_{3}\oplus \mathbb Z_{3}.
\label{eq:finalanswer}
\end{equation}

  \bibliography{RefEHB.bib}
  
  \bibliographystyle{JHEP}
  \end{document}